\documentclass[journal,onecolumn,draftcls,11pt]{IEEEtran}%
\usepackage{amsfonts}
\usepackage{amsmath}
\usepackage{amssymb}
\usepackage{graphicx}
\usepackage{cite}%
\newtheorem{theorem}{Theorem}

\newtheorem{case}{Case}

\newtheorem{definition}{Definition}

\newtheorem{lemma}{Lemma}

\newtheorem{proposition}{Proposition}
\newtheorem{remark}{Remark}

\begin{document}

\title{Opportunistic Communications in Fading Multiaccess Relay Channels}
\author{Lalitha Sankar \IEEEmembership{Member,~IEEE},~Yingbin Liang
\IEEEmembership{Member,~IEEE},~N. B. Mandayam
\IEEEmembership{Senior~Member,~IEEE}, and~H. Vincent Poor
\IEEEmembership{Fellow,~IEEE} \thanks{L. Sankar and H. V. Poor are with the
Department of Electrical\ Engineering, Princeton University, Princeton, NJ
08544, USA. Y. Liang is with the University of Hawaii, Honolulu, HI 96822,
USA. N.~B.~Mandayam is with the WINLAB, Rutgers University, North Brunswick,
NJ 08902, USA. A part of this work was done when L. Sankar was with the
WINLAB, Rutgers University and Y. Liang was with Princeton University.
}\thanks{The work of L.~Sankar, (previously Sankaranarayanan), Y. Liang, and
H. V. Poor was supported by the National Science Foundation under Grants
ANI-03-38807 and CNS-06-25637. The work of N.~B.~Mandayam was supported in
part by the National\ Science Foundation under
Grant~No.~{\scriptsize ITR-0205362}. The material in this paper was presented
in part at the IEEE\ International Symposium on Information Theory, Nice,
France, Jun. 2007.}}
\pubid{~}
\specialpapernotice{~}
\maketitle

\begin{abstract}
The problem of optimal resource allocation is studied for ergodic fading
\textit{orthogonal} multiaccess relay channels (MARCs) in which the users
(sources) communicate with a destination with the aid of a half-duplex relay
that transmits on a channel orthogonal to that used by the transmitting
sources. Under the assumption that the instantaneous fading state information
is available at all nodes, the maximum sum-rate and the optimal user and relay
power allocations (policies) are developed for a decode-and-forward (DF)
relay. With the observation that a DF relay results in two multiaccess
channels, one at the relay and the other at the destination, a single known
lemma on the sum-rate of two intersecting polymatroids is used to determine
the DF\ sum-rate and the optimal user and relay policies. The lemma also
enables a broad topological classification of fading MARCs into one of three
types. The first type is the set of \textit{partially clustered }MARCs where a
user is clustered either with the relay or with the destination such that the
users waterfill on their bottle-neck links to the distant receiver. The second
type is the set of \textit{clustered }MARCs where all users are either
proximal to the relay or to the destination such that opportunistic multiuser
scheduling to one of the receivers is optimal. The third type consists of
\textit{arbitrarily clustered} MARCs which are a combination of the first two
types, and for this type it is shown that the optimal policies are
opportunistic non-waterfilling solutions. The analysis is extended to develop
the rate region of a $K$-user orthogonal half-duplex MARC. Finally, cutset
outer bounds are used to show that DF achieves the capacity region for a class
of clustered orthogonal half-duplex MARCs.

\end{abstract}

\begin{keywords}
Multiple-access relay channel (MARC), decode-and-forward, ergodic capacity.
\end{keywords}

\IEEEpeerreviewmaketitle

\section{Introduction}

Node cooperation in multi-terminal wireless networks has been shown to improve
performance by providing increased robustness to channel variations and by
enabling energy savings (see
\cite{cap_theorems:SEA01,cap_theorems:LWT01,cap_theorems:KGG_IT,cap_theorems:AGS01,cap_theorems:HMZ01,cap_theorems:Liang_Veeravalli02,cap_theorems:SKM05}
and the references therein). A specific example of relay cooperation in
multi-terminal networks is the multi-access relay channel\ (MARC). The MARC is
a network in which several users (source nodes) communicate with a single
destination with the aid of a relay \cite{cap_theorems:KvW01}. The coding
strategies developed for the relay channel \cite{cap_theorems:CEG01} extend
readily to the MARC \cite{cap_theorems:SKM02a}. For example, the strategy of
\cite[Theorem~1]{cap_theorems:CEG01}, now often called
\textit{decode-and-forward} (DF), has a relay that decodes user messages
before forwarding them to the destination
\cite{cap_theorems:KGG_IT,cap_theorems:SKM_journal01}. Similarly, the strategy
in \cite[Theorem~6]{cap_theorems:CEG01}, now often called
\textit{compress-and-forward} (CF), has the relay quantize its output symbols
and transmit the resulting quantized bits to the destination
\cite{cap_theorems:SKM02a}.

We consider a MARC with a half-duplex wireless relay that transmits and
receives in orthogonal channels. Specifically, we model a MARC with a
half-duplex relay as an \textit{orthogonal} MARC in which the relay receives
on a channel over which all the sources transmit, and transmits to the
destination on an orthogonal channel\footnote{Yet another class of orthogonal
single-source half-duplex relay channels is defined in
\cite{cap_theorems:ElGZahedi01} where the source and relay transmit on
orthogonal bands. The source transmits in both bands, one of which is received
at the relay and the other is received at the destination, such that the relay
also transmits on the band received at the destination. In contrast to
\cite{cap_theorems:ElGZahedi01}, we assume that all sources transmit in only
one of orthogonal bands and the relay transmits in the other. Furthermore, we
assume that signals in both bands are received at the destination. \ Later in
the sequel we briefly discuss the general model where the sources transmit on
both bands.}. This channel models a relay-inclusive uplink in a variety of
networks such as wireless LAN, cellular, and sensor networks. The study of
wireless relay channels and networks has focused on several performance
aspects, including capacity
\cite{cap_theorems:CEG01,cap_theorems:KGG_IT,cap_theorems:SEA01}, diversity
\cite{cap_theorems:LWT01,cap_theorems:AGS01,cap_theorems:Yuksel_Erkip01},
outage
\cite{cap_theorems:GunduzErkip_RA,cap_theorems:AvestiTse01,cap_theorems:LSKNBM_02}%
, and cooperative coding
\cite{cap_theorems:RLiu_Spaso_01,cap_theorems:Stefanov_Erkip}. Equally
pertinent is the problem of resource allocation in fading wireless channels
where both source and relay nodes can allocate their transmit power to enhance
a desired performance metric when the fading state information is available.
Resource allocation for a variety of relay channels and networks has been
studied in several papers, including
\cite{cap_theorems:HMZ01,cap_theorems:MY01,cap_theorems:MY02,cap_theorems:YaoCaiGian,cap_theorems:GunduzErkip_RA}%
. A common assumption in all these papers is that the source and relay nodes
are subject to a total power constraint.

For a wireless fading relay channel, i.e., a single-user specialization of a
fading MARC, the problem of resource allocation when the source and relay
nodes are subject to individual power constraints in studied in
\cite{cap_theorems:Liang_Veeravalli02} (see also
\cite{cap_theorems:LiangVP_ResAllocJrnl}). The authors formulate the problem
as a \textit{max-min} optimization. They draw parallels with the classical
minimax optimization in hypothesis testing to show that, depending on the
joint fading statistics, the resource allocation problem results in one of
three solutions. The three solutions broadly correspond to three types of
channel topologies, namely, source-relay clustering, relay-destination
clustering, and the non-clustered (arbitrary) topology.

Resource allocation in multiuser relay networks has been studied recently in
\cite{cap_theorems:Mesbah_01,cap_theorems:OOyman,cap_theorems:Yener_Relay}.
The authors in \cite{cap_theorems:Mesbah_01} and
\cite{cap_theorems:Yener_Relay} consider a specific orthogonal model where the
sources time-duplex their transmissions and are aided in their transmissions
by a half-duplex relay, while in \cite{cap_theorems:OOyman} the optimal
multiuser scheduling is determined under the assumption of a non-fading
backhaul channel between the relay and destination. In contrast, in this
paper, we consider a more general multiaccess channel with a half-duplex relay
and model all inter-node wireless links as ergodic fading channels with
perfect fading information available at all nodes. Assuming a DF relay, we
develop the optimal source and relay power allocations and present the
conditions under which opportunistic time-duplexing of the users is optimal.

The orthogonal MARC is a multiaccess generalization of the orthogonal relay
channel studied in \cite{cap_theorems:Liang_Veeravalli02}; however, the
optimal DF policies developed in \cite{cap_theorems:Liang_Veeravalli02} do not
extend readily to maximize the DF sum-rate of the MARC. This is because unlike
the single-user case, in order to determine the DF sum-rate for the MARC, we
need to consider the intersection of the two multiaccess rate regions that
result from decoding at both the relay and the destination. Here, we exploit
the polymatroid properties of these multiaccess regions and use a single known
lemma on the sum-rate of two intersecting polymatroids \cite[chap.
46]{cap_theorems:Schrijver01} to develop inner (DF) and outer bounds on the
sum-rate and the rate region and specify the sub-class of orthogonal MARCs for
which the DF bounds are tight.

A lemma in \cite[chap. 46]{cap_theorems:Schrijver01} enables us to classify
polymatroid intersections broadly into two sets, namely, the sets of
\textit{active }and \textit{inactive cases}. An active or an inactive case
result when, in the region of intersection, the constraints on the $K$-user
sum-rate at both receivers are active or inactive, respectively. In the sequel
we show that inactive cases suggest \textit{partially clustered }topologies
where a subset of users is clustered closer to one of the receivers while the
complementary subset is closer to the remaining receiver. On the other hand,
active cases can result from specific \textit{clustered }topologies such as
those in which all sources and the relay are clustered or those in which the
relay and the destination are clustered, or more generally, from
\textit{arbitrarily clustered} topologies that are either a combination of the
two clustered models or of a clustered and a partially clustered model. For
both the active and inactive cases, the polymatroid intersection lemma yields
closed form expressions for the sum-rates which in turn allow one to develop
the sum-rate optimal power allocations (policies).

We first study the two-user orthogonal MARC and develop the DF sum-rate
maximizing power policies. Using the polymatroid intersection lemma we show
that the fading-averaged DF sum-rate is achieved by either one of five
disjoint cases, two inactive and three active, or by a \textit{boundary case}
that lies at the boundary of an active and an inactive case. We develop the
sum-rate for all cases and show that the sum-rate maximizing DF power policy
either: 1) exploits the multiuser fading diversity to opportunistically
schedule users analogously to the fading MAC
\cite{cap_theorems:Knopp_Humblet,cap_theorems:TH01} though the optimal
multiuser policies are not necessarily water-filling solutions, or 2) involves
simultaneous water-filling over two independent point-to-point links. Using
similar techniques, we also develop the two-user DF rate region.

Next, we generalize the two-user sum-rate and rate region analysis to the
$K$-user channel and show that the inactive, active, and boundary cases
correspond to partially clustered, clustered, and arbitrarily clustered
topologies, respectively. Finally, we develop the cutset outer bounds on the
sum-capacity of an ergodic fading orthogonal and non-orthogonal $K$-user
Gaussian MARC. We show that DF achieves the sum-capacity for a class of
half-duplex MARCs in which the sources and relay are clustered such that the
outer bound on the $K$-user sum-rate at the destination dominates all other
sum-rate outer bounds. We also show that DF achieves the capacity region when
the cutset bounds at the destination are the dominant bounds for all rate
points on the boundary of the outer bound rate region.

In the course of developing the main results of this paper, we also show that
DF achieves the capacity region of a class of \textit{degraded} discrete
memoryless and Gaussian non-fading orthogonal MARCs\ where the received signal
at the destination is physically degraded with respect to that at the relay
conditioned on the transmit signal at the relay. The relatively few capacity
results known for specific classes of full-duplex single-user relay channels,
such as those for degraded relay channels \cite[Theorem 5]{cap_theorems:CEG01}
and for a class of orthogonal relay channels \cite{cap_theorems:ElGZahedi01},
have not been straightforward to extend to the MARC. The result developed here
is the first in which the entire capacity region is given for a class of
degraded MARCs. In contrast, in \cite{cap_theorems:LSMandPoor_01} it is shown
that DF achieves the sum-capacity of a class of full-duplex degraded
Gaussian\ MARCs for which the polymatroid intersections at the relay and
destination belong to the active set.

The paper is organized as follows. In Section \ref{Section 2}, we present the
channel models and introduce polymatroids and a lemma on their intersections.
In Section \ref{section 3} we develop the DF rate region for ergodic fading
orthogonal MARCs. In\ Section \ref{Sec_4} we develop the power policies that
maximize the DF sum-rate for a two-user MARC. We extend the analysis to the
$K$-user orthogonal MARC as well as to non-orthogonal models in Section
\ref{Sec_5}. In Section \ref{Sec_OB}, we present outer bounds and illustrate
our results numerically. We summarize our contributions in Section
\ref{Sec_End}.

\section{\label{Section 2}Channel Model and Preliminaries}

\subsection{Orthogonal Half-Duplex MARC}

\begin{figure}
[ptb]
\begin{center}
\includegraphics[
height=1.6855in,
width=3.1168in
]%
{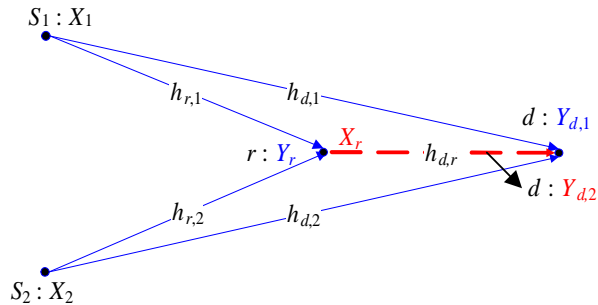}%
\caption{A two-user orthogonal MARC.}%
\label{Fig_1_model}%
\end{center}
\end{figure}

A $K$-user MARC consists of $K$ source nodes numbered $1,2,\ldots,K$, a relay
node $r\,$,$\ $and a destination node $d$. We write $\mathcal{K}=\left\{
1,2,\ldots,K\right\}  $ to denote the set of sources, $\mathcal{T}%
=\mathcal{K}\cup\left\{  r\right\}  $ to denote the set of transmitters, and
$\mathcal{D}=\left\{  r,d\right\}  $ to denote the set of receivers. In an
orthogonal MARC, the sources transmit to the relay and destination on one
channel, say channel 1, while the half-duplex relay transmits to the
destination on an orthogonal channel 2 as shown in Fig. \ref{Fig_1_model}.
Thus, a fraction $\theta$ of the total bandwidth resource is allocated to
channel 1 while the remaining fraction $\overline{\theta}$ $=$ $1-\theta$ is
allocated to channel 2. In the fraction $\theta$, the source $k$, for all
$k\in\mathcal{K}$, transmits the signal $X_{k}$ while the relay and the
destination receive $Y_{r}$ and $Y_{d,1}$ respectively. In the fraction
$\overline{\theta}$, the relay transmits $X_{r}$ and the destination receives
$Y_{d,2}$ where the sources precede the relay in the transmission order. In
each symbol time (channel use), we thus have
\begin{align}
&
\begin{array}
[c]{cc}%
Y_{r}=\sum_{k=1}^{K}H_{r,k}X_{k}+Z_{r}, &
\end{array}
\label{GMARC_defn1}\\
&
\begin{array}
[c]{cc}%
Y_{d,1}=\sum_{k=1}^{K}H_{d,k}X_{k}+Z_{d,1}, & \text{and}%
\end{array}
\label{GMARC_defn2}\\
&
\begin{array}
[c]{cc}%
Y_{d,2}=H_{d,r}X_{r}+Z_{d,2}, &
\end{array}
\label{GMARC_defn3}%
\end{align}
where $Z_{r},$ $Z_{d,1},$ and $Z_{d,2}$ are independent circularly symmetric
complex Gaussian noise random variables with zero means and unit variances. We
write \underline{$H$} to denote a vector of fading gains, $H_{k,m}$, for all
$k$ $\in$ $\mathcal{D}$ and $m$ $\in$ $\mathcal{T}$, $k$ $\not =$ $m$, such
that \underline{$h$} is a realization for a given channel use of a jointly
stationary and ergodic (not necessarily Gaussian) fading process $\left\{
\underline{H}\right\}  $. Note that the channel gains $H_{k,m}$, for all
$k,m$, are not assumed to be independent. We assume that the fraction $\theta$
is fixed \textit{a priori} and is known at all nodes. Since the relay is
assumed to be causal, we note that the signal $X_{r}$ at the relay in each
channel use depends causally only on the $Y_{r}$ received in the previous
channel uses.

Over $n$ uses of the channel, the source and relay transmit sequences
$\left\{  X_{k,i}\right\}  $ and $\left\{  X_{r,i}\right\}  $, respectively,
are constrained in power according to%
\begin{equation}
\left.  \sum\limits_{i=1}^{n}\left\vert X_{k,i}\right\vert ^{2}\leq
n\overline{P}_{k}\right.  ,\text{ for all }k\in\mathcal{T}\text{.}
\label{GMARC_Pwr_def0}%
\end{equation}
Since the sources and relay know the fading states of the links on which they
transmit, they can allocate their transmitted signal power according to the
channel state information. We write $P_{k}(\underline{H})$ to denote the power
allocated at the $k^{th}$ transmitter, for all $k\in\mathcal{T}$, as a
function of the channel states $\underline{H}$. For an ergodic fading channel,
(\ref{GMARC_Pwr_def0}) then simplifies to
\begin{equation}
\left.  \mathbb{E}\left[  P_{k}(\underline{H})\right]  \leq\overline{P}%
_{k}\right.  \text{ for all }k\in\mathcal{T} \label{GMARC_Pwr_defn}%
\end{equation}
where the expectation in (\ref{GMARC_Pwr_defn}) is over the distribution of
\underline{$H$}. We write $\underline{P}\left(  \underline{H}\right)  $ to
denote a vector of power allocations with entries $P_{k}(\underline{H})$ for
all $k$ $\in$ ${\mathcal{T}}$, and define $\mathcal{P}$ to be the set of all
$\underline{P}\left(  \underline{H}\right)  $ whose entries satisfy
(\ref{GMARC_Pwr_defn}). Throughout the sequel, we refer to the fractions
$\theta$ and $1-\theta$ as the first and second fractions, respectively.%

%TCIMACRO{\FRAME{ftbpFU}{3.1168in}{1.6855in}{0pt}{\Qcb{A two-user orthogonal
%MARC.}}{\Qlb{Fig_1_model}}{channel_model_twousers_alt.eps}%
%width 3.1168in;  height 1.6855in;  depth 0pt;  original-width 2.7345in;
%original-height 1.4659in;  cropleft "0";  croptop "1";  cropright "1";
%filename 'Channel_model_twousers_alt.eps';file-properties "XNPEU";}}}%
%BeginExpansion
%EndExpansion

\subsection{Polymatroids}

In the sequel, we use the properties of polymatroids to develop the ergodic
sum-rate results. Polymatroids have been used to develop capacity
characterizations for a variety of multiple-access channel models including
the MARC (see for e.g.,
\cite{cap_theorems:Han01,cap_theorems:TH01,cap_theorems:SKM_journal01}).
Furthermore, in \cite{cap_theorems:Han01}, Han demonstrates that for certain
multi-terminal channels, polymatroid intersections need to be considered. To
the best of our knowledge, this is the first work where the polymatroid
intersection lemma has been used to explicitly characterize sum-rates and
sum-capacity, where possible. We review the following definition of a polymatroid.

\begin{definition}
\label{Def_PolyM}Let $\mathcal{K}=\left\{  1,2,\ldots,K\right\}  $ and
$f=2^{K}\rightarrow\mathfrak{R}_{+}$ be a set function. The polyhedron
\begin{equation}
\mathcal{B}\left(  f\right)  \equiv\left\{  \left(  R_{1},R_{2},\ldots
,R_{K}\right)  :R_{\mathcal{S}}\leq f\left(  \mathcal{S}\right)  ,\text{ for
all }\mathcal{S}\subseteq\mathcal{K}\text{, }R_{k}\geq0\right\}
\end{equation}
is a polymatroid if $f$ satisfies
\end{definition}

\begin{enumerate}
\item $f\left(  \emptyset\right)  =0$ (normalization)

\item $f\left(  \mathcal{S}\right)  \leq f\left(  \mathcal{P}\right)  $ if
$\mathcal{S\subset P}$ (monotonicity)

\item $f\left(  \mathcal{S}\right)  +f\left(  \mathcal{P}\right)  \geq
f\left(  \mathcal{S\cup P}\right)  +f\left(  \mathcal{S\cap P}\right)  $ (submodularity).
\end{enumerate}

\begin{remark}
The submodularity property in Definition \ref{Def_PolyM} above is equivalent
to requiring, for all $k_{1},k_{2}$ in $\mathcal{K}$ with $k_{1}\neq k_{2}$,
$k_{1}\notin\mathcal{S}$, $k_{2}\notin\mathcal{S}$, that $f$ satisfies
\cite[Ch.~44]{cap_theorems:Schrijver01}%
\begin{equation}
f(\mathcal{S}\cup\{k_{1}\})+f(\mathcal{S}\cup\{k_{2}\})\geq f(\mathcal{S}%
)+f(\mathcal{S}\cup\{k_{1},k_{2}\}.
\end{equation}
This property is used in \cite{cap_theorems:SKM_journal01} to show that the
rate regions achieved at both the relay and the destination in a full-duplex
MARC are polymatroids.
\end{remark}

We use the following lemma on polymatroid intersections to develop optimal
inner and outer bounds on the sum-rate for $K$-user half-duplex MARCs.

\begin{lemma}
[{\cite[p. 796, Cor. 46.1c]{cap_theorems:Schrijver01}}]%
\label{Lemma_PolyIntersect}Let $R_{\mathcal{S}}\leq f_{1}\left(
\mathcal{S}\right)  $ and $R_{\mathcal{S}}\leq f_{2}\left(  \mathcal{S}%
\right)  $, for all $\mathcal{S}\subseteq\mathcal{K}$, be two polymatroids
such that $f_{1}$ and $f_{2}$ are nondecreasing submodular set functions on
$\mathcal{K}$ with $f_{1}\left(  \emptyset\right)  =f_{2}\left(
\emptyset\right)  =0$. Then
\begin{equation}
\max R_{\mathcal{K}}=\min\limits_{\mathcal{S}\subseteq\mathcal{K}}\left(
f_{1}\left(  \mathcal{S}\right)  +f_{2}\left(  \mathcal{K}\backslash
\mathcal{S}\right)  \right)  . \label{DG_Lemma_RK}%
\end{equation}

\end{lemma}

Lemma \ref{Lemma_PolyIntersect} states that the maximum $K$-user sum-rate
$R_{\mathcal{K}}$ that results from the intersection of two polymatroids,
$R_{\mathcal{S}}\leq f_{1}\left(  \mathcal{S}\right)  $ and $R_{\mathcal{S}%
}\leq f_{2}\left(  \mathcal{S}\right)  $, is given by the minimum of the two
$K$-user sum-rate planes $f_{1}\left(  \mathcal{K}\right)  $ and $f_{2}\left(
\mathcal{K}\right)  $ only if both sum-rates are at most as large as the sum
of the orthogonal rate planes $f_{1}\left(  \mathcal{S}\right)  $ and
$f_{2}\left(  \mathcal{K}\backslash\mathcal{S}\right)  $, for all
$\emptyset\not =\mathcal{S}\subset\mathcal{K}$. We refer to the resulting
intersection as belonging to the set of \textit{active cases}.

When there exists at least one $\emptyset\not =\mathcal{S}\subset\mathcal{K}$
for which the above condition is not true, an \textit{inactive case} is said
to result. For such cases, the maximum sum-rate in (\ref{DG_Lemma_RK}) is the
sum of two orthogonal rate planes achieved by two complementary subsets of
users. As a result, the $K$-user sum-rate bounds $f_{1}(\mathcal{K})$ and
$f_{2}(\mathcal{K})$ are no longer active for this case, and thus, the region
of intersection is no longer a polymatroid with $2^{K}-1$ faces. For a
$K$-user MARC, there are $2^{K}-2$ possible inactive cases.

The intersection of two polymatroids can also result in a \textit{boundary
case} when for any $\mathcal{S\subset K}$, $f_{1}\left(  \mathcal{S}\right)
+f_{2}\left(  \mathcal{K}\backslash\mathcal{S}\right)  $ is equal to one or
both of the $K$-user sum-rate planes. The orthogonality of the planes
$f_{1}\left(  \mathcal{S}\right)  $ and $f_{2}\left(  \mathcal{K}%
\backslash\mathcal{S}\right)  $ implies that no two inactive cases have a
boundary and thus a boundary case always arises between an inactive and an
active case. Note that by definition, a boundary case is also an active case
though for ease of exposition, throughout the sequel we explicitly distinguish
between them. From (\ref{DG_Lemma_RK}), there are three possible active cases
corresponding to the three cases in which the sum-rate plane at one of the
receivers is smaller than, larger than, or equal to that at the other. In
fact, the case in which the sum-rates are equal is also a boundary case
between the other two active cases. Thus, there are a total of $\left(
2^{K}-1\right)  $ boundary cases for each active case.

In summary, the \textit{inactive set }consists of all intersections for which
the constraints on the two sum-rates are not active, i.e., no rate tuple on
the sum-rate plane achieved at one of the receivers lies within or on the
boundary of the rate region achieved at the other receiver. On the other hand,
the intersections for which there exists at least one such rate tuple such
that the two sum-rates constraints are active belong to the \textit{active
set}. Thus, by definition, the active set also includes those \textit{boundary
cases} between the active and inactive cases for which there is exactly one
such rate pair.\ 

\section{\label{section 3}Two-User Orthogonal MARC: Ergodic DF\ Rate Region}

The DF\ rate region for a discrete memoryless MARC and a full-duplex relay is
developed in \cite[Appendix A]{cap_theorems:KGG_IT} (see
\cite{cap_theorems:SKM_journal01} for a detailed proof). For this model,
$X_{k}$, $k\in\mathcal{T}$, denotes the transmit signals at the sources and
relay and $Y_{r}$ and $Y_{d}$, denote the received signals at the relay and
destination, respectively. The rate region is achieved using block Markov
encoding and backward decoding. The following proposition summarizes the DF
rate region.

\begin{proposition}
[{\cite[Appendix I]{cap_theorems:SKM_journal01}}]\label{Prop_MARC_DF}The DF
rate region is the union of the set of rate tuples $(R_{1},R_{2},\ldots,$
$R_{K})$ that satisfy, for all $\mathcal{S}\subseteq\mathcal{K}$,%
\begin{equation}
R_{\mathcal{S}}\leq\min\left\{  I(X_{\mathcal{S}};Y_{r}|X_{\mathcal{S}^{c}%
}V_{\mathcal{K}}X_{r}U),I(X_{\mathcal{S}}X_{r};Y_{d}|X_{\mathcal{S}^{c}%
}V_{\mathcal{S}^{c}}U)\right\}  \label{Prop_DF_rateregion}%
\end{equation}
where the union is over all distributions that factor as%
\begin{equation}
p(u)\cdot\left(
%TCIMACRO{\tprod \nolimits_{k=1}^{K}}%
%BeginExpansion
{\textstyle\prod\nolimits_{k=1}^{K}}
%EndExpansion
p(v_{k}|u)p(x_{k}|v_{k},u)\right)  \cdot p(x_{r}|v_{\mathcal{K}},u)\cdot
p(y_{r},y_{d}|x_{\mathcal{T}}). \label{Prop_inp_dist}%
\end{equation}

\end{proposition}

\begin{remark}
The \textit{time-sharing} random variable $U$ ensures that the region of
Theorem \ref{Prop_MARC_DF} is convex.
\end{remark}

\begin{remark}
The independent auxiliary random variables $V_{k}$, $k=1,2,\ldots,K$, help the
sources cooperate with the relay.
\end{remark}

In \cite{cap_theorems:SKM02a} (see also \cite[Proposition 2.5]%
{cap_theorems:LalithaSankar}), the DF\ rate bounds for a discrete memoryless
MARC with a half-duplex relay are developed. For the orthogonal MARC model
studied, since the sources and relay transmit on orthogonal channels, the need
for auxiliary random variables $V_{k}$, for all $k$, that model the coherent
combining gains is eliminated. Under the assumption that the transmit
(bandwidth) fractions $\theta$ and $1-\theta$ at the users and relay,
respectively, are known at all nodes, the following proposition summarizes the
DF rate region for the orthogonal half-duplex MARC.

\begin{proposition}
The DF rate region of a orthogonal MARC is the union of the set of rate tuples
$(R_{1},R_{2},\ldots,$ $R_{K})$ that satisfy, for all $\mathcal{S}%
\subseteq\mathcal{K}$,
\begin{equation}
R_{\mathcal{S}}\leq\min\left\{  \theta I(X_{\mathcal{S}};Y_{r}|X_{\mathcal{S}%
^{c}}U),\theta I(X_{\mathcal{S}};Y_{d,1}|X_{\mathcal{S}^{c}}U)+\overline
{\theta}I(X_{r};Y_{d,2}|U)\right\}  \label{DF_HD_Rates}%
\end{equation}
where the union is over all distributions that factor as%
\begin{equation}
p(u)\cdot\left(  \theta\left[
%TCIMACRO{\tprod \nolimits_{k=1}^{K}}%
%BeginExpansion
{\textstyle\prod\nolimits_{k=1}^{K}}
%EndExpansion
p(x_{k}|u)\right]  \cdot p(y_{r},y_{d}|x_{\mathcal{K}})+\overline{\theta
}p(x_{r}|u)\cdot p(y_{d}|x_{r})\right)  . \label{DF_HD_dist}%
\end{equation}

\end{proposition}

\begin{definition}
A parallel MARC is a collection of $M$ MARCs, for which the inputs and outputs
of parallel channel (sub-channel) $j$, $j=1,2,\ldots,M$, are $X_{k,j}$,
$k\in\mathcal{K}\cup\left\{  r\right\}  $ and $\left(  Y_{r,j},Y_{d,j}\right)
$, respectively, such that conditioned on its inputs, the outputs of each
sub-channel are independent of the inputs and outputs of other sub-channels.
\end{definition}

\begin{theorem}
\label{Th_Par_DF}For the parallel MARC, the DF rate region is the union of the
set of rate tuples $(R_{1},R_{2},\ldots,$ $R_{K})$ that satisfy, for all
$\mathcal{S}\subseteq\mathcal{K}$,%
\begin{multline}
R_{\mathcal{S}}\leq\min\left\{  \sum\nolimits_{m=1}^{M}I(X_{\mathcal{S}%
,m};Y_{r,m}|X_{\mathcal{S}^{c},m}V_{\mathcal{K},m}X_{r,m}U_{m}),\right.
\label{DF_par_rates}\\
\left.  \sum\nolimits_{m=1}^{M}I(X_{\mathcal{S}.,m}X_{r,m};Y_{d,m}%
|X_{\mathcal{S}^{c},m}V_{\mathcal{S}^{c},m}U_{m})\right\}
\end{multline}
where the union is over all distributions that factor as%
\begin{equation}%
%TCIMACRO{\tprod \nolimits_{m=1}^{M}}%
%BeginExpansion
{\textstyle\prod\nolimits_{m=1}^{M}}
%EndExpansion
\left(  p(u_{m})\cdot%
%TCIMACRO{\tprod \nolimits_{k=1}^{K}}%
%BeginExpansion
{\textstyle\prod\nolimits_{k=1}^{K}}
%EndExpansion
p(v_{k,m},x_{k,m}|u_{m})p(y_{r,m},y_{d,m}|x_{\mathcal{T},m})\right)  .
\end{equation}

\end{theorem}

\begin{proof}
The inner bounds in (\ref{DF_par_rates}) are obtained by setting $U=\left(
U_{1}\text{ }U_{2}\text{ }\ldots\text{ }U_{M}\right)  $, $V_{k}=\left(
V_{k,1}\text{ }V_{k,2}\text{ }\ldots\text{ }V_{k,M}\right)  $, $X_{k}=\left(
X_{k,1}\text{ }X_{k,2}\text{ }\ldots\text{ }X_{k,M}\right)  $, $Y_{r}=\left(
Y_{r,1}\text{ }Y_{r,2}\text{ }\ldots\text{ }Y_{r,M}\right)  $, and
$Y_{d}=\left(  Y_{d,1}\text{ }Y_{d,2}\text{ }\ldots\text{ }Y_{d,M}\right)  $,
in (\ref{Prop_DF_rateregion}) and choosing $\left(  U_{m},V_{\mathcal{K}%
,m},X_{\mathcal{K},m}\right)  $ to be independent for all $m$.
\end{proof}

For the (half-duplex) orthogonal Gaussian MARC with a fixed $\underline{H}$
and $\theta$ that is assumed known at all nodes, we consider Gaussian
signaling with zero mean and variance $P_{k}$ at transmitter $k$ such that
$X_{k}\sim C\mathcal{N}(0,P_{k})$, for all $k\in\mathcal{K}$. Thus, from
(\ref{DF_HD_Rates}) the DF rate region includes the set of all rate pairs
$(R_{1},R_{2})$ that satisfy%
\begin{equation}
R_{k}\leq\min\left\{  \theta C\left(  \frac{\left\vert H_{d,k}\right\vert
^{2}P_{k}}{\theta}\right)  +\overline{\theta}C\left(  \frac{\left\vert
H_{d,r}\right\vert ^{2}P_{r}}{\overline{\theta}}\right)  ,\theta C\left(
\frac{\left\vert H_{r,k}\right\vert ^{2}P_{k}}{\theta}\right)  \right\}
,k=1,2 \label{GMARC_R1_fixh}%
\end{equation}
and
\begin{equation}
R_{1}+R_{2}\leq\min\left\{  \theta C\left(  \sum\limits_{k=1}^{2}%
\frac{\left\vert H_{d,k}\right\vert ^{2}P_{k}}{\theta}\right)  +\overline
{\theta}C\left(  \frac{\left\vert H_{d,r}\right\vert ^{2}P_{r}}{\overline
{\theta}}\right)  ,\theta C\left(  \sum\limits_{k=1}^{2}\frac{\left\vert
H_{r,k}\right\vert ^{2}P_{k}}{\theta}\right)  \right\}  .
\label{GMARC_R12_fixh}%
\end{equation}

For a stationary and ergodic process $\left\{  \underline{H}\right\}  $, the
channel in (\ref{GMARC_defn1})-(\ref{GMARC_defn3}) can be modeled as a set of
parallel Gaussian orthogonal MARCs, one for each fading instantiation
$\underline{H}$. For a power policy $\underline{P}(\underline{H})$, the DF
rate bounds for this ergodic fading channel are obtained from Theorem
\ref{Th_Par_DF} by averaging the bounds in (\ref{GMARC_R1_fixh}) and
(\ref{GMARC_R12_fixh}) over all channel realizations. The ergodic fading DF
rate region, $\mathcal{R}_{DF}$, achieved over all $\underline{P}%
(\underline{H})\in\mathcal{P}$, for a fixed bandwidth fraction $\theta$, is
summarized by the following theorem.

\begin{theorem}
\label{DF_Th1}The DF rate region $\mathcal{R}_{DF}$ of an ergodic fading
orthogonal Gaussian MARC is
\begin{equation}
\mathcal{R}_{DF}=\bigcup\limits_{\underline{P}\in\mathcal{P}}\left\{
\mathcal{R}_{r}\left(  \underline{P}\right)  \cap\mathcal{R}_{d}\left(
\underline{P}\right)  \right\}  \label{GMARC_R_DF}%
\end{equation}
where, for all $\mathcal{S}\subseteq\mathcal{K}$, we have
\begin{equation}
\mathcal{R}_{r}\left(  \underline{P}\right)  =\left\{  \left(  R_{1}%
,R_{2}\right)  :R_{\mathcal{S}}\leq\mathbb{E}\left[  \theta C\left(
\frac{\sum\limits_{k\in\mathcal{S}}\left\vert H_{r,k}\right\vert ^{2}%
P_{k}(\underline{H})}{\theta}\right)  \right]  \right\}  \label{DF_Rates_rel}%
\end{equation}
and%
\begin{equation}
\mathcal{R}_{d}\left(  \underline{P}\right)  =\left\{  \left(  R_{1}%
,R_{2}\right)  :R_{\mathcal{S}}\leq\mathbb{E}\left[  \theta C\left(
\frac{\sum\limits_{k\in\mathcal{S}}\left\vert H_{d,k}\right\vert ^{2}%
P_{k}(\underline{H})}{\theta}\right)  +\overline{\theta}C\left(
\frac{\left\vert H_{d,r}\right\vert ^{2}P_{r}(\underline{H})}{\overline
{\theta}}\right)  \right]  \right\}  . \label{DF_Rates_dest}%
\end{equation}

\end{theorem}

\begin{proof}
The proof follows from the observation that the channel in (\ref{GMARC_defn1}%
)-(\ref{GMARC_defn3}) can be modeled as a set of parallel Gaussian orthogonal
MARCs, one for each fading instantiation $\underline{H}$. Thus, from Theorem
\ref{Th_Par_DF}, for Gaussian inputs and for each $\underline{P}(\underline
{H})\in\mathcal{P}$, the regions $\mathcal{R}_{r}\left(  \underline
{P}(\underline{H})\right)  $ and $\mathcal{R}_{d}\left(  \underline
{P}(\underline{H})\right)  $ are given by the bounds in (\ref{DF_Rates_rel})
and (\ref{DF_Rates_dest}), respectively. The DF rate region, $\mathcal{R}%
_{DF}$, is given by the union of such intersections, one for each
$\underline{P}(\underline{H})\in\mathcal{P}$. The convexity of $\mathcal{R}%
_{DF}$ follows from the convexity of the set $\mathcal{P}$ and the concavity
of the $\log$ function. Consider two rate tuples $\left(  R_{1}^{\prime}%
,R_{2}^{\prime}\right)  $ and $\left(  R_{1}^{\prime\prime},R_{2}%
^{\prime\prime}\right)  $ that result from the policies \underline{$P$%
}$^{\prime}(\underline{H})$ and \underline{$P$}$^{\prime\prime}(\underline
{H})$, respectively. For any $\lambda>0$ such that $\overline{\lambda
}=1-\lambda$, and for all $k=1,2$, from (\ref{DF_Rates_rel}), we bound
$R_{k}=\lambda R_{k}^{\prime}+\left(  1-\lambda\right)  R_{k}^{\prime\prime}$
achieved at the relay as%
\begin{align}
R_{k}  &  \leq\lambda\theta\mathbb{E}\left[  C\left(  \frac{\left\vert
H_{r,k}\right\vert ^{2}P_{k}^{\prime}(\underline{H})}{\theta}\right)  \right]
+\overline{\lambda}\theta\mathbb{E}\left[  C\left(  \frac{\left\vert
H_{r,k}\right\vert ^{2}P_{k}^{\prime\prime}(\underline{H})}{\theta}\right)
\right] \\
&  \leq\theta\mathbb{E}\left[  C\left(  \frac{\left\vert H_{r,k}\right\vert
^{2}\left(  \lambda P_{k}^{\prime}(\underline{H})+\overline{\lambda}%
P_{k}^{\prime\prime}(\underline{H})\right)  }{\theta}\right)  \right]
\label{Concavity_proof}\\
&  \leq\theta\mathbb{E}\left[  C\left(  \frac{\left\vert H_{r,k}\right\vert
^{2}P_{k}(\underline{H})}{\theta}\right)  \right]  \label{Concavity_proof2}%
\end{align}
where (\ref{Concavity_proof}) follows from Jensen's inequality and
(\ref{Concavity_proof2}) follows from the convexity of the set $\mathcal{P}$
such that $\underline{P}(\underline{H})=\left(  \lambda\underline{P}^{\prime
}(\underline{H})+\overline{\lambda}\underline{P}^{\prime\prime}(\underline
{H})\right)  \in\mathcal{P}$. Thus, we see that the bound on $R_{k}$ is
achievable. One can similarly bound the sum-rate $R_{1}+R_{2}$ achieved at the
relay thus proving that the tuple $\left(  R_{1},R_{2}\right)  \in
\mathcal{R}_{r}$. The same approach also allows us to show that $\left(
R_{1},R_{2}\right)  \in\mathcal{R}_{d}$, thus proving that $R_{DF}$ is convex.
\end{proof}

\begin{proposition}
$\mathcal{R}_{r}\left(  \underline{P}(\underline{H})\right)  $ and
$\mathcal{R}_{d}\left(  \underline{P}(\underline{H})\right)  $ are polymatroids.
\end{proposition}

\begin{proof}
In \cite[Sec. IV.B]{cap_theorems:SKM_journal01}, it is shown that for each
choice of the input distribution in (\ref{Prop_inp_dist}), the DF rate region
in (\ref{Prop_DF_rateregion}) is an intersection of two polymatroids, one
resulting from the bounds at the relay and the other from the bounds at the
destination. For the orthogonal MARC, the bounds in (\ref{DF_HD_Rates}),
relative to (\ref{Prop_DF_rateregion}), involve a weighted sum of mutual
information expressions; using the same approach as in \cite[Sec.
IV.B]{cap_theorems:SKM_journal01}, the submodularity of these expressions can
be verified in a straightforward manner.
\end{proof}

\begin{remark}
The DF rate region given by (\ref{GMARC_R1_fixh}) and (\ref{GMARC_R12_fixh})
is achieved using block Markov encoding at the sources. For the ergodic fading
model, the rates in Theorem \ref{DF_Th1} are obtained assuming that all fading
instantiations are seen in each such block.
\end{remark}

In the following section, we develop sum-rate optimal DF power policies.

\section{\label{Sec_4}Two-User Orthogonal MARC: DF Sum-Rate Optimal Power
Policy}

For ease of notation, throughout the sequel, we write $R_{\mathcal{A},j}$ to
denote the sum-rate bound on the users in $\mathcal{A}$ and $\mathcal{R}%
_{\mathcal{A},j}^{\min}$ to denote the sum-rate obtained by successively
decoding the users in $\mathcal{K}\backslash\mathcal{A}$ before decoding those
in $\mathcal{A}$ at receiver $j=r,d$. For the two-user case, $R_{\mathcal{K}%
,j}$ and $R_{\mathcal{A},j}$, for all $\mathcal{A}\subset\mathcal{K}$ are
given by the sum-rate and single-user bounds in (\ref{DF_Rates_rel}) and
(\ref{DF_Rates_dest}) at the relay and destination, respectively. The rate
$\mathcal{R}_{\mathcal{A},j}^{\min}$ $=$ $R_{\mathcal{K},j}-R_{\mathcal{K}%
\backslash\mathcal{A},j}$, for all $\mathcal{A}\subset\mathcal{K}$, is
obtained by successively decoding the users in $\mathcal{K}\backslash
\mathcal{A}$ before decoding those in $\mathcal{A}$ at the corner points of
the regions $\mathcal{R}_{r}$ and $\mathcal{R}_{d}$.

\begin{figure*}[tbp] \centering
%EndExpansion%
%TCIMACRO{\FRAME{itbpFU}{5.3125in}{2.3644in}{0in}{\Qcb{{}}}{}%
%{case1_case2_marc.eps}{\special{ language "Scientific Word";  type "GRAPHIC";
%display "USEDEF";  valid_file "F";  width 5.3125in;  height 2.3644in;
%depth 0in;  original-width 4.817in;  original-height 2.7423in;  cropleft "0";
%filename 'Case1_Case2_MARC.eps';file-properties "XNPEU";}}}%
%BeginExpansion
{\parbox[b]{5.3125in}{\begin{center}
\includegraphics[
height=2.3644in,
width=5.3125in
]%
{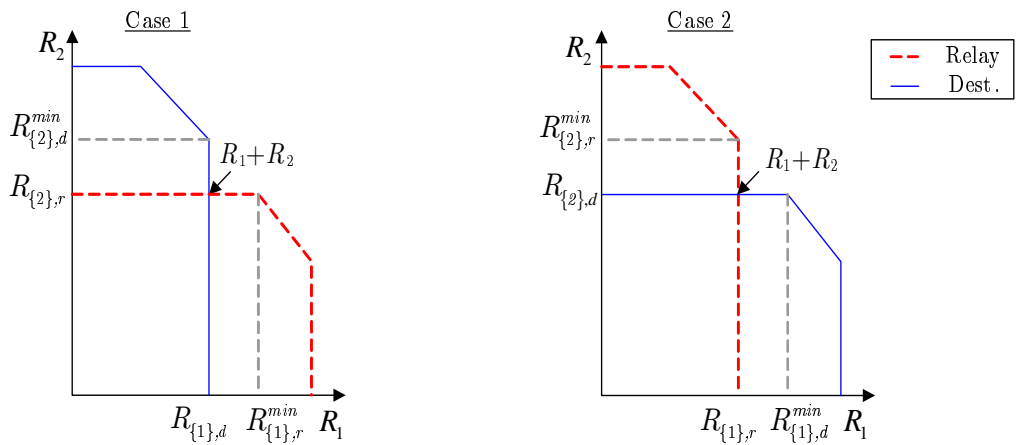}%
\\
{}%
\end{center}}}%
%EndExpansion
\caption{Rate regions $R_{r}(P)$ and $R_{d}(P)$ and sum-rate for case 1 and case 2.}\label{Fig_Case12}%
%TCIMACRO{\TeXButton{E}{\end{figure*}}}%
%BeginExpansion
\end{figure*}

\begin{figure*}[tbp] \centering
%EndExpansion%
%TCIMACRO{\FRAME{itbpFU}{5.9101in}{2.1577in}{0in}{\Qcb{{}}}{}%
%{case3abc_marc.eps}{\special{ language "Scientific Word";  type "GRAPHIC";
%display "USEDEF";  valid_file "F";  width 5.9101in;  height 2.1577in;
%depth 0in;  original-width 0pt;  original-height 0pt;  cropleft "0";
%filename 'Case3abc_MARC.eps';file-properties "XNPEU";}}}%
%BeginExpansion
{\parbox[b]{5.9101in}{\begin{center}
\includegraphics[
height=2.1577in,
width=5.9101in
]%
{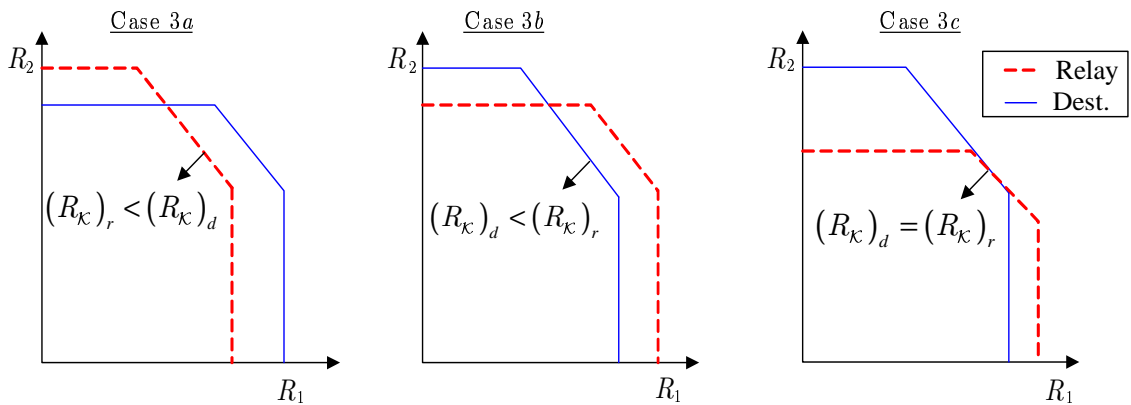}%
\\
{}%
\end{center}}}%
%EndExpansion
\caption{Rate regions $R_{r}(P)$ and $R_{d}(P)$ and sum-rate for cases $3a$, $3b$, and $3c$.}\label{Fig_Case3abc}%
%TCIMACRO{\TeXButton{E}{\end{figure*}}}%
%BeginExpansion
\end{figure*}

The region $\mathcal{R}_{DF}$ in (\ref{GMARC_R_DF}) is a union of the
intersections of the regions $\mathcal{R}_{r}(\underline{P}(\underline{H}))$
and $\mathcal{R}_{d}(\underline{P}(\underline{H}))$ achieved at the relay and
destination respectively, where the union is over all $\underline
{P}(\underline{H})\in$ $\mathcal{P}$. Since $\mathcal{R}_{DF}$ is convex, each
point on the boundary of $\mathcal{R}_{DF}$ is obtained by maximizing the
weighted sum $\mu_{1}R_{1}$ $+$ $\mu_{2}R_{2}$ over all $\underline
{P}(\underline{H})\in\mathcal{P}$, and for all $\mu_{1}>0$, $\mu_{2}>0$.
Specifically, we determine the optimal policy $\underline{P}^{\ast}%
(\underline{H})$ that maximizes the sum-rate $R_{1}+R_{2}$ when $\mu_{1}$ $=$
$\mu_{2}$ $=$ $1$. Observe from (\ref{GMARC_R_DF}) that every point on the
boundary of $\mathcal{R}_{DF}$ results from the intersection of the
polymatroids (pentagons) $\mathcal{R}_{r}(\underline{P}(\underline{H}))$ and
$\mathcal{R}_{d}(\underline{P}(\underline{H}))$ for some $\underline
{P}(\underline{H})$. In Figs. \ref{Fig_Case12} and \ref{Fig_Case3abc} we
illustrate the five possible choices for the sum-rate resulting from such an
intersection for a two-user MARC of which two belong to the inactive set and
three to the active set.

\begin{figure*}[tbp] \centering
%EndExpansion%
%TCIMACRO{\FRAME{itbpFU}{5.7294in}{2.5304in}{0in}{\Qcb{{}}}{}%
%{casesbc_1and3.eps}{\special{ language "Scientific Word";  type "GRAPHIC";
%display "USEDEF";  valid_file "F";  width 5.7294in;  height 2.5304in;
%depth 0in;  original-width 0pt;  original-height 0pt;  cropleft "0";
%filename 'CasesBC_1and3.eps';file-properties "XNPEU";}}}%
%BeginExpansion
{\parbox[b]{5.7294in}{\begin{center}
\includegraphics[
height=2.5304in,
width=5.7294in
]%
{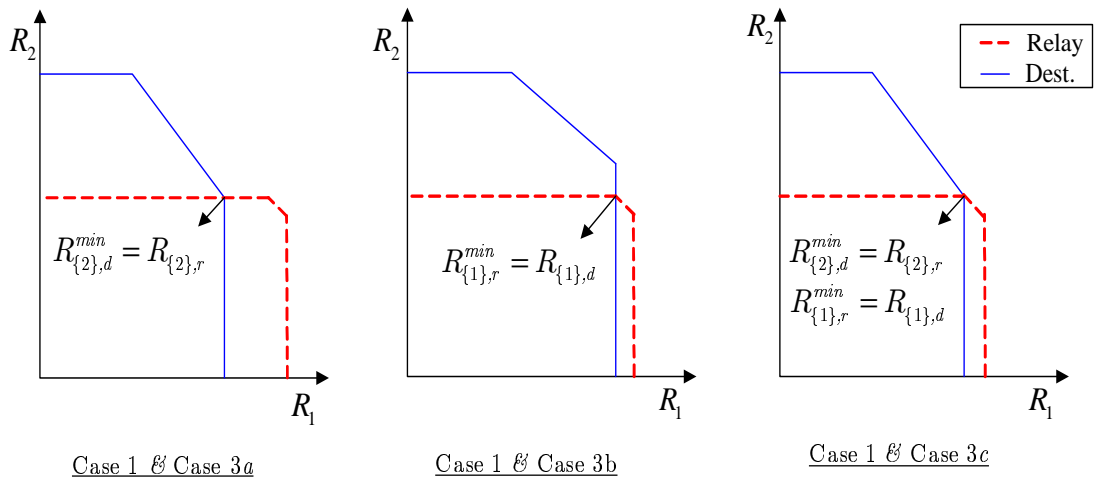}%
\\
{}%
\end{center}}}%
%EndExpansion
\caption{Rate regions $R_{r}(P)$ and $R_{d}(P)$ for cases (1,3a), (1,3b), and (1,3c).}\label{Fig_BC13}%
%TCIMACRO{\TeXButton{E}{\end{figure*}}}%
%BeginExpansion
\end{figure*}

\begin{figure*}[tbp] \centering
%EndExpansion%
%TCIMACRO{\FRAME{itbpFU}{5.7432in}{2.5201in}{0in}{\Qcb{{}}}{}%
%{casesbc_2and3.eps}{\special{ language "Scientific Word";  type "GRAPHIC";
%display "USEDEF";  valid_file "F";  width 5.7432in;  height 2.5201in;
%depth 0in;  original-width 0pt;  original-height 0pt;  cropleft "0";
%filename 'CasesBC_2and3.eps';file-properties "XNPEU";}}}%
%BeginExpansion
{\parbox[b]{5.7432in}{\begin{center}
\includegraphics[
height=2.5201in,
width=5.7432in
]%
{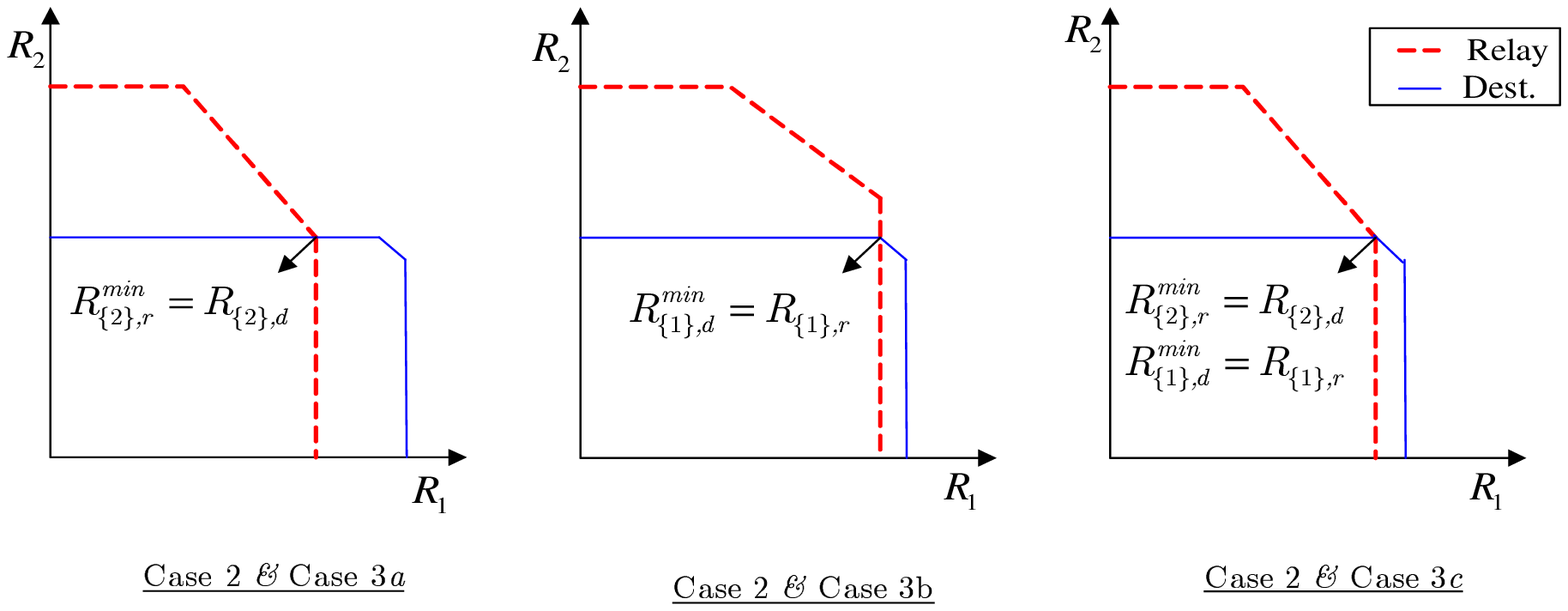}%
\\
{}%
\end{center}}}%
%EndExpansion
\caption{Rate regions $R_{r}(P)$ and $R_{d}(P)$ for cases (2,3a), (2,3b), and (2,3c).}\label{Fig_BC23}%
%TCIMACRO{\TeXButton{E}{\end{figure*}}}%
%BeginExpansion
\end{figure*}

The inactive set\textit{ }consists of cases $1$ and $2$ in which user $1$
achieves a significantly larger rate at the relay and destination,
respectively, than it does at the other receiver; and vice-versa for user $2$.
The active set includes cases $3a$, $3b$, and $3c$ shown in Fig.
\ref{Fig_Case12} in which the sum-rate at relay $r$ is smaller, larger, or
equal, respectively, to that achieved at the destination $d$. The three
boundary cases between case $1$ and the three active cases are shown in Fig.
\ref{Fig_BC13} while the remaining three between case $2$ and the active cases
are shown in Fig. \ref{Fig_BC23}. We denote a boundary case as case $\left(
l,n\right)  $, $l=1,2,$ $n=3a,3b,3c$.

We write $\mathcal{B}_{i}\subseteq\mathcal{P}$ and $\mathcal{B}_{l,n}%
\subseteq\mathcal{P}$ to denote the set of power policies that achieve case
$i$, $i=1,2,3a,3b,3c$, and case $\left(  l,n\right)  $, $l=1,2$, $n=3a,3b,3c$,
respectively. We show in the sequel that the optimization is simplified when
the conditions for each case are defined such that the sets $\mathcal{B}_{i}$
and $\mathcal{B}_{l,n}$ are disjoint for all $i,l,$ and $n$, and thus, are
either open or half-open sets such that no two sets share a boundary. Observe
that cases $1$ and $2$ do not share a boundary since such a transition (see
Fig. \ref{Fig_Case12}) requires passing through case $3a$ or $3b$ or $3c$.
Finally, note that Fig. \ref{Fig_Case3abc} illustrates two specific
$\mathcal{R}_{r}$ and $\mathcal{R}_{d}$ regions for $3a$, $3b$, and $3c$. For
ease of exposition, we write $\mathcal{B}_{3}$ $=$ $\mathcal{B}_{3a}%
\cup\mathcal{B}_{3b}\cup\mathcal{B}_{3c}$, where $\mathcal{B}_{i}$,
$i=3a,3b,3c$.%

%TCIMACRO{\TeXButton{B}{\begin{figure*}[tbp] \centering}}%
%BeginExpansion
%EndExpansion
%

%TCIMACRO{\TeXButton{B}{\begin{figure*}[tbp] \centering}}%
%BeginExpansion
%EndExpansion
%

%TCIMACRO{\TeXButton{B}{\begin{figure*}[tbp] \centering}}%
%BeginExpansion
%EndExpansion
%

%TCIMACRO{\TeXButton{B}{\begin{figure*}[tbp] \centering}}%
%BeginExpansion
%EndExpansion

In general, the occurrence of any one of the disjoint cases depends on both
the channel statistics and the policy $\underline{P}(\underline{H})$. Since it
is not straightforward to know \textit{a priori} the power allocations that
achieve a certain case, we maximize the sum-capacity for each case over all
allocations in $\mathcal{P}$ and write $\underline{P}^{(i)}(\underline{H})$
and $\underline{P}^{(l,n)}(\underline{H})$ to denote the optimal solution for
case $i$ and case $(l,n)$, respectively. Explicitly including boundary cases
ensures that the sets $\mathcal{B}_{i}$ and $\mathcal{B}_{l,n}$ are disjoint
for all $i$ and $(l,n)$, i.e., these sets are either open or half-open sets
such that no two sets share a power policy in common. This in turn simplifies
the convex optimization as follows.

Let $\underline{P}^{(i)}(\underline{H})$ be the optimal policy maximizing the
sum-rate for case $i$ over all $\underline{P}(\underline{H})\in\mathcal{P}$.
The optimal $\underline{P}^{(i)}(\underline{H})$ must satisfy the conditions
for case $i$, i.e., $\underline{P}^{(i)}(\underline{H})\in$ $\mathcal{B}_{i}$.
If the conditions are satisfied, we prove the optimality of \underline{$P$%
}$^{\left(  i\right)  }(\underline{H})$ using the fact that the rate function
for each case is concave. On the other hand, when \underline{$P$}$^{\left(
i\right)  }(\underline{H})\not \in $ $\mathcal{B}_{i}$, it can be shown that
$R_{1}+R_{2}$ achieves its maximum outside $\mathcal{B}_{i}$. The proof again
follows from the fact that $R_{1}+R_{2}$ for all cases is a concave function
of \underline{$P$}$(\underline{H})$ for all $\underline{P}(\underline{H})$
$\in$ $\mathcal{P}$. Thus, when \underline{$P$}$^{\left(  i\right)
}(\underline{H})$ $\not \in $ $\mathcal{B}_{i}$, for every \underline{$P$%
}$(\underline{H})\in$ $\mathcal{B}_{i}$ there exists a \underline{$P$%
}$^{\prime}(\underline{H})$ $\in$ $\mathcal{B}_{i}$ with a larger sum-rate.
Combining this with the fact that the sum-rate expressions are continuous
while transitioning from one case to another at the boundary of the open set
$\mathcal{B}_{i}$, ensures that the maximal sum-rate is achieved by some
$\underline{P}(\underline{H})$ $\not \in $ $\mathcal{B}_{i}$. Similar
arguments justify maximizing the optimal policy for each case over all
$\mathcal{P}$. Due to the concavity of the rate functions, only one
$\underline{P}^{\left(  i\right)  }(\underline{H})$ or $\underline{P}^{\left(
l,n\right)  }(\underline{H})$ will satisfy the conditions for its case. The
optimal $\underline{P}^{\ast}(\underline{H})$ is given by this $\underline
{P}^{\left(  i\right)  }(\underline{H})$ or $\underline{P}^{\left(
l,n\right)  }(\underline{H})$.

The optimization problem for case $i$ or case $\left(  l,n\right)  $ is given
by
\begin{equation}
\frame{$%
\begin{array}
[c]{l}%
\max\limits_{\underline{P}\in\mathcal{P}}S^{\left(  i\right)  }\text{ or }%
\max\limits_{\underline{P}\in\mathcal{P}}S^{\left(  l,n\right)  }\\%
\begin{array}
[c]{cc}%
s.t. & \mathbb{E}\left[  P_{k}(\underline{H})\right]  \leq\overline{P}%
_{k}\text{, }k=1,2,r
\end{array}
\\%
\begin{array}
[c]{cc}%
\text{ \ \ \ \ \ } & P_{k}(\underline{H})\geq0\text{, }k=1,2,r
\end{array}
\end{array}
$} \label{DF_OptProb}%
\end{equation}
where
\begin{equation}
\frame{$%
\begin{array}
[c]{l}%
S^{\left(  1\right)  }=R_{\left\{  1\right\}  ,d}+R_{\left\{  2\right\}  ,r}\\
S^{\left(  2\right)  }=R_{\left\{  1\right\}  ,r}+R_{\left\{  2\right\}  ,d}\\
S^{\left(  i\right)  }=R_{\mathcal{K},j}\text{ for }\left(  i,j\right)
=\left(  3a,r\right)  ,\left(  3b,d\right) \\
S^{\left(  3c\right)  }=R_{\mathcal{K},r}\text{ }s.t.\text{ }R_{\mathcal{K}%
,r}=R_{\mathcal{K},d}.\\
S^{\left(  l,n\right)  }=S^{\left(  l\right)  }\text{ }s.t.\text{ }S^{\left(
l\right)  }=S^{\left(  n\right)  }.
\end{array}
$} \label{DF_Jdef}%
\end{equation}

The optimal policy for each case is determined using Lagrange multipliers and
the \textit{Karush}-\textit{Kuhn}-\textit{Tucker} (KKT) conditions
\cite[5.5.3]{cap_theorems:BVbook01}. A detailed analysis is developed in the
Appendix and we summarize the KKT conditions and the optimal policies for all
cases below. From (\ref{DF_Jdef}), the KKT conditions for each case~$x$,
$x=i,\left(  l,n\right)  $, for all $i$ and $\left(  l,n\right)  $ is given as%
\begin{equation}%
\begin{array}
[c]{cc}%
f_{k}^{\left(  x\right)  }\left(  \underline{P}(\underline{h})\right)
-\nu_{k}\ln2\leq0, & \text{with equality for }P_{k}\left(  \underline
{h}\right)  >0\text{, }k=1,2,r
\end{array}
\label{DF_KKT}%
\end{equation}
where $\nu_{k}$, for all $k=1,2,r$, are dual variables associated with the
power constraints in (\ref{DF_OptProb}). Specializing the KKT conditions for
each case, we obtain the optimal policies for each case as summarized below
following which we list the conditions that the optimal policy for each case
needs to satisfy.

\textit{Case }$1:$ The functions $f_{k}^{\left(  1\right)  }\left(
\underline{P}(\underline{h})\right)  $, $k=1,2,r,$ in (\ref{DF_KKT}) for case
1 are
\begin{align}
&
\begin{array}
[c]{cc}%
f_{k}^{\left(  1\right)  }\left(  \underline{P}(\underline{h})\right)
=\frac{\left\vert h_{m,k}\right\vert ^{2}}{\left(  1+\left\vert h_{m,k}%
\right\vert ^{2}\left.  P_{k}\left(  \underline{h}\right)  \right/
\theta\right)  } & (k,m)=(1,d),(2,r)\text{ }%
\end{array}
\label{DF_f1_k}\\
&
\begin{array}
[c]{cc}%
f_{r}^{\left(  1\right)  }\left(  \underline{P}(\underline{h})\right)
=\frac{\left\vert h_{d,r}\right\vert ^{2}}{\left(  1+\left\vert h_{d,r}%
\right\vert ^{2}\left.  P_{k}\left(  \underline{h}\right)  \right/
\overline{\theta}\right)  }. &
\end{array}
\label{DF_fr}%
\end{align}
It is straightforward to verify that these KKT conditions simplify to%
\begin{equation}%
\begin{array}
[c]{ll}%
P_{k}^{\left(  1\right)  }\left(  \underline{h}\right)  =\left(  \frac{\theta
}{\nu_{k}\ln2}-\frac{\theta}{\left\vert h_{m,k}\right\vert ^{2}}\right)  ^{+}
& (k,m)=(1,d),(2,r)
\end{array}
\label{DF_Pk_1}%
\end{equation}
and
\begin{equation}
P_{r}^{\left(  1\right)  }\left(  \underline{h}\right)  =\left(
\frac{\overline{\theta}}{\nu_{r}\ln2}-\frac{\overline{\theta}}{\left\vert
h_{d,r}\right\vert ^{2}}\right)  ^{+}.\text{ }%
\end{equation}

\textit{Case }$2:$ From (\ref{DF_Jdef}), since $S^{\left(  2\right)  }$ can be
obtained from $S^{\left(  1\right)  }$ by interchanging the user indexes $1$
and $2$, the functions $f_{k}^{\left(  2\right)  }\left(  \underline
{P}(\underline{h})\right)  $, and hence, the KKT conditions for this case can
be obtained by replacing the superscript $\left(  1\right)  $ by $\left(
2\right)  $ and using the pairs $(k,m)=(1,d),(2,r)$ in (\ref{DF_f1_k}%
)-(\ref{DF_Pk_1}). The resulting optimal policies are
\begin{equation}%
\begin{array}
[c]{ll}%
P_{r}^{\left(  2\right)  }\left(  \underline{h}\right)  =P_{r}^{\left(
1\right)  }\left(  \underline{h}\right)  & \text{for all }\underline{h}\text{,
and}\\
P_{k}^{\left(  2\right)  }\left(  \underline{h}\right)  =\left(  \frac{\theta
}{\nu_{k}\ln2}-\frac{\theta}{\left\vert h_{m,k}\right\vert ^{2}}\right)  ^{+}
& (k,m)=(1,r),(2,d).
\end{array}
\end{equation}
Case $3a:$ The functions $f_{k}^{\left(  3a\right)  }\left(  \underline
{P}(\underline{h})\right)  $, $k=1,2,$ satisfying the KKT conditions in
(\ref{DF_KKT}) are
\begin{equation}%
\begin{array}
[c]{cc}%
f_{k}^{\left(  3a\right)  }\left(  \underline{P}(\underline{h})\right)
=\left\vert h_{r,k}\right\vert ^{2}\left/  \left(  1+\sum\limits_{k=1}%
^{2}\left\vert h_{r,k}\right\vert ^{2}\left.  P_{k}\left(  \underline
{h}\right)  \right/  \theta\right)  \right.  & k=1,2.
\end{array}
\label{DF_fk3a}%
\end{equation}
Since this case maximizes the multiaccess sum-rate at the relay, the optimal
user policies are multiuser opportunistic water-filling solutions given by
\begin{equation}%
\begin{array}
[c]{ll}%
\frac{\left\vert h_{r,1}\right\vert ^{2}}{v_{1}}>\frac{\left\vert
h_{r,2}\right\vert ^{2}}{\nu_{2}} & P_{1}^{\left(  3a\right)  }\left(
\underline{h}\right)  =\left(  \frac{\theta}{\nu_{1}\ln2}-\frac{\theta
}{\left\vert h_{r,1}\right\vert ^{2}}\right)  ^{+},P_{2}^{\left(  3a\right)
}=0\\
\frac{\left\vert h_{r,1}\right\vert ^{2}}{v_{1}}\leq\frac{\left\vert
h_{r,2}\right\vert ^{2}}{\nu_{2}} & P_{1}^{\left(  3a\right)  }\left(
\underline{h}\right)  =0,P_{2}^{\left(  3a\right)  }=\left(  \frac{\theta}%
{\nu_{2}\ln2}-\frac{\theta}{\left\vert h_{r,2}\right\vert ^{2}}\right)  ^{+}%
\end{array}
\label{DF_C1_P}%
\end{equation}
where without loss of generality, the users are time-duplexed even when their
scaled fading states in (\ref{DF_C1_P}) are the same. While the relay power
does not explicitly appear in the optimization, since this case results when
the sum-rate is smaller than that at the destination, choosing the optimal
relay policy to maximize the sum-rate at the destination, i.e., $P_{r}%
^{\left(  3a\right)  }\left(  \underline{H}\right)  =P_{r}^{\left(  1\right)
}\left(  \underline{H}\right)  $, suffices.

Case $3b:$ The functions $f_{k}^{\left(  3b\right)  }\left(  \underline
{P}(\underline{h})\right)  $, $k=1,2,$ satisfying the KKT conditions in
(\ref{DF_KKT}) can be obtained from (\ref{DF_fk3a}) by replacing the subscript
`$r$' by `$d$' in (\ref{DF_fk3a}) while $f_{r}^{\left(  3b\right)  }\left(
\underline{P}(\underline{h})\right)  =f_{r}^{\left(  3a\right)  }\left(
\underline{P}(\underline{h})\right)  =f_{r}^{\left(  1\right)  }\left(
\underline{P}(\underline{h})\right)  $. Thus, this case maximizes the
multiaccess sum-rate at the destination and the optimal user policies are
multiuser opportunistic water-filling solutions given by
\begin{equation}%
\begin{array}
[c]{ll}%
\frac{\left\vert h_{d,1}\right\vert ^{2}}{v_{1}}>\frac{\left\vert
h_{d,2}\right\vert ^{2}}{\nu_{2}} & P_{1}^{\left(  3a\right)  }\left(
\underline{h}\right)  =\left(  \frac{\theta}{\nu_{1}\ln2}-\frac{\theta
}{\left\vert h_{d,1}\right\vert ^{2}}\right)  ^{+},P_{2}^{\left(  3a\right)
}=0\\
\frac{\left\vert h_{r,1}\right\vert ^{2}}{v_{1}}\leq\frac{\left\vert
h_{d,2}\right\vert ^{2}}{\nu_{2}} & P_{1}^{\left(  3a\right)  }\left(
\underline{h}\right)  =0,P_{2}^{\left(  3a\right)  }=\left(  \frac{\theta}%
{\nu_{2}\ln2}-\frac{\theta}{\left\vert h_{d,2}\right\vert ^{2}}\right)  ^{+}%
\end{array}
\end{equation}
while the optimal relay policy is a water-filling solution $P_{r}^{\left(
3a\right)  }\left(  \underline{H}\right)  =P_{r}^{\left(  1\right)  }\left(
\underline{H}\right)  $.

Case $3c:$ The functions $f_{k}^{\left(  3c\right)  }\left(  \underline
{P}(\underline{h})\right)  $, $k=1,2,r,$ satisfying the KKT conditions in
(\ref{DF_KKT}) are given as
\begin{align}
&
\begin{array}
[c]{cc}%
f_{k}^{\left(  3c\right)  }\left(  \underline{P}(\underline{h})\right)
=\left(  1-\alpha\right)  f_{k}^{\left(  3a\right)  }\left(  \underline
{P}(\underline{h})\right)  +\alpha f_{k}^{\left(  3b\right)  }\left(
\underline{P}(\underline{h})\right)  , & k=1,2,
\end{array}
\\
&
\begin{array}
[c]{cc}%
f_{r}^{\left(  3c\right)  }\left(  \underline{P}(\underline{h})\right)
=\alpha f_{r}^{\left(  3b\right)  }\left(  \underline{P}(\underline
{h})\right)  , & k=r,
\end{array}
\end{align}
where the Lagrange multiplier $\alpha$ accounts for the boundary condition
\begin{equation}
R_{\mathcal{K},d}\left(  \underline{P}\left(  \underline{H}\right)  \right)
=R_{\mathcal{K},r}\left(  \underline{P}\left(  \underline{H}\right)  \right)
\label{DF3cCond}%
\end{equation}
and the optimal policy $\underline{P}^{\left(  3b\right)  }\left(
\underline{H}\right)  \in\mathcal{B}_{3c}$ satisfies this condition where
$\mathcal{B}_{3c}$ is the set of $\underline{P}\left(  \underline{H}\right)  $
that satisfy (\ref{DF3cCond}). Thus, this case maximizes the multiaccess
sum-rate at both the relay and the destination. In the Appendix, using the KKT
conditions we show that the optimal user policies are opportunistic in form
and are given by
\begin{equation}%
\begin{array}
[c]{ll}%
f_{1}^{\left(  3c\right)  }/\nu_{1}>f_{2}^{\left(  3c\right)  }/\nu_{2} &
P_{1}^{\left(  3c\right)  }\left(  \underline{h}\right)  =\left(  \text{root
of }F_{1}^{\left(  3c\right)  }|_{P_{2}=0}\right)  ^{+},P_{2}^{\left(
3c\right)  }\left(  \underline{h}\right)  =0\\
f_{1}^{\left(  3c\right)  }/\nu_{1}\leq f_{2}^{\left(  3c\right)  }/\nu_{2} &
P_{1}^{\left(  3c\right)  }\left(  \underline{h}\right)  =0,P_{2}^{\left(
3c\right)  }\left(  \underline{h}\right)  =\left(  \text{root of }%
F_{2}^{\left(  3c\right)  }|_{P_{1}=0}\right)  ^{+}.
\end{array}
\label{DF_3c_P}%
\end{equation}
Analogous to cases $3a$ and $3b$, the scheduling conditions in (\ref{DF_3c_P})
depend on both the channel states and the water-filling levels $\nu_{k}$ at
both users. However, the conditions in (\ref{DF_3c_P}) also depend on the
power policies, and thus, the optimal solutions are no longer water-filling
solutions. In the Appendix we show that the optimal user policies can be
computed using an \textit{iterative non-water-filling algorithm} which starts
by fixing the power policy of one user, computing that of the other, and
vice-versa until the policies converge to the optimal policy. The iterative
algorithm is computed for increasing values of $\alpha\in\left(  0,1\right)  $
until the optimal policy satisfies (\ref{DF3cCond}) at the optimal
$\alpha^{\ast}$. The proof of convergence is detailed in the Appendix.

Boundary Cases $\left(  l,n\right)  :$ A boundary case $\left(  l,n\right)  $
results when
\begin{equation}%
\begin{array}
[c]{cc}%
S^{\left(  l\right)  }=S^{\left(  n\right)  } & l=1,2,\text{ and }n=3a,3b,3c.
\end{array}
\label{DF_BC_Cond}%
\end{equation}
Recall that $S^{\left(  l\right)  }$ and $S^{\left(  n\right)  }$ are
sum-rates for an inactive case $l$, and an active case $n$, respectively.
Thus, in addition to the constraints in (\ref{DF_OptProb}), the maximization
problem for these cases includes the additional constraint in
(\ref{DF_BC_Cond}). For all except the two cases where $n=3c$, the equality
condition in (\ref{DF_OptProb}) is represented by a Lagrange multiplier
$\alpha$. The two cases with $n=3c$ have two Lagrange multipliers $\alpha_{1}$
and $\alpha_{2}$ to also account for the condition $S^{\left(  3a\right)
}=S^{\left(  3b\right)  }$.

For the different boundary cases, the functions $f_{k}^{\left(  l,n\right)
}\left(  \underline{P}(\underline{h})\right)  $, $k=1,2,$ satisfying the KKT
conditions in (\ref{DF_KKT}) are given as
\begin{align}
&
\begin{array}
[c]{cc}%
f_{k}^{\left(  l,n\right)  }\left(  \underline{P}(\underline{h})\right)
=\left(  1-\alpha\right)  f_{k}^{\left(  l\right)  }\left(  \underline
{P}(\underline{h})\right)  +\alpha f_{k}^{\left(  n\right)  }\left(
\underline{P}(\underline{h})\right)  , & k=1,2,n\not =3c
\end{array}
\\
&
\begin{array}
[c]{cc}%
f_{k}^{\left(  l,3c\right)  }\left(  \underline{P}(\underline{h})\right)
=\left(  1-\alpha_{1}-\alpha_{2}\right)  f_{k}^{\left(  l\right)  }\left(
\underline{P}(\underline{h})\right)  +\alpha_{2}f_{k}^{\left(  3a\right)
}\left(  \underline{P}(\underline{h})\right)  +\alpha_{1}f_{k}^{\left(
3b\right)  }\left(  \underline{P}(\underline{h})\right)  , & k=1,2,
\end{array}
\\
&
\begin{array}
[c]{cc}%
f_{r}^{\left(  l,n\right)  }\left(  \underline{P}(\underline{h})\right)
=\alpha f_{r}^{\left(  l\right)  }\left(  \underline{P}(\underline{h})\right)
, & n=3a
\end{array}
\\
&
\begin{array}
[c]{cc}%
f_{r}^{\left(  l,n\right)  }\left(  \underline{P}(\underline{h})\right)
=\alpha f_{r}^{\left(  l\right)  }\left(  \underline{P}(\underline{h})\right)
+\left(  1-\alpha\right)  f_{r}^{\left(  n\right)  }\left(  \underline
{P}(\underline{h})\right)  , & n=3b
\end{array}
\\
&
\begin{array}
[c]{cc}%
f_{r}^{\left(  l,n\right)  }\left(  \underline{P}(\underline{h})\right)
=\alpha_{1}f_{r}^{\left(  l\right)  }\left(  \underline{P}(\underline
{h})\right)  +\left(  1-\alpha_{1}-\alpha_{2}\right)  f_{r}^{\left(
3b\right)  }\left(  \underline{P}(\underline{h})\right)  , & n=3c
\end{array}
\end{align}
For ease of exposition and brevity, we summarize the KKT conditions and the
optimal policies for case $\left(  1,3a\right)  $. In the Appendix, using the
KKT conditions we show that the optimal user policies $P_{k}^{\left(
1,3a\right)  }(\underline{H})$ are opportunistic in form and are given by
\begin{equation}%
\begin{array}
[c]{ll}%
\frac{f_{1}^{\left(  1,3a\right)  }}{\nu_{1}}>\frac{f_{2}^{\left(
1,3a\right)  }}{\nu_{2}} & P_{1}\left(  \underline{h}\right)  =\left(
\text{root of }F_{1}^{\left(  1,3a\right)  }|_{P_{2}=0}\right)  ^{+}%
,P_{2}\left(  \underline{h}\right)  =0\\
\frac{f_{1}^{\left(  1,3a\right)  }}{\nu_{1}}\leq\frac{f_{2}^{\left(
1,3a\right)  }}{\nu_{2}} & P_{1}\left(  \underline{h}\right)  =0,P_{2}\left(
\underline{h}\right)  =\left(  \text{root of }F_{2}^{\left(  1,3a\right)
}|_{P_{1}=0}\right)  ^{+}%
\end{array}
\end{equation}
where $F_{k}^{\left(  1,3a\right)  }=f_{k}^{\left(  1,3a\right)  }-\nu_{k}%
\ln2,$ $k=1,2.$ As in case $3c$, the optimal policies take an opportunistic
non-waterfilling form and in fact can be obtained by an \textit{iterative
non-water-filling algorithm} as described for case $3c$. The optimal
$P_{r}^{\left(  1,3a\right)  }\left(  \underline{H}\right)  =\alpha
P_{r}^{\left(  1\right)  }\left(  \underline{H}\right)  $ is a water-filling solution.

The optimal policies for all other boundary cases can be obtained similarly as
detailed in the Appendix. In general, for all boundary cases, the optimal user
policies are opportunistic non-water-filling solutions while that for the
relay are water-filling solutions. Finally, the sum-rate maximizing policy for
any case is the optimal policy only if it satisfies the conditions for that
case. The conditions for the cases are%

\begin{align}
&
\begin{array}
[c]{cl}%
\underline{\text{Case}\ 1}: &
\begin{array}
[c]{ccc}%
R_{\left\{  1\right\}  ,d}<R_{\left\{  1\right\}  ,r}^{\min} & \text{and} &
R_{\left\{  2\right\}  ,r}<R_{\left\{  2\right\}  ,d}^{\min}%
\end{array}
\end{array}
\label{Case1_Cond}\\
&
\begin{array}
[c]{cc}%
\underline{\text{Case}\ 2}: &
\begin{array}
[c]{ccc}%
R_{\left\{  1\right\}  ,r}<R_{\left\{  1\right\}  ,d}^{\min} & \text{and} &
R_{\left\{  2\right\}  ,d}<R_{\left\{  2\right\}  ,r}^{\min}%
\end{array}
\end{array}
\label{Case2_Cond}\\
&
\begin{array}
[c]{cc}%
\underline{\text{Case}\ 3a}: & \left(  R_{\mathcal{K}}\right)  _{r}<\left(
R_{\mathcal{K}}\right)  _{d}%
\end{array}
\label{Case3a_Cond}\\
&
\begin{array}
[c]{cc}%
\underline{\text{Case}\ 3b}: & \left(  R_{\mathcal{K}}\right)  _{r}>\left(
R_{\mathcal{K}}\right)  _{d}%
\end{array}
\label{Case3b_Cond}\\
&
\begin{array}
[c]{cc}%
\underline{\text{Case}\ 3c}: & \left(  R_{\mathcal{K}}\right)  _{r}=\left(
R_{\mathcal{K}}\right)  _{d}%
\end{array}
\label{Case3c_Cond}%
\end{align}%
\begin{align}
&
\begin{array}
[c]{cc}%
\underline{\text{Case}\ \left(  1,3a\right)  }: & R_{\mathcal{K}%
,r}=R_{\left\{  1\right\}  ,d}+R_{\left\{  2\right\}  ,r}\text{ }<\left(
R_{\mathcal{K}}\right)  _{d}%
\end{array}
\label{Case13a_Cond}\\
&
\begin{array}
[c]{cc}%
\underline{\text{Case}\ \left(  2,3a\right)  }: & R_{\mathcal{K}%
,r}=R_{\left\{  1\right\}  ,r}+R_{\left\{  2\right\}  ,d}<\left(
R_{\mathcal{K}}\right)  _{d}%
\end{array}
\label{Case23a_Cond}\\
&
\begin{array}
[c]{cc}%
\underline{\text{Case}\ \left(  1,3b\right)  }: & R_{\mathcal{K}%
,d}=R_{\left\{  1\right\}  ,d}+R_{\left\{  2\right\}  ,r}<\left(
R_{\mathcal{K}}\right)  _{r}%
\end{array}
\label{Case13b_Cond}\\
&
\begin{array}
[c]{cc}%
\underline{\text{Case}\ \left(  2,3b\right)  }: & R_{\mathcal{K}%
,d}=R_{\left\{  1\right\}  ,r}+R_{\left\{  2\right\}  ,d}<\left(
R_{\mathcal{K}}\right)  _{r}%
\end{array}
\label{Case23b_Cond}\\
&
\begin{array}
[c]{cc}%
\underline{\text{Case}\ \left(  1,3c\right)  }: & R_{\mathcal{K}%
,r}=R_{\mathcal{K},d}=R_{\left\{  1\right\}  ,d}+R_{\left\{  2\right\}  ,r}%
\end{array}
\label{Case13c_Cond}\\
&
\begin{array}
[c]{cc}%
\underline{\text{Case}\ \left(  2,3c\right)  }: & R_{\mathcal{K}%
,r}=R_{\mathcal{K},d}=R_{\left\{  1\right\}  ,r}+R_{\left\{  2\right\}  ,d}%
\end{array}
\label{Case23c_Cond}%
\end{align}
where in fading state $\underline{H}$, (\ref{Case1_Cond})-(\ref{Case23c_Cond})
are evaluated for $X_{k}\sim\mathcal{CN}\left(  0,P_{k}^{\left(  x\right)
}\left(  \underline{H}\right)  /\theta\right)  $, $k=1,2$, and $X_{r}%
\sim\mathcal{CN}\left(  0,P_{r}^{\left(  x\right)  }\left(  \underline
{H}\right)  /\overline{\theta}\right)  $ for $x=i,\left(  l,n\right)  $.
\newline The following theorem summarizes the form of \underline{$P$}$^{\ast}$
and presents an algorithm to compute it.

\begin{theorem}
\label{DF_Th_OptP}The optimal policy \underline{$P$}$^{\ast}(\underline{H})$
maximizing the DF sum-rate of a two-user ergodic fading orthogonal MARC is
obtained by computing $\underline{P}^{(i)}(\underline{H})$ and $\underline
{P}^{(l,n)}(\underline{H})$ starting with the inactive cases~$1$ and $2$,
followed by the boundary cases $(l,n)$, and finally the active cases $3a,$
$3b,$ and $3c$ until for some case the corresponding $\underline{P}%
^{(i)}(\underline{H})$ or $\underline{P}^{(l,n)}(\underline{H})$ satisfies the
case conditions. The optimal \underline{$P$}$^{\ast}(\underline{H})$ is given
by the optimal $\underline{P}^{(i)}(\underline{H})$ or $\underline{P}%
^{(l,n)}(\underline{H})$ that satisfies its case conditions and falls into one
of the following three categories:

\textit{Inactive Cases}: The optimal policy for the two users is such that one
user water-fills over its link to the relay while the other water-fills over
its link to the destination. The optimal relay policy $P_{r}^{\ast}%
(\underline{H})$ is water-filling over its direct link to the destination.

\textit{Cases }$\left(  3a,3b,3c\right)  $: The optimal user policy
$P_{k}^{\ast}(\underline{H})$, for all $k\in\mathcal{K}$, is opportunistic
water-filling over its link to the relay for case $3a$ and to the destination
for case $3b$. For case $3c$, $P_{k}^{\ast}(\underline{H})$, for all
$k\in\mathcal{K}$, takes an opportunistic non-waterfilling form and depends on
the channel gains of user $k$ at both receivers. The optimal relay policy
$P_{r}^{\ast}(\underline{H})$ is water-filling over its direct link to the destination.

\textit{Boundary Cases}: The optimal user policy $P_{k}^{\ast}(\underline{H}%
)$, for all $k\in\mathcal{K}$, takes an opportunistic non-water-filling form.
The optimal relay policy $P_{r}^{\ast}(\underline{H})$ is water-filling over
its direct link to the destination.\newline
\end{theorem}

\begin{proof}
The closed form expressions for the optimal policies for each case are
developed in the Appendix. The need for an order in evaluating \underline{$P$%
}$^{\ast}(\underline{H})$ is due to the following reasons. Since every case
results from an intersection of two polymatroids, the conditions in
(\ref{Case3a_Cond})-(\ref{Case3c_Cond}) hold for all cases. Thus, all feasible
power policies satisfy one of these three conditions as a result of which
these conditions do not allow a clear distinction between the cases. In
contrast, the conditions for cases $1$ and $2$ in (\ref{Case1_Cond}) and
(\ref{Case2_Cond}), respectively, are mutually exclusive. For the boundary
cases, since every boundary case $\left(  l,n\right)  $ results from the
intersection of an active case $n=3a,3b,3c,$ with an inactive case $l=1,2,$
and is itself an active case, one of its conditions corresponds to the
condition for case $n$, $n=3a,3b$,$3c$ in (\ref{Case3a_Cond}%
)-(\ref{Case3c_Cond}). Additionally, in the Appendix we show that the boundary
condition $S^{\left(  l\right)  }=S^{\left(  n\right)  }$ implies that only
one of the two inequality conditions of case $l$ holds strictly while the
other simplifies to an equality. An immediate implication of these two
conditions is that the boundary cases are mutually exclusive and the set of
power policies satisfying them are also disjoint from those satisfying cases
$1$ and $2$. Thus, to determine the optimal \underline{$P$}$^{\ast}%
(\underline{H})$, one can start with any one of the mutually exclusive
inactive and boundary cases. If the optimal policy for any one of these cases
satisfies its case conditions, then, \underline{$P$}$^{\ast}(\underline{H})$
is given by that policy. However, if all these cases are eliminated, i.e.,
none of their optimal policies satisfy the appropriate case conditions, the
optimal policies for remaining three cases $3a$, $3b$, and $3c$ can be
computed one at a time. From (\ref{Case3a_Cond})-(\ref{Case3c_Cond}), cases
$3a$, $3b$, and $3c$ are mutually exclusive, i.e., their feasible power sets
are disjoint, and thus, the optimal policy, satisfies the conditions for only
one of three active cases.
\end{proof}

\begin{remark}
The conditions for cases $3a$, $3b$, and $3c$ can also be redefined to include
the negation of all the conditions for the other cases. This in turn
eliminates the need for an order in computing the optimal policy; however, the
number of conditions that need to be checked to verify if the optimal policy
satisfies the conditions for cases $3a$ or $3b$ or $3c$ remain unchanged
relative to the algorithm in Theorem \ref{DF_Th_OptP}.
\end{remark}

We now discuss in detail the optimal power policies at the sources and the
relay for the different cases.

\textit{Optimal\ Relay Policy}: In the orthogonal model we consider, the relay
transmits directly to the destination on a channel orthogonal to the source
transmissions. Thus, the relay to destination link can be viewed as a fading
point to point link. In fact, in all cases the optimal relay policy involves
water-filling over the fading states analogous to a fading point to point link
(see \cite{cap_theorems:Goldsmith_01}). However, the exact solution, including
scale factors, depends on the case considered.

The optimal cooperation strategy at the relay also depends on the case
studied. For instance, consider case $1$ where users 1 and 2 achieve
significantly larger rates (relative to the other receiver) at the relay and
destination, respectively. Thus, the sum-rate is the sum of the rates achieved
over the bottle-neck links from user 1 to the destination and from user 2 to
the relay; i.e., it is the sum of the single-user rate user 1 achieves at the
destination and the rate user 2 achieves at the relay. The single-user rate
achieved by user 1 at the destination requires the relay to completely
cooperate with user 1, i.e., the relay uses its power $P_{r}\left(
\underline{H}\right)  $ to forward only the message from user $1$ in every
fading state in which it transmits. As shown in the intersection for case $1$
in Fig. \ref{Fig_Case12}, this is due to the fact that since user $1$ achieves
a significantly larger rate at the relay than does user $2$, the sum-rate is
maximized when the relay allocates its resources entirely to cooperating with
user $1$. Finally, for case $2$, the relay cooperates entirely with user $2$.

For the active cases, $3a$ and $3b$, the sum-rate may be achieved by an
infinite number of feasible points on one or both of the sum-rate planes; the
optimal cooperative strategy at the relay will differ for each such point.
Thus, for a corner point the relay transmits a message from only one of the
users while for all non-corner points the relay transmits both messages.

For the boundary cases including case $3c$, the requirement of an equality
(boundary) condition results in the introduction of an additional parameter.
Thus, for case $3c$, the parameter $\alpha$ is introduced to satisfy the
equality constraint on the sum-rates achieved at the relay and destination.
Similarly for cases $\left(  1,3a\right)  $, $\left(  2,3a\right)  $, $\left(
1,3b\right)  $, and $\left(  2,3b\right)  $, the parameter $\alpha$ is chosen
to ensure that the optimal power policies at the users and relay satisfy the
equality constraint for that case. Finally, for cases $\left(  1,3c\right)  $
and $\left(  2,3c\right)  $, the requirement of satisfying two boundary
conditions requires the introduction of the two parameters, $\alpha_{1}$ and
$\alpha_{2}$, one for each condition.

\textit{Optimal User Policies}: As with the relay, the optimal policies for
the two users depend on the case considered. For cases $1$ and $2$, the
optimal policies are water-filling solutions, i.e., each user allocates power
optimally as if it were transmitting to only that receiver at which it
achieves a lower rate. In fact, the conditions for case \thinspace1 in
(\ref{Case1_Cond}) suggest a network geometry in which source 1 and the relay
are physically proximal enough to form a \textit{cluster} and source 2 and the
destination form another cluster; and vice-versa for case 2. This clustering
and the resulting water-filling over the bottle-neck links is the reason why
the relay forwards the message of only that user physically proximal to it,
namely, only user $1$ and only user $2$ for cases $1$ and $2$, respectively.

For case $3a$, the optimal policies at the two users maximize the sum-rate
achieved at the relay (the smaller of sum-rates achieved at the two
receivers). These policies are the same as those achieving the sum-capacity of
a two-user multiple-access channel with the relay as the intended receiver
(see \cite{cap_theorems:Knopp_Humblet,cap_theorems:TH01}). Thus, the optimal
policy for each user involves water-filling over its fading states to the
relay. The solution also exploits the multiuser diversity to opportunistically
schedule the users in each use of the channel.

Analogously, the optimal policies for case $3b$ require multiuser
water-filling over the user links to the destination. For both cases, if the
channel gains have a joint fading distribution with a continuous density, the
sum-rate maximization simplifies to scheduling only one user in each fading
instantiation (parallel channel). Thus, the users time-share their channel use
and the maximum sum-rate is achieved by a unique point on the boundary of the
rate region (see \cite[III.D]{cap_theorems:TH01}).

The optimal policies for case $3c$ require the users to allocate power such
that the sum-rates achieved at both the relay and the destination are the
same. This constraint has the effect that it preserves the opportunistic
scheduling since the sum-rate involves the multiaccess sum-rate bounds at both
receivers. However, the solutions are no longer waterfilling due to the fact
that the equality (boundary) condition results in the function $f_{k}^{(3c)}$
being a weighted sum of the functions $f_{k}^{(3a)}$ and $f_{k}^{(3b)}$ for
cases $3a$ and $3b$, respectively.

Finally, the requirement of satisfying one or more boundary conditions also
affects the nature of the optimal policies for all the $\left(  l,n\right)  $
cases. Thus, for these boundary cases, since the sum-rate planes are active,
i.e., the functions $f_{k}^{\left(  l,n\right)  }$ involve the multiple-access
sum-rate bounds, the optimal power policies result in an opportunistic
scheduling of the users. However, as with case $3c$, here too the optimal
policies are no longer water-filling since the boundary conditions result in
the functions $f_{k}^{\left(  l,n\right)  }$ being weighted sums of the
functions for cases $l$ and $n$.

\begin{remark}
The case conditions in (\ref{Case1_Cond})-(\ref{Case23c_Cond}) require
averaging over the channel states; thus, the case that maximizes the sum-rate
depends on the average power constraints and the channel statistics (including
network topology).
\end{remark}

\begin{remark}
The optimal policy for each source for cases $1,$ $2$, $3a$, and $3b$ depends
on the channel gains at only one of the receivers. However, the optimal policy
for the boundary cases, including case $3c$, depends on the instantaneous
channel states at both receivers. Furthermore, all the cases exploiting the
multiuser diversity require a centralized protocol to coordinate the
opportunistic scheduling of users.
\end{remark}

\section{\label{Sec_5}Two-User DF\ Rate Region:\ Optimal Power Policies}

In Theorem \ref{DF_Th1}, the DF rate region $\mathcal{R}_{DF}$ is shown to be
a union of the intersections of the regions $\mathcal{R}_{r}(\underline
{P}(\underline{H}))$ and $\mathcal{R}_{d}(\underline{P}(\underline{H}))$
achieved at the relay and destination, respectively, where the union is over
all $\underline{P}(\underline{H})\in$ $\mathcal{P}$. Furthermore, since
$\mathcal{R}_{DF}$ is convex, each point on the boundary of $\mathcal{R}_{DF}$
is obtained by maximizing the weighted sum $\mu_{1}R_{1}$ $+$ $\mu_{2}R_{2}$
over all $\underline{P}(\underline{H})\in\mathcal{P}$, and for all $\mu_{1}%
>0$, $\mu_{2}>0$. In fact, for every $\left(  \mu_{1},\mu_{2}\right)  $, the
rate tuple $\left(  R_{1},R_{2}\right)  $ that maximizes the weighted sum
$\mu_{1}R_{1}+\mu_{2}R_{2}$ results from an intersection of two rate polymatroids.

Thus, analogously to the sum-rate analysis for $\mu_{1}=\mu_{2}=1$, for
arbitrary $\left(  \mu_{1},\mu_{2}\right)  $, $\mu_{1}R_{1}+\mu_{2}R_{2}$, is
maximized by either one of two inactive cases, or by one of nine active cases
of which six are boundary\ cases. To find the rate tuple maximizing $\mu
_{1}R_{1}+\mu_{2}R_{2}$, we use the classic result in linear programming that
the maximum value of a linear function constrained over a feasible bounded
polyhedron is achieved at a vertex of the polyhedron \cite[Chapter
1.2.2]{cap_theorems:BVbook01}. Thus, for any $\underline{P}(\underline{H})$,
the $\left(  R_{1},R_{2}\right)  $-tuple maximizing $\mu_{1}R_{1}+\mu_{2}%
R_{2}$ is given by a vertex of a $\mathcal{R}_{r}\left(  \underline
{P}(\underline{H})\right)  \cap\mathcal{R}_{d}\left(  \underline{P}%
(\underline{H})\right)  $ polyhedron at which $\mu_{1}R_{1}+\mu_{2}R_{2}$ is a
tangent. Recall that for the two inactive cases, the polymatroid intersections
result in rectangles, and thus, there is a unique vertex maximizing $\mu
_{1}R_{1}+\mu_{2}R_{2}$. The intersection is also a rectangle for the six
boundary cases since these active cases are such that only one point on the
sum-rate plane is included in the region of intersection. On the other hand,
for cases $3a$, $3b$, and $3c$, the intersection of $K$-dimensional
polymatroids results in a $K$-dimensional polyhedron. Thus, for these cases,
when $\mu_{1}<\mu_{2}$, $\mu_{1}R_{1}+\mu_{2}R_{2}$ is maximized by that
vertex where user $1$ is decoded before user $2$, i.e., at the vertex where
user $2$ achieves its maximal single-user rate.

For simplicity, we present the results for cases $1$, $3a$, and $\left(
1,3a\right)  .$ The results for the other cases follow naturally from
discussions for these three cases. Without loss of generality, we let $\mu
_{1}<\mu_{2}$; the analysis for $\mu_{1}>\mu_{2}$ follows in an analogous manner.

\textit{Case 1}: From Fig. \ref{Fig_Case12}, the weighted sum $\mu_{1}%
R_{1}+\mu_{2}R_{2}$ for this case is given by%
\begin{equation}
\mu_{1}R_{\left\{  1\right\}  ,d}+\mu_{2}R_{\left\{  2\right\}  ,r}.
\end{equation}
Since $\mu_{1}$ and $\mu_{2}$ are independent of the transmit powers, the
optimization problem is the same as that for the sum-rate case. Thus, at the
maximal rate, users $1$ and $2$ waterfill on their bottleneck links to the
destination and relay, respectively.

The analysis for case $2$ is the same as that for case $1$ except now user $1$
and $2$ waterfill on their bottleneck links to the relay and destination, respectively.

\textit{Case 3a}: The weighted sum $\mu_{1}R_{1}+\mu_{2}R_{2}$ is maximized by
the vertex with coordinates%
\begin{align}
R_{1}  &  =R_{\mathcal{K},r}-R_{2}\label{Region_min1}\\
R_{2}  &  =\min\left(  R_{\left\{  2\right\}  ,r},R_{\left\{  2\right\}
,d}\right)  . \label{Region_min2}%
\end{align}
The maximization of $\mu_{1}R_{1}+\mu_{2}R_{2}$ thus simplifies to
\begin{equation}
\max_{\underline{P}\in\mathcal{P}}\left(  \mu_{1}R_{\mathcal{K},r}+\left(
\mu_{2}-\mu_{1}\right)  \min\left(  R_{\left\{  2\right\}  ,r},R_{\left\{
2\right\}  ,d}\right)  \right)  \label{Region_3a}%
\end{equation}
\newline where the optimal $\underline{P}^{\left(  3a\right)  }(\underline
{H})$ satisfies the conditions in (\ref{Case3a_Cond}) for this case. As in the
appendix, there are three possible disjoint solutions to the max-min
optimization in (\ref{Region_3a}) resulting from either $R_{\left\{
2\right\}  ,r}$ being smaller, larger, or equal to $R_{\left\{  2\right\}
,d}$. We discuss the optimal policies for each of these sub-cases separately below.

\begin{enumerate}
\item $R_{\left\{  2\right\}  ,r}<R_{\left\{  2\right\}  ,d}$: For this
sub-case, the vertex of interest in (\ref{Region_min1}) and (\ref{Region_min2}%
) is achieved by the MAC bounds at the same receiver. Thus, the maximization
for these cases simplifies to that developed in for an ergodic fading MAC in
\cite{cap_theorems:TH01}. The optimal power policies involve water-filling and
opportunistic scheduling of the users and water-filling at the relay over its
direct link to the destination.

\item $R_{\left\{  2\right\}  ,r}>R_{\left\{  2\right\}  ,d}$: The
maximization here simplifies to
\begin{equation}
\max_{\underline{P}\in\mathcal{P}}\left(  \mu_{1}R_{\mathcal{K},r}+\left(
\mu_{2}-\mu_{1}\right)  R_{\left\{  2\right\}  ,d}\right)  .
\label{Region_C3_2}%
\end{equation}
\ As with the Lagrangian expressions for the boundary cases in the Appendix,
here too, the weighted sum of rates in (\ref{Region_C3_2}) is an appropriately
weighted mixture of sum and single-user rates achieved at the relay and
destination, respectively. Thus, analogously to the boundary cases, one can
verify in a straightforward manner that the optimal policies maximizing
(\ref{Region_C3_2}) at both users are non-waterfilling solutions with
opportunistic scheduling based on relative fading states while that at the
relay requires waterfilling over its direct link to the destination. Note that
the optimal policies at both the users and the relay depend on the values
chosen for $\mu_{1}$ and $\mu_{2}$.

\item $R_{\left\{  2\right\}  ,r}=R_{\left\{  2\right\}  ,d}$: Subject to
average power and positivity constraints, the maximization here simplifies to
\begin{equation}%
\begin{array}
[c]{l}%
\max_{\underline{P}\in\mathcal{P}}\left(  \mu_{1}R_{\mathcal{K},r}+\left(
\mu_{2}-\mu_{1}\right)  R_{\left\{  2\right\}  ,r}\right) \\
\text{s.t. }R_{\left\{  2\right\}  ,r}=R_{\left\{  2\right\}  ,d}.
\end{array}
\label{Region_C3_3}%
\end{equation}
The maximization in (\ref{Region_C3_3}) subject to the equality constraint
results in a Lagrangian with a weighted mixture of single-user rates achieved
at the relay and destination. Thus, the optimal user and relay policies have a
form similar to that discussed in the previous sub-case in which the sum-rate
and single-user rate are achieved at different receivers,
\end{enumerate}

\begin{remark}
The three sub-cases for case $3a$ studied above are differentiated by
additional constraints relating the single-user rates at the relay and the
destination. This in turn implies that the region $\mathcal{B}_{3a}$ will be
divided into three mutually exclusive subsets, where the condition for each
sub-case is satisfied in only one of the subsets.
\end{remark}

\textit{Boundary case }$\left(  1,3a\right)  $: Recall that a boundary case
$\left(  l,n\right)  $ results from satisfying the conditions for the active
case $n$ and satisfying the conditions for the inactive case $l$ as a mixture
of equalities and inequalities. The resulting rate region belongs to the set
of active cases but has one unique sum-rate point such that the intersection
of pentagons results in a rectangle (see Figs. \ref{Fig_BC13} and
\ref{Fig_BC23}). Thus, the weighted optimization $\mu_{1}R_{1}+\mu_{2}R_{2}$
for case $\left(  1,3a\right)  $ simplifies to
\begin{equation}%
\begin{array}
[c]{l}%
\mu_{1}R_{\left\{  1\right\}  ,d}+\mu_{2}R_{\left\{  2\right\}  ,r}\\
\text{s.t. }\left(  R_{\mathcal{K}}\right)  _{r}=R_{\left\{  1\right\}
,d}+R_{\left\{  2\right\}  ,r}.
\end{array}
\label{Region_13a}%
\end{equation}
Note that the constraint in (\ref{Region_13a}) is the same as that for the
boundary case $\left(  1,3a\right)  $ in (\ref{Case13a_Cond}). Thus, the
constrained maximization problem in (\ref{Region_13a}) is analogous to the
sum-rate maximization for the boundary cases and admits a similar
non-waterfilling opportunistic solution for the user power policies and a
waterfilling solution at the relay.

\begin{remark}
The discussion here for cases $3a$ and $\left(  1,3a\right)  $ also applies to
the other active (including boundary) cases. In each such case, the optimal
policies depend on all the Lagrange dual variables, with each variable
reflecting a specific constraint.
\end{remark}

\section{\label{Sec_6}$K$-User Generalization}

\subsection{$K$-user Sum-Rate Analysis}

We use Lemma \ref{Lemma_PolyIntersect} to extend the analysis in the previous
sections to the $K$-user case. Recall that $\mathcal{R}_{DF}$ is given by a
union of the intersection of polymatroids, where the union is over all power
policies. From Lemma \ref{Lemma_PolyIntersect}, we have that the maximal
$K$-user sum-rate tuple is achieved by an intersection that either belongs to
active set or to the inactive set. We write $l=1,2,\ldots,2^{K}-2$, to index
the $2^{K}-2$ non-empty subsets of $\mathcal{K}$. For a $K$-user MARC, there
are $\left(  2^{K}-2\right)  $ possible intersections of the inactive kind
with sum-rate $J^{\left(  l\right)  }$ given by%
\begin{equation}%
\begin{array}
[c]{ll}%
\underline{\text{case}\ l}:S^{\left(  l\right)  }=R_{\mathcal{S}%
,r}+R_{\mathcal{K}\backslash\mathcal{S},d} & l=1,2,\ldots,2^{K}-2\\
s.t.\text{ }R_{\mathcal{S},r}<R_{\mathcal{S},d}^{\min}\text{ and
}R_{\mathcal{K}\backslash\mathcal{S},d}<R_{\mathcal{K}\backslash\mathcal{S}%
,d}^{\min} &
\end{array}
\label{DF_K_InA}%
\end{equation}
where $R_{\mathcal{A},j}$ and $R_{\mathcal{A},j}^{\min}$ are as defined
in\ Section \ref{section 3} and for $j=r,d,$ are given by the bounds in
(\ref{DF_Rates_rel}) and (\ref{DF_Rates_dest}), respectively. The sum-rates
$J^{\left(  i\right)  }$, for the active cases $i=3a,3b,3c$, are
\begin{align}
S^{\left(  i\right)  }  &  =R_{\mathcal{K},j}\text{ for }\left(  i,j\right)
=\left(  3a,r\right)  ,\left(  3b,d\right) \label{DF_K_Case3}\\
S^{\left(  3c\right)  }  &  =R_{\mathcal{K},r}\text{ }s.t.\text{
}R_{\mathcal{K},r}=R_{\mathcal{K},d}.
\end{align}
Finally, the sum-rate $J^{\left(  l,n\right)  }$, for the boundary cases
totaling $3\left(  2^{K}-2\right)  $ and enumerated as cases $\left(
l,n\right)  $, $l=1,2,\ldots,2^{K}-2$, $n=3a,3b,3c$, are
\begin{align}
&
\begin{array}
[c]{l}%
\underline{\text{case}~\left(  l,3a\right)  }:S^{\left(  l,3a\right)
}=R_{\mathcal{K},r}\\
s.t.\text{ }R_{\mathcal{K},r}=R_{\mathcal{S},r}+R_{\mathcal{K}\backslash
\mathcal{S},d}\text{ for case }l
\end{array}
\label{DF_K_l3a}\\
&
\begin{array}
[c]{l}%
\underline{\text{case}~\left(  l,3b\right)  }:S^{\left(  l,3b\right)
}=R_{\mathcal{K},d}\\
s.t.\text{ }R_{\mathcal{K},d}=R_{\mathcal{S},r}+R_{\mathcal{K}\backslash
\mathcal{S},d}\text{ for case }l
\end{array}
\label{DF_K_l3b}\\
&
\begin{array}
[c]{l}%
\underline{\text{case}~\left(  l,3c\right)  }:S^{\left(  l,3c\right)
}=R_{\mathcal{K},r}\\
s.t.\text{ }R_{\mathcal{K},r}=R_{\mathcal{K},d}=R_{\mathcal{S},r}%
+R_{\mathcal{K}\backslash\mathcal{S},d}\text{ for case }l
\end{array}
\label{DF_K_l3c}%
\end{align}
where the subset $\mathcal{S}$ is chosen to correspond to the appropriate case
$l$.

\begin{remark}
The constraint for case $l$ in (\ref{DF_K_InA}) can be obtained directly from
the requirement that the $K$-user sum-rate constraints at the two receivers
are larger than that for case $l$ (see (\ref{DG_Lemma_RK})).
\end{remark}

The $K$-user sum-rate optimization problem for cases $i$ and $\left(
l,n\right)  $ can be written as%
\begin{equation}
\frame{$%
\begin{array}
[c]{l}%
\max\limits_{\underline{P}\in\mathcal{P}}S^{\left(  i\right)  }\text{ or }%
\max\limits_{\underline{P}\in\mathcal{P}}S^{\left(  l,n\right)  }\\%
\begin{array}
[c]{cc}%
s.t. & \mathbb{E}\left[  P_{k}(\underline{H})\right]  \leq\overline{P}%
_{k}\text{, }k=1,2,r
\end{array}
\\%
\begin{array}
[c]{cc}%
\text{ \ \ \ \ \ } & P_{k}(\underline{H})\geq0\text{, }k=1,2,r.
\end{array}
\end{array}
$} \label{DF_Kuser_Opt}%
\end{equation}
An inactive case $l$ results when the conditions for that case in
(\ref{DF_K_InA}) are satisfied. A boundary case results when the conditions
for one of the cases in (\ref{DF_K_l3a})-(\ref{DF_K_l3c}) is satisfied for the
appropriate $\left(  l,n\right)  $ case. Finally, case~$3a$ or $3b$ or $3c$
results when the conditions for neither the inactive nor the boundary cases
are satisfied.

As in Section \ref{Sec_4}, the optimization for each case involves writing the
Lagrangian and the KKT conditions. The optimal policy $\underline{P}^{\ast
}(\underline{H})$ satisfies the conditions for only one of the cases. For
brevity, we summarize the details below.

\begin{itemize}
\item \textit{Inactive cases}: The Lagrangian for these cases involves a sum
of the DF bounds at the relay in (\ref{DF_Rates_rel}) for users in
$\mathcal{S}$, for each non-empty $\mathcal{S}$, and the DF bounds at the
destination in (\ref{DF_Rates_dest}) for the remaining users in $\mathcal{K}%
\backslash\mathcal{S}$ such that for $i=1,2,\ldots,2^{K}-2$,%
\begin{align}
\mathcal{L}^{\left(  i\right)  }  &  =\mathbb{E}\left[  \theta C\left(
\sum_{k\in\mathcal{S}}\left\vert H_{r,k}\right\vert ^{2}\frac{P_{k}%
(\underline{H})}{\theta}\right)  +\theta C\left(  \sum_{k\in\mathcal{K}%
\backslash\mathcal{S}}\left\vert H_{d,k}\right\vert ^{2}\frac{P_{k}%
(\underline{H})}{\theta}\right)  +\overline{\theta}C\left(  \left\vert
H_{d,r}\right\vert ^{2}\frac{P_{r}(\underline{H})}{\overline{\theta}}\right)
\right] \nonumber\\
&  \text{ \ \ \ \ }-\sum\nolimits_{k\in\mathcal{T}}\nu_{k}\mathbb{E}\left[
P_{k}(\underline{H})-\overline{P}_{k}\right]  +\sum\nolimits_{k\in\mathcal{T}%
}\lambda_{k}P_{k}(\underline{H})
\end{align}
where $\nu_{k}$, for all $k$, are the dual variables associated with the power
constraints in (\ref{GMARC_Pwr_defn}), and $\lambda_{k}\geq0$ are the dual
variables associated with the positivity constraints ($P_{k}\geq0$) on $P_{k}%
$. Writing the KKT conditions, it is straightforward to verify that the
optimal policy for a user in $\mathcal{S}$ is a function of the channel gains
at the relay while that for a user in $\mathcal{K}\backslash\mathcal{S}$ is a
function of the channel gains only at the destination. In fact, when
$\mathcal{S}$ or $\mathcal{K}\backslash\mathcal{S}$ are singleton sets, the
optimal policy for the user in $\mathcal{S}$ or $\mathcal{K}\backslash
\mathcal{S}$ is simply water-filling over its bottle-neck link to either the
relay (if $\mathcal{S}$) or the destination (if $\mathcal{K}\backslash
\mathcal{S}$). More generally, when $\mathcal{S}$ or $\mathcal{K}%
\backslash\mathcal{S}$ are not singleton sets, the optimal policy is an
opportunistic water-filling solution. Finally, the relay's policy is
water-filling over its direct link to the destination.

\item \textit{Cases }$3a$\textit{, }$3b$\textit{, and }$3c$: The Lagrangian
for these three cases is given by%
\begin{equation}
\mathcal{L}^{\left(  i\right)  }=%
\begin{array}
[c]{cc}%
S^{\left(  i\right)  }-\sum\nolimits_{k\in\mathcal{T}}\nu_{k}\mathbb{E}\left[
P_{k}(\underline{H})-\overline{P}_{k}\right]  +\sum\nolimits_{k\in\mathcal{T}%
}\lambda_{k}P_{k}(\underline{H}) & i=3a,3b,3c
\end{array}
\label{LagK3abc}%
\end{equation}
where
\begin{equation}
S^{\left(  i\right)  }=\left\{
\begin{array}
[c]{ll}%
\mathbb{E}\left[  \theta C\left(  \sum_{k\in\mathcal{K}}\left\vert
H_{r,k}\right\vert ^{2}P_{k}(\underline{H})/\theta\right)  \right]  & i=3a\\
\mathbb{E}\left[  \theta C\left(  \sum_{k\in\mathcal{K}}\left\vert
H_{r,k}\right\vert ^{2}P_{k}(\underline{H})/\theta\right)  +\overline{\theta
}C\left(  \left\vert H_{d,r}\right\vert ^{2}P_{r}(\underline{H})/\overline
{\theta}\right)  \right]  & i=3b\\
\alpha S^{\left(  3a\right)  }+\left(  1-\alpha\right)  S^{\left(  3b\right)
} & i=3c
\end{array}
\right.  \label{SK_3abc}%
\end{equation}
where the dual variable $\alpha$ is associated with the boundary condition
$S^{\left(  3a\right)  }=S^{\left(  3b\right)  }$ for case $3c$. From
(\ref{LagK3abc}) and (\ref{SK_3abc}), for case $3a$, since the dominant bounds
are the MAC bounds at the relay, the optimal user policies involve
opportunistic water-filling over their links to the relay. The optimal
policies take a similar form for case $3b$, except now since the dominant
bounds are the MAC bounds at the destination, each user opportunistically
waterfills over its link to the destination. Finally, for case $3c$, the KKT
conditions satisfied by $P_{k}(\underline{H})$, for all $k$, are%
\begin{equation}%
\begin{array}
[c]{cc}%
\frac{\alpha\left\vert h_{r,k}\right\vert ^{2}}{C\left(  \sum_{k\in
\mathcal{K}}\left\vert H_{r,k}\right\vert ^{2}\frac{P_{k}(\underline{h}%
)}{\theta}\right)  }+\frac{\left(  1-\alpha\right)  \left\vert h_{d,k}%
\right\vert ^{2}}{C\left(  \sum_{k\in\mathcal{K}}\left\vert H_{d,k}\right\vert
^{2}\frac{P_{k}(\underline{h})}{\theta}\right)  }\leq\nu_{k} & \text{with
equality if }P_{k}(\underline{h})>0.
\end{array}
\end{equation}
Thus, as in the two-user analysis for case $3c$ in the Appendix, the optimal
user policies are no longer water-filling but involve opportunistic scheduling
of the users to exploit the multiuser diversity. In fact, the optimal policy
for each user depend on its channel gains to both the relay and the
destination and can be computed using the iterative algorithm detailed in the
Appendix. Finally, for all cases, the optimal relay policy is a water-filling solution.

\item \textit{Boundary cases }$\left(  l,n\right)  $: The Lagrangian for these
cases is given by
\begin{align}
\mathcal{L}^{\left(  l,n\right)  }  &  =\alpha S^{\left(  l\right)  }+\left(
1-\alpha\right)  S^{\left(  n\right)  }-\sum\nolimits_{k\in\mathcal{T}}\nu
_{k}\mathbb{E}\left[  P_{k}(\underline{H})-\overline{P}_{k}\right]
+\sum\nolimits_{k\in\mathcal{T}}\lambda_{k}P_{k}(\underline{H}),\nonumber\\
&
\begin{array}
[c]{cc}
& \text{
\ \ \ \ \ \ \ \ \ \ \ \ \ \ \ \ \ \ \ \ \ \ \ \ \ \ \ \ \ \ \ \ \ \ \ \ \ \ \ \ \ \ \ \ \ \ \ \ \ \ \ \ \ \ \ \ \ \ \ \ \ \ \ \ \ \ }%
l=1,2,n=3a,3b
\end{array}
\\
\mathcal{L}^{\left(  l,3c\right)  }  &  =\alpha_{1}S^{\left(  l\right)
}+\alpha_{2}S^{\left(  3a\right)  }+\left(  1-\alpha_{1}-\alpha_{2}\right)
S^{\left(  3b\right)  }-\sum\nolimits_{k\in\mathcal{T}}\nu_{k}\mathbb{E}%
\left[  P_{k}(\underline{H})-\overline{P}_{k}\right] \nonumber\\
&
\begin{array}
[c]{ccc}
& +\sum\nolimits_{k\in\mathcal{T}}\lambda_{k}P_{k}(\underline{H}), & l=1,2
\end{array}
\end{align}
where $\alpha$ is the dual variable associated with the boundary condition
$S^{\left(  l\right)  }=S^{\left(  n\right)  }$ for $n\not =3c$ and
$\alpha_{1}$ and $\alpha_{2}$ are dual variables associated with the boundary
conditions $S^{\left(  l\right)  }=S^{\left(  3a\right)  }$ and $S^{\left(
3a\right)  }=S^{\left(  3b\right)  }$. Here again the optimal solution for
each is no longer water-filling and depends on the channel gains to both the
relay and the destination. Furthermore, as with case $3c$, here too the
optimal user policies exploit the multiuser diversity to opportunistically
schedule the user transmissions. Finally, the optimal relay policy for all
boundary cases is a water-filling solution over its direct link to the destination.
\end{itemize}

\begin{theorem}
The optimal power policy $\underline{P}^{\ast}(\underline{H})$ that maximizes
the DF sum-rate of an $K$-user ergodic fading orthogonal Gaussian\ MARC is
obtained by computing $\underline{P}^{(i)}(\underline{H})$ and $\underline
{P}^{(l,n)}(\underline{H})$ starting with the inactive cases~$1,2,\ldots
,2^{K}-2,$ followed by the boundary cases $(l,n)$, and finally the active
cases $3a,$ $3b,$ and $3c$ until for some case the corresponding
$\underline{P}^{(i)}(\underline{H})$ or $\underline{P}^{(l,n)}(\underline{H})$
satisfies the case conditions. The optimal \underline{$P$}$^{\ast}%
(\underline{H})$ is given by the optimal $\underline{P}^{(i)}(\underline{H})$
or $\underline{P}^{(l,n)}(\underline{H})$ that satisfies its case conditions
and falls into one of the following three categories:

\textit{Inactive Cases}: The optimal user policy $P_{k}^{\ast}(\underline{H}%
)$, for all $k\in\mathcal{K}$, is multi-user water-filling over its
bottle-neck (rate limiting) link to the relay or the destination. The optimal
relay policy $P_{r}^{\ast}(\underline{H})$ is water-filling over its direct
link to the destination.

\textit{Active Cases }$\left(  3a,3b,3c\right)  $: The optimal user policy
$P_{k}^{\ast}(\underline{H})$, for all $k\in\mathcal{K}$, is opportunistic
water-filling over its link to the relay for case $3a$ and to the destination
for case $3b$. For case $3c$, $P_{k}^{\ast}(\underline{H})$, for all
$k\in\mathcal{K}$, takes an opportunistic non-waterfilling form. The optimal
relay policy $P_{r}^{\ast}$ is water-filling over the relay-destination link.

\textit{Boundary Cases}: The optimal user policy $P_{k}^{\ast}(\underline{H}%
)$, for all $k\in\mathcal{K}$, takes an opportunistic non-water-filling form.
The optimal relay policy $P_{r}^{\ast}(\underline{H})$ is water-filling over
its direct link to the destination.\newline
\end{theorem}

Based on the optimal DF policies, one can conclude that the topology of the
network affects the form of the solution with the classic multiuser
opportunistic waterfilling solutions applicable only for the sources-relay or
the relay-destination clustered models. For all other partially clustered or
non-clustered networks, the solutions are a combination of single- and
multi-user water-filling and non-waterfilling but opportunistic solutions.

\subsection{\label{DF_KRegion}$K$-user Rate Region}

Analogous to the two-user analysis, one can also generalize the sum-rate
analysis above to develop the optimal policies for all points on the boundary
of the $K$-user DF rate region. For brevity, we outline the approach below.

We start with the observation that the DF rate region,$\ \mathcal{R}_{DF}$, is
convex, and thus, every point on the boundary of $\mathcal{R}_{DF}$ is
obtained by maximizing the weighted sum $%
%TCIMACRO{\tsum \nolimits_{k\in\mathcal{K}}}%
%BeginExpansion
{\textstyle\sum\nolimits_{k\in\mathcal{K}}}
%EndExpansion
\mu_{k}R_{k}$, $\mu_{k}>0$ for all $k$. As noted earlier, each point on the
boundary of $\mathcal{R}_{DF}$ is obtained by an intersection of two
polymatroids for some $\underline{P}(\underline{H})$. Thus, analogously to the
sum-rate analysis for $\mu_{k}=1$ for all $k$, for arbitrary $\left(  \mu
_{1},\mu_{2},\ldots,\mu_{K}\right)  $, $%
%TCIMACRO{\tsum \nolimits_{k\in\mathcal{K}}}%
%BeginExpansion
{\textstyle\sum\nolimits_{k\in\mathcal{K}}}
%EndExpansion
\mu_{k}R_{k}$, is maximized by either by an inactive or an active case.

Since the maximum value of $%
%TCIMACRO{\tsum \nolimits_{k\in\mathcal{K}}}%
%BeginExpansion
{\textstyle\sum\nolimits_{k\in\mathcal{K}}}
%EndExpansion
\mu_{k}R_{k}$ over a feasible bounded polyhedron is achieved at a vertex of
the polyhedron, for any $\underline{P}(\underline{H})$, the $\left(
R_{1},R_{2},\ldots,R_{K}\right)  $-tuple maximizing $%
%TCIMACRO{\tsum \nolimits_{k\in\mathcal{K}}}%
%BeginExpansion
{\textstyle\sum\nolimits_{k\in\mathcal{K}}}
%EndExpansion
\mu_{k}R_{k}$ is given by a vertex of a $\mathcal{R}_{r}\left(  \underline
{P}(\underline{H})\right)  \cap\mathcal{R}_{d}\left(  \underline{P}%
(\underline{H})\right)  $ polyhedron at which $%
%TCIMACRO{\tsum \nolimits_{k\in\mathcal{K}}}%
%BeginExpansion
{\textstyle\sum\nolimits_{k\in\mathcal{K}}}
%EndExpansion
\mu_{k}R_{k}$ is a tangent. For the $2^{K}-2$ inactive cases, the polymatroid
intersections are polytopes with constraints on the multiaccess rates of all
users in $\mathcal{S}$ and $\mathcal{K}\backslash\mathcal{S}$ at the relay and
destination, respectively. Since bounds on the multiaccess rates of $l$ users
result in a polymatroid with $l!$ vertices, the intersection of the two
orthogonal sum-rate planes will result in a polytope with $\left(  \left\vert
\mathcal{S}\right\vert !\right)  \left(  \left\vert \mathcal{K}\backslash
\mathcal{S}\right\vert !\right)  $ vertices of which an appropriate vertex
will maximize $%
%TCIMACRO{\tsum \nolimits_{k\in\mathcal{K}}}%
%BeginExpansion
{\textstyle\sum\nolimits_{k\in\mathcal{K}}}
%EndExpansion
\mu_{k}R_{k}$. Each of the $3\left(  2^{K}-2\right)  $ boundary cases are also
characterized by an intersection with $\left(  \left\vert \mathcal{S}%
\right\vert !\right)  \left(  \left\vert \mathcal{K}\backslash\mathcal{S}%
\right\vert !\right)  $ vertices since these active cases are such that only
one point on the sum-rate plane is included in the region of intersection.
Finally, for cases $3a$, $3b$, and $3c$, the intersection of $K$-dimensional
polymatroids results in a $K$-dimensional polyhedron.

In general, the intersection of two polymatroids is not a polymatroid, and
thus, unlike polymatroids, greedy algorithms do not maximize the weighted sum
of rates. This in turn implies that closed form expressions are not in general
possible and determining the optimal power policies requires convex
programming techniques. However, for specific clustered geometries, we present
closed form results.

We write \underline{$\mu$} to denote a vector of weights with entries $\mu
_{k}$, for all $k$. Let $\pi$ be a permutation corresponding to a decreasing
order of the entries of \underline{$\mu$} such that $\pi\left(  k\right)  $ is
the $k^{th}$ entry of $\pi$ and $\pi\left(  j:k\right)  =\left\{  \pi\left(
j\right)  ,\pi\left(  j+1\right)  ,\ldots,\pi\left(  k\right)  \right\}  $.
Thus, $%
%TCIMACRO{\tsum \nolimits_{k\in\mathcal{K}}}%
%BeginExpansion
{\textstyle\sum\nolimits_{k\in\mathcal{K}}}
%EndExpansion
\pi\left(  k\right)  R_{\pi\left(  k\right)  }$ is maximized by a vertex whose
rate tuple $\left(  R_{1},R_{2},\text{ }\ldots,\text{ }R_{K}\right)  $ is such
that $R_{\pi\left(  1\right)  }>R_{\pi\left(  2\right)  }>\ldots>R_{\pi\left(
K\right)  }$, i.e., the decoding order at the vertex is the reverse of the
order of entries of \underline{$\mu$}.

For simplicity, as with the two-user analysis, we summarize the results for
cases $l$, $3a$, and $\left(  l,3a\right)  $, for all $l\in\{1,2,\ldots
,2^{K}-2\}.$ The results for the other cases follow naturally from discussions
for these cases.

\textit{Case }$l:$ This case results when the sum-rate plane at the relay for
users in $\mathcal{S\subset K}$ intersects the sum-rate plane at the
destination for the complementary users in $\mathcal{K}\backslash\mathcal{S}$.
For a permutation $\pi$ with decreasing order of the entries in \underline
{$\mu$}, let $\pi_{\mathcal{A}}$ be the decreasing order of the entries of
\underline{$\mu$} for the users in $\mathcal{A}\subset\mathcal{K}$ . The
weighted rate-sum can be expanded as
\begin{equation}%
%TCIMACRO{\tsum \nolimits_{k=1}^{K}}%
%BeginExpansion
{\textstyle\sum\nolimits_{k=1}^{K}}
%EndExpansion
\pi\left(  k\right)  R_{\pi\left(  k\right)  }=%
%TCIMACRO{\tsum \nolimits_{k\in\mathcal{S}}}%
%BeginExpansion
{\textstyle\sum\nolimits_{k\in\mathcal{S}}}
%EndExpansion
\pi_{\mathcal{S}}\left(  k\right)  R_{\pi_{\mathcal{S}}\left(  k\right)  }+%
%TCIMACRO{\tsum \nolimits_{k\in\mathcal{S}\backslash\mathcal{K}}}%
%BeginExpansion
{\textstyle\sum\nolimits_{k\in\mathcal{S}\backslash\mathcal{K}}}
%EndExpansion
\pi_{\mathcal{K}\backslash\mathcal{S}}\left(  k\right)  R_{\pi\left(
k\right)  }. \label{DFKRSum}%
\end{equation}
where (\ref{DFKRSum}) is maximized by choosing the rates $R_{\pi\left(
k\right)  }$ for all $k$ as%
\begin{align}
&
\begin{array}
[c]{cc}%
R_{\pi_{\mathcal{S}}\left(  1\right)  }=\mathbb{E}\left[  C\left(
H_{r,\pi_{\mathcal{S}}\left(  1\right)  }P_{\pi_{\mathcal{S}}\left(  1\right)
}\right)  \right]  &
\end{array}
\label{ClRS1}\\
&
\begin{array}
[c]{cc}%
R_{\pi_{\mathcal{S}}\left(  k\right)  }=\mathbb{E}\left[  C\left(
\frac{\left\vert H_{r,\pi_{\mathcal{S}}\left(  k\right)  }\right\vert
^{2}P_{\pi_{\mathcal{S}}\left(  k\right)  }}{1+%
%TCIMACRO{\tsum \nolimits_{j=1}^{k-1}}%
%BeginExpansion
{\textstyle\sum\nolimits_{j=1}^{k-1}}
%EndExpansion
\left\vert H_{d,\pi_{\mathcal{S}}\left(  j\right)  }\right\vert ^{2}%
P_{\pi_{\mathcal{S}}\left(  j\right)  }}\right)  \right]  &
k=2,3,..,\left\vert \mathcal{S}\right\vert
\end{array}
\end{align}
and
\begin{align}
&
\begin{array}
[c]{cc}%
R_{\pi_{\mathcal{K}\backslash\mathcal{S}}\left(  1\right)  }=\mathbb{E}\left[
C\left(  \left\vert H_{d,\pi_{\mathcal{K}\backslash\mathcal{S}}\left(
1\right)  }\right\vert ^{2}P_{\pi_{\mathcal{K}\backslash\mathcal{S}}\left(
1\right)  }\right)  \right]  &
\end{array}
\\
&
\begin{array}
[c]{cc}%
R_{\pi_{\mathcal{K}\backslash\mathcal{S}}\left(  k\right)  }=\mathbb{E}\left[
C\left(  \frac{\left\vert H_{d,\pi_{\mathcal{K}\backslash\mathcal{S}}\left(
k\right)  }\right\vert ^{2}P_{\pi_{\mathcal{K}\backslash\mathcal{S}}\left(
k\right)  }}{1+%
%TCIMACRO{\tsum \nolimits_{j=1}^{k-1}}%
%BeginExpansion
{\textstyle\sum\nolimits_{j=1}^{k-1}}
%EndExpansion
\left\vert H_{d,\pi_{\mathcal{K}\backslash\mathcal{S}}\left(  j\right)
}\right\vert ^{2}P_{\pi_{\mathcal{K}\backslash\mathcal{S}}\left(  j\right)  }%
}\right)  \right]  & k=2,3,..,\left\vert \mathcal{K}\backslash\mathcal{S}%
\right\vert .
\end{array}
\label{ClRKSk}%
\end{align}
Thus, the users in $\mathcal{S}$ and $\mathcal{K}\backslash\mathcal{S}$ are
decoded in the increasing order of their weights at the relay and destination
respectively. The optimal power and rate allocation for the users in
$\mathcal{S}$ and $\mathcal{K}\backslash\mathcal{S}$ are the multiuser
opportunistic water-filling solutions at the relay and destination,
respectively, and can be computed using a \textit{utility function }approach
developed in \cite[II.C]{cap_theorems:TH01}.

\textit{Case 3a}: The polytope resulting from the intersection of two
polymatroids is defined by the constraints
\begin{align}
&
\begin{array}
[c]{cc}%
R_{\mathcal{S}}\leq\min\left\{  \mathbb{E}\left[  C\left(
%TCIMACRO{\tsum \nolimits_{k\in\mathcal{S}}}%
%BeginExpansion
{\textstyle\sum\nolimits_{k\in\mathcal{S}}}
%EndExpansion
\left\vert H_{r,k}\right\vert ^{2}P_{k}\right)  \right]  ,\mathbb{E}\left[
C\left(
%TCIMACRO{\tsum \nolimits_{k\in\mathcal{S}}}%
%BeginExpansion
{\textstyle\sum\nolimits_{k\in\mathcal{S}}}
%EndExpansion
\left\vert H_{d,k}\right\vert ^{2}P_{k}\right)  \right]  \right\}  & \text{for
all }\mathcal{S\subset K}%
\end{array}
\label{PT1}\\
&
\begin{array}
[c]{ccc}%
\text{and} & R_{\mathcal{K}}\leq\mathbb{E}\left[  C\left(
%TCIMACRO{\tsum \nolimits_{k\in\mathcal{K}}}%
%BeginExpansion
{\textstyle\sum\nolimits_{k\in\mathcal{K}}}
%EndExpansion
\left\vert H_{r,k}\right\vert ^{2}P_{k}\right)  \right]  . &
\end{array}
\label{PT2}%
\end{align}
However, since the polytope given by (\ref{PT1}) and (\ref{PT2}) above is in
general not a polymatroid, greedy algorithms cannot be used to maximize the
weighted sums and thus developing closed form solutions for this case is not
possible in general. However, the optimal policies maximizing the weighted sum
of rates can be computed in strongly polynomial time\footnote{An algorithm is
said to run in strongly polynomial time when the algorithm run time is
independent of the numerical data size and is dependent only on the inherent
dimensions of the problem. In contrast polynomial time algorithms are
characterized by run times that are polynomial not in the size of the input
but the numerical value of the input which may be exponentially large.}
\cite[Theorem 47.4]{cap_theorems:Schrijver01}.

\begin{remark}
For the special case where the bounds at the relay are smaller than the bounds
at the destination for all $\mathcal{S}$, i.e., $\mathcal{R}_{r}%
\subset\mathcal{R}_{d},$ the optimal user policies are multiuser water-filling
solutions developed in \cite[II.C]{cap_theorems:TH01} with the relay as the
receiver. Note that this condition implies that all possible subset of users
achieve better rates at the destination than at the relay. This can happen
when either all users are clustered closer to the destination or when the
relay has a relatively high SNR link to the destination sufficient enough to
achieve rate gains for all users at the destination.
\end{remark}

\begin{remark}
Similarly, for case $3b,$ for the special case in which $\mathcal{R}%
_{d}\subset\mathcal{R}_{r}$, the optimal user policies are multiuser
water-filling solutions with the destination as the receiver. In the following
section we show that DF achieves the capacity region when case $3b$ holds for
all points on the boundary of the outer bound rate region. In fact, this
condition implies that all possible subset of users achieve better rates at
the relay than they do at the destination which in turn suggests a geometry
where all subsets of users are clustered closer to the relay than to the
destination. The optimal relay policy in all cases is a waterfilling solution
over its link to the destination.
\end{remark}

\textit{Boundary case }$\left(  l,3a\right)  $: Recall that a boundary case
$\left(  l,n\right)  $ results when the $K$-user sum-rate for the active case
$n$ is equal to that for the inactive case $l$. The resulting region of
intersection, analogous to the inactive cases, is a polytope with $\left(
\mathcal{S}!\right)  \left(  \mathcal{K}\backslash\mathcal{S}!\right)  $
vertices. The weighted optimization $%
%TCIMACRO{\tsum \nolimits_{k\in\mathcal{K}}}%
%BeginExpansion
{\textstyle\sum\nolimits_{k\in\mathcal{K}}}
%EndExpansion
\pi\left(  k\right)  R_{\pi\left(  k\right)  }$ for case $\left(  l,3a\right)
$ simplifies to
\begin{equation}%
\begin{array}
[c]{l}%
%TCIMACRO{\tsum \nolimits_{k\in\mathcal{K}}}%
%BeginExpansion
{\textstyle\sum\nolimits_{k\in\mathcal{K}}}
%EndExpansion
\pi\left(  k\right)  R_{\pi\left(  k\right)  }\\
\text{s.t. }\left(  R_{\mathcal{K}}\right)  _{r}=%
%TCIMACRO{\tsum \nolimits_{k\in\mathcal{K}}}%
%BeginExpansion
{\textstyle\sum\nolimits_{k\in\mathcal{K}}}
%EndExpansion
\pi\left(  k\right)  R_{\pi\left(  k\right)  }=%
%TCIMACRO{\tsum _{k\in\mathcal{S}}}%
%BeginExpansion
{\textstyle\sum_{k\in\mathcal{S}}}
%EndExpansion
R_{\pi_{\mathcal{S}}\left(  k\right)  }+%
%TCIMACRO{\tsum \nolimits_{k\in\mathcal{S}\backslash\mathcal{K}}}%
%BeginExpansion
{\textstyle\sum\nolimits_{k\in\mathcal{S}\backslash\mathcal{K}}}
%EndExpansion
R_{\pi\left(  k\right)  }%
\end{array}
\label{KRegion_13a}%
\end{equation}
where $R_{\pi_{\mathcal{S}}\left(  k\right)  }$ and $R_{\pi_{\mathcal{K}%
\backslash\mathcal{S}}\left(  k\right)  }$ are given by (\ref{ClRS1}%
)-(\ref{ClRKSk}). Here again, given the complexity of the optimization, closed
form solutions are difficult to obtain. However, as before, one can compute
the optimal policies and the rate tuple maximizing (\ref{KRegion_13a}) in
polynomial time using combinatorial methods.

\begin{remark}
The discussion here for cases $3a$ and $\left(  1,3a\right)  $ also applies to
the other active (including boundary) cases. In each such case, the optimal
policies depend on all the Lagrange dual variables, with each variable
reflecting a specific constraint.
\end{remark}

\section{\label{Sec_OB}Outer Bounds}

An outer bound on the capacity region $C_{MARC}$ of a $K$-user full-duplex
MARC is presented in \cite{cap_theorems:SKM02a} (see also \cite[Th.
1]{cap_theorems:LSMandPoor_01})\ using cut-set bounds as applied to the case
of independent sources and we summarize it below.

\begin{proposition}
[{\cite[Th. 1]{cap_theorems:LSMandPoor_01}}]\label{Prop_OB}The capacity region
$\mathcal{C}_{\text{MARC}}$ is contained in the union of the set of rate
tuples $(R_{1},R_{2},\ldots,R_{K})$ that satisfy, for all $\mathcal{S}%
\subseteq\mathcal{K}$,
\begin{equation}
R_{\mathcal{S}}\leq\min\left\{  I(X_{\mathcal{S}};Y_{r},Y_{d}|X_{\mathcal{S}%
^{c}},X_{r},U),I(X_{\mathcal{S}},X_{r};Y_{d}|X_{\mathcal{S}^{c}},U)\right\}
\label{MARC_OB_cutset}%
\end{equation}
where the union is over all distributions that factor as%
\begin{equation}
p(u)\cdot\left(  \prod\nolimits_{k=1}^{K}p(x_{k}|u)\allowbreak\right)  \cdot
p(x_{r}|\allowbreak x_{\mathcal{K}}\allowbreak,u)\cdot p(y_{r},y_{d}%
|x_{\mathcal{K}},x_{r}). \label{GMARC_converse_inpdist}%
\end{equation}

\end{proposition}

\begin{remark}
The \textit{time-sharing} random variable $U\in\mathcal{U}$ ensures that the
region in (\ref{MARC_OB_cutset}) is convex. One can apply Caratheodory's
theorem \cite{cap_theorems:Eggbook01} to this $K$-dimensional convex region to
bound the cardinality of $\mathcal{U}$ as $\left\vert \mathcal{U}\right\vert
\leq K+1$.
\end{remark}

\begin{proposition}
For the orthogonal MARC the cutset bounds in (\ref{MARC_OB_cutset}) specialize
as
\begin{equation}
R_{\mathcal{S}}\leq%
\begin{array}
[c]{cc}%
\min\left\{  \theta I(X_{\mathcal{S}};Y_{r}Y_{d,1}|X_{\mathcal{S}^{c}%
},U),\theta I(X_{\mathcal{S}};Y_{d,1}|X_{\mathcal{S}^{c}},U)+\overline{\theta
}I(X_{r};Y_{d,2}|U)\right\}  & \text{for all }\mathcal{S}\subseteq\mathcal{K}%
\end{array}
\label{Orth_MARC_OB}%
\end{equation}
where the union is taken over all distributions that factor as%
\begin{equation}
p(u)\cdot\left[  \theta\cdot\left(  \prod\nolimits_{k=1}^{K}p(x_{k}%
|u)\allowbreak\right)  \cdot p(y_{r}y_{d}|x_{\mathcal{K}})+\overline{\theta
}\cdot p(x_{r}|\allowbreak u)\cdot p(y_{d}|x_{r})\right]  .
\label{Orth_OB_dist}%
\end{equation}

\end{proposition}

\begin{remark}
The above bounds can also be obtained by using a mode variable $M_{r}$ to
denote the half-duplex listen and transmit states at the relay such that
$M_{r}$ is in the listen and transmit states with probabilities $\theta$ and
$1-\theta$, respectively (see \cite{cap_theorems:GKR01}). The instantaneous
relay mode is assumed known at all nodes, such that (\ref{Orth_MARC_OB})
results from conditioning the bounds in (\ref{MARC_OB_cutset}) on $M_{r}$, and
(\ref{Orth_OB_dist}) from replacing $X_{r}$ with $\left(  X_{r},M_{r}\right)
$ in (\ref{GMARC_converse_inpdist}) and expanding the resulting joint distribution.
\end{remark}

\begin{remark}
The joint distribution for the cutset bounds in (\ref{Orth_OB_dist}) is the
same as that for DF\ in (\ref{DF_HD_dist}). This is in contrast to the
full-duplex MARC where in general, the two distributions (and bounds) are not
the same.
\end{remark}

\begin{theorem}
For a degraded orthogonal discrete memoryless MARC where $X_{\mathcal{S}%
}-Y_{r}-Y_{d}$ form a Markov chain, DF achieves the capacity region of a
degraded orthogonal MARC.
\end{theorem}

\begin{proof}
The proof follows directly from applying the Markov property $X_{\mathcal{S}%
}-Y_{r}-Y_{d}$ to the cutset bounds in (\ref{Orth_MARC_OB}) and comparing the
resulting bounds with those for DF in (\ref{DF_HD_Rates}). Note that for the
full-duplex degraded MARC, the inner and outer bounds are not the same in
general. In fact, for the degraded Gaussian (full-duplex)\ MARC, it has been
recently shown in \cite{cap_theorems:LSMandPoor_01} that DF achieves the
sum-capacity when the intersection of the two polymatroids at the relay and
destination belongs to the set of active cases.
\end{proof}

For an orthogonal Gaussian MARC with fixed $\underline{H}$ and $\theta$, using
a conditional entropy theorem, one can show that Gaussian signals maximize the
bounds in (\ref{Orth_MARC_OB}). Thus, substituting $X_{k}\sim\mathcal{CN}%
\left(  0,P_{k}/\theta\right)  $, $k=1,2$, and $X_{r}\sim\mathcal{CN}\left(
0,P_{r}/\overline{\theta}\right)  $ in (\ref{Orth_MARC_OB}), we have
\begin{equation}
R_{\mathcal{S}}\leq\min\left(  \theta\log\left\vert I+\sum\limits_{k\in
\mathcal{S}}\mathbf{G}_{k}\left.  P_{k}\right/  \theta\right\vert ,\theta
C\left(  \sum\limits_{k\in\mathcal{S}}\left\vert H_{d,k}\right\vert
^{2}\left.  P_{k}\right/  \theta\right)  +\overline{\theta}C\left(  \left\vert
H_{d,r}\right\vert ^{2}\left.  P_{r}\right/  \overline{\theta}\right)
\right)
\end{equation}
where
\begin{equation}
\mathbf{G}_{k}=\left[
\begin{array}
[c]{cc}%
H_{r,k} & H_{d,k}%
\end{array}
\right]  ^{T}\left[
\begin{array}
[c]{cc}%
H_{r,k}^{\ast} & H_{d,k}^{\ast}%
\end{array}
\right]
\end{equation}
and $H_{\left(  \cdot\right)  }^{\ast}$ is the complex conjugate of
$H_{\left(  \cdot\right)  }$. Using the fact that the ergodic channel is a
collection of parallel non-fading channels, the capacity region of an ergodic
fading orthogonal Gaussian MARC is given by the following theorem.

\begin{theorem}
\label{Th_R_OB}The capacity region $\mathcal{C}_{O-MARC}$ of an ergodic fading
orthogonal Gaussian MARC is contained in%
\begin{equation}
\mathcal{R}_{OB}=\bigcup\limits_{\underline{P}\in\mathcal{P}}\left\{
\mathcal{R}_{1}\left(  \underline{P}\right)  \cap\mathcal{R}_{2}\left(
\underline{P}\right)  \right\}  \label{ROB}%
\end{equation}
where, for all $\mathcal{S}\subseteq\mathcal{K}$, we have
\begin{equation}
\mathcal{R}_{1}\left(  \underline{P}\right)  =\left\{  \left(  R_{1}%
,R_{2}\right)  :R_{\mathcal{S}}\leq\mathbb{E}\left[  \theta\log\left\vert
I+\sum_{k\in\mathcal{S}}\mathbf{G}_{k}\left.  P_{k}(\underline{H})\right/
\theta\right\vert \right]  \right\}  \label{OB_Rr}%
\end{equation}
and%
\begin{equation}
\mathcal{R}_{2}\left(  \underline{P}\right)  =\left\{  \left(  R_{1}%
,R_{2}\right)  :R_{\mathcal{S}}\leq\mathbb{E}\left[  \theta C\left(
\sum\limits_{k\in\mathcal{S}}\left\vert H_{d,k}\right\vert ^{2}\left.
P_{k}(\underline{H})\right/  \theta\right)  +\overline{\theta}C\left(
\left\vert H_{d,r}\right\vert ^{2}\left.  P_{r}(\underline{H})\right/
\overline{\theta}\right)  \right]  \right\}  . \label{OB_Rd}%
\end{equation}

\end{theorem}

\begin{remark}
Comparing outer bounds in (\ref{OB_Rd}) with the DF bounds in
(\ref{DF_Rates_dest}), we see that the bounds at the destination are the same
in both cases. However, unlike the DF bound at only the relay in
(\ref{DF_Rates_rel}), the cutset bounds in (\ref{OB_Rr}) is a SIMO bound with
single-antenna transmitters and both the relay and the destination acting as a
multi-antenna receiver.
\end{remark}

The expressions in (\ref{OB_Rr}) and (\ref{OB_Rd}) are concave functions of
$P_{k}(\underline{H})$, for all $k$, and thus, the region $\mathcal{R}_{OB}$
is convex. Thus, as in Theorem \ref{DF_Th1}, the region $\mathcal{R}_{OB}$ in
(\ref{ROB}) is a union of the intersections of the regions $\mathcal{R}%
_{1}(\underline{P}(\underline{H}))$ and $\mathcal{R}_{2}(\underline
{P}(\underline{H}))$, where the union is taken over all $\underline
{P}(\underline{H})\in$ $\mathcal{P}$ and each point on the boundary of
$\mathcal{R}_{DF}$ is obtained by maximizing the weighted sum $\mu_{1}R_{1}$
$+$ $\mu_{2}R_{2}$ over all $\underline{P}(\underline{H})\in\mathcal{P}$, and
for all $\mu_{1}>0$, $\mu_{2}>0$. In \cite{cap_theorems:LSMandPoor_01}, it is
shown that the rate polytopes satisfying (\ref{MARC_OB_cutset}) are
polymatroids. Since, the polytopes in (\ref{OB_Rr}) and (\ref{OB_Rd}) are
obtained from (\ref{MARC_OB_cutset}) for the special case of orthogonal
signaling, one can verify in a straightforward manner using Definition
\ref{Def_PolyM} that these are polymatroids as well.

\subsection{Optimal Sum-rate Policies and Sum-capacity}

Since $\mathcal{R}_{OB}$ is obtained completely as a union of the intersection
of polymatroids, one for each choice of power policy, Lemma
\ref{Lemma_PolyIntersect} can be applied to explicitly characterize the outer
bounds on the sum-rate. Thus, the maximum sum-rate tuple is achieved by an
intersection that belongs to either the active set or to the inactive set. Let
$l=1,2,\ldots,2^{K}-2$, index the $2^{K}-2$ non-empty subsets of $\mathcal{K}%
$. For a $K$-user MARC, there are $\left(  2^{K}-2\right)  $ possible
intersections of the inactive kind with sum-rate $J^{\left(  l\right)  }$
given by%
\begin{equation}%
\begin{array}
[c]{ll}%
\underline{\text{case}\ l}:J^{\left(  l\right)  }=R_{\mathcal{S}%
,1}+R_{\mathcal{K}\backslash\mathcal{S},2} & l=1,2,\ldots,2^{K}-2\\
s.t.\text{ }R_{\mathcal{S},1}<R_{\mathcal{S},2}^{\min}\text{ and
}R_{\mathcal{K}\backslash\mathcal{S},2}<R_{\mathcal{K}\backslash\mathcal{S}%
,1}^{\min} &
\end{array}
\label{OBK_InA_Cases}%
\end{equation}
where $R_{\mathcal{A},j}$ and $R_{\mathcal{A},j}^{\min}$ are as defined
in\ Section \ref{section 3} and for $j=1,2,$ are given by the bounds in
(\ref{OB_Rr}) and (\ref{OB_Rd}), respectively. The sum-rates $J^{\left(
i\right)  }$, $i=$ $3a,3b,3c$, are
\begin{align}
J^{\left(  i\right)  }  &  =R_{\mathcal{K},j}\text{ for }\left(  i,j\right)
=\left(  3a,1\right)  ,\left(  3b,2\right) \\
J^{\left(  3c\right)  }  &  =R_{\mathcal{K},1}\text{ }s.t.\text{
}R_{\mathcal{K},1}=R_{\mathcal{K},2}.
\end{align}
Finally, the sum-rate $J^{\left(  l,n\right)  }$, for the $3\left(
2^{K}-2\right)  $ boundary cases, enumerated as cases $\left(  l,n\right)  $,
$l=1,2,\ldots,2^{K}-2$, $n=3a,3b,3c$, are
\begin{align}
&
\begin{array}
[c]{l}%
\underline{\text{case}~\left(  l,3a\right)  }:J^{\left(  l,3a\right)
}=R_{\mathcal{K},1}\\
s.t.\text{ }R_{\mathcal{K},1}=R_{\mathcal{S},1}+R_{\mathcal{K}\backslash
\mathcal{S},2}\text{ for case }l
\end{array}
\label{OBK_l3a}\\
&
\begin{array}
[c]{l}%
\underline{\text{case}~\left(  l,3b\right)  }:J^{\left(  l,3b\right)
}=R_{\mathcal{K},2}\\
s.t.\text{ }R_{\mathcal{K},2}=R_{\mathcal{S},1}+R_{\mathcal{K}\backslash
\mathcal{S},2}\text{ for case }l
\end{array}
\label{OBK_l3b}\\
&
\begin{array}
[c]{l}%
\underline{\text{case}~\left(  l,3c\right)  }:J^{\left(  l,3c\right)
}=R_{\mathcal{K},1}\\
s.t.\text{ }R_{\mathcal{K},1}=R_{\mathcal{K},2}=R_{\mathcal{S},1}%
+R_{\mathcal{K}\backslash\mathcal{S},2}\text{ for case }l
\end{array}
\label{OBK_l3c}%
\end{align}
where the subset $\mathcal{S}$ is chosen to correspond to the appropriate case
$l$.

The $K$-user sum-rate optimization problem for case $i$ and case $\left(
l,n\right)  $ is
\begin{equation}
\frame{$%
\begin{array}
[c]{l}%
\max\limits_{\underline{P}\in\mathcal{P}}J^{\left(  i\right)  }\text{ or }%
\max\limits_{\underline{P}\in\mathcal{P}}J^{\left(  l,n\right)  }\\%
\begin{array}
[c]{cc}%
s.t. & \mathbb{E}\left[  P_{k}(\underline{H})\right]  \leq\overline{P}%
_{k}\text{, }k=1,2,r
\end{array}
\\%
\begin{array}
[c]{cc}%
\text{ \ \ \ \ \ } & P_{k}(\underline{H})\geq0\text{, }k=1,2,r.
\end{array}
\end{array}
$}%
\end{equation}
An inactive case $l$ results when the conditions for that case in
(\ref{OBK_InA_Cases}) are satisfied. A boundary case results when one of the
conditions in (\ref{OBK_l3a})-(\ref{OBK_l3c}) is satisfied for the appropriate
$\left(  l,n\right)  $ case. Finally, case~$3a$ or $3b$ or $3c$ results when
the conditions for neither the inactive nor the boundary cases are satisfied.

As in Section \ref{Sec_4}, the optimization for each case involves writing the
Lagrangian and the KKT conditions. The optimal policy $\underline{P}^{\left(
ob\right)  }(\underline{H})$ satisfies the conditions for only one of the
cases. For brevity and to avoid repetition, we summarize the details below.

\begin{itemize}
\item \textit{Inactive cases}: The Lagrangian for these cases involves a sum
of the MIMO\ cutset bounds in (\ref{OB_Rr}) for users in $\mathcal{S}$, for
some $\mathcal{S}$, and the cutset bounds at the destination in (\ref{OB_Rd})
for the remaining users in $\mathcal{K}\backslash\mathcal{S}$. Thus, the
optimal policy for a user in $\mathcal{S}$ is a function of the channel gains
at both the relay and destination while that for a user in $\mathcal{K}%
\backslash\mathcal{S}$ is a function of the channel gains only at the
destination. For $K>2$, using the results in \cite[Theorem 1]%
{cap_theorems:Yu_Rhee} for ergodic fading SIMO-MAC channels, the policies for
the users in $\mathcal{S}$ are water-filling and allow at most $l^{2}=4$ users
to transmit simultaneously, where $l$ is the number of antennas at the
receiver. Furthermore, the optimal user policies can be obtained using an
iterative water-filling approach \cite{cap_theorems:Yu_Rhee02}. On the other
hand, the SISO-MAC bounds for the users in $\mathcal{K}\backslash\mathcal{S}$
result in a multiuser opportunistic water-filling solution. Finally, the
relay's policy is water-filling over its direct link to the destination.

\item \textit{Cases }$3a$\textit{, }$3b$\textit{, and }$3c$: For case $3a$,
the dominant bounds are the SIMO cut-set bounds, and thus, as discussed for
the inactive cases, the optimal policy is water-filling for each user such
that a maximum of 4 users can transmit simultaneously. On the other hand for
case $3b$, the dominant bounds are the cooperative bounds at the destination
and the optimal policy for each user is an opportunistic water-filling
solution. Finally, for case $3c$, as one would expect, the optimal policies
are no longer water-filling. In all cases, the optimal relay policy is a
water-filling solution.

\item \textit{Boundary cases }$\left(  l,n\right)  $: The Lagrangian for these
cases is a weighted sum of the sum-rates for one of cases $3a$, $3b$, or $3c$
and one of the inactive cases. Here again the optimal solution for each is no
longer water-filling and depends on the channel gains to both the relay and
the destination. As with the other cases, here too the optimal relay policy
for all boundary cases is a water-filling solution.
\end{itemize}

Comparing these optimal policies with that for DF, we have the following
capacity theorem.

\begin{theorem}
The sum-capacity of a $K$-user ergodic fading orthogonal Gaussian MARC is
achieved by DF when the optimal policy $\underline{P}^{\left(  ob\right)
}(\underline{H})$ for the cutset bounds satisfies the conditions for case $3b$
and for no other case.
\end{theorem}

\begin{proof}
The proof follows from comparing the expressions $J^{\left(  \cdot\right)  }$
for all cases in (\ref{DF_Jdef}) and (\ref{OBK_InA_Cases})-(\ref{OBK_l3c}) for
the inner and outer bounds, respectively. For all cases where the SIMO cut-set
bound dominates the sum-rate, the cutset bounds do not match the DF bounds.
Thus, when the optimal policy $\underline{P}^{\left(  ob\right)  }%
(\underline{H})$ satisfies the conditions for case $3b$, where the sum-rate
bounds at the destination dominate, DF achieves capacity.
\end{proof}

\begin{remark}
Recall that case $3b$ corresponds to a clustered geometry in which the relay
is clustered with all sources such that the cooperative multiaccess link from
the sources and the relay to the destination is the bottleneck link.
\end{remark}

\begin{remark}
The set of power policies, $\mathcal{B}^{\left(  i\right)  }$ and
$\mathcal{B}^{\left(  l,n\right)  }$, are defined by the conditions in
(\ref{OBK_InA_Cases})-(\ref{OBK_l3c}). Note that these conditions are in
general not the same as those for DF. Thus, the set $\mathcal{B}^{\left(
3b\right)  }$ for the cut-set outer bound will in general not be exactly the
same as that for the inner DF bound. However, when case $3b$ maximizes the
cut-set outer bounds the optimal DF $P^{\ast}(\underline{H})=P^{\left(
ob\right)  }(\underline{H})=P^{\left(  3b\right)  }(\underline{H})$ belongs to
$\mathcal{B}^{\left(  3b\right)  }$ for both bounds.
\end{remark}

\subsection{Outer Bounds Rate Region: Optimal Policy and Capacity\ Region}

One can similarly write the rate expressions and the KKT conditions for every
point on the boundary of $\mathcal{R}_{OB}.$ Such an analysis will be similar
to that for the $K$-user orthogonal MARC under DF developed in Section
\ref{DF_KRegion}. From Theorem \ref{Th_R_OB}, every point $\sum_{k\in
\mathcal{K}}\mu_{k}R_{k}$ on $\mathcal{R}_{OB}$ results from an intersection
of two polymatroids. For those cases in which the intersection is an inactive
case, both the SIMO cut-set bound at the relay and destination and the
cooperative cut-set bound at the destination are involved, and thus, one
cannot achieve capacity. This is also true for the boundary cases. For cases
$3a$, $3b$, and $3c$, in which the polymatroid intersection also has $2^{K}-1$
constraints, and hence, $K!$ corner points on the dominant $K$-user sum-rate
face, $\sum_{k\in\mathcal{K}}\mu_{k}R_{k}$ is maximized by a corner point of
the resulting polytope. Since any polytope that results from some or all of
the SIMO bounds will be larger than the corresponding DF inner bounds, the
cut-set bounds are tight only when $\mathcal{R}_{2}\left(  \underline
{P}^{\left(  ob\right)  }(\underline{H}\mathbf{,}\mu_{1},\mu_{2})\right)
\subset\mathcal{R}_{1}\left(  \underline{P}^{\left(  ob\right)  }%
(\underline{H}\mathbf{,}\mu_{1},\mu_{2})\right)  $ where $\underline
{P}^{\left(  ob\right)  }(\underline{H}\mathbf{,}\mu_{1},\mu_{2})$ denotes the
power policy maximizing $\sum_{k\in\mathcal{K}}\mu_{k}R_{k}$. We summarize
this observation in the following theorem.

\begin{theorem}
\label{Th_OBRateReg}The capacity region $\mathcal{C}_{O-MARC}$ of an ergodic
orthogonal Gaussian MARC is achieved by DF when for every point $\sum
_{k\in\mathcal{K}}\mu_{k}R_{k}$ on $\mathcal{R}_{OB}$ achieved by
$\underline{P}^{\left(  ob\right)  }(\underline{H}\mathbf{,}\mu_{1},\mu_{2}%
)$,
\begin{equation}
\mathcal{R}_{2}\left(  \underline{P}^{\left(  ob\right)  }(\underline
{H}\mathbf{,}\mu_{1},\mu_{2})\right)  \subset\mathcal{R}_{1}\left(
\underline{P}^{\left(  ob\right)  }(\underline{H}\mathbf{,}\mu_{1},\mu
_{2})\right)
\end{equation}
such that $\underline{P}^{\left(  ob\right)  }(\underline{H}\mathbf{,}\mu
_{1},\mu_{2})=\underline{P}^{\left(  3b\right)  }(\underline{H}\mathbf{,}%
\mu_{1},\mu_{2})$. Thus, $\mathcal{C}_{O-MARC}$ is given by%
\begin{equation}
\mathcal{C}_{O-MARC}=\mathcal{R}_{2}\left(  \underline{P}^{\left(  3b\right)
}(\underline{H}\mathbf{,}\mu_{1},\mu_{2})\right)  =\mathcal{R}_{d}\left(
\underline{P}^{\left(  3b\right)  }(\underline{H}\mathbf{,}\mu_{1},\mu
_{2})\right)  .
\end{equation}

\end{theorem}

\begin{proposition}
[{\cite[Theorem 9]{cap_theorems:KGG_IT}}]For the case in which $\underline{H}$
has uniform phase fading and the channel state information is not known at the
transmitters such that $P_{k}^{\left(  ob\right)  }(\underline{H}%
\mathbf{)}=P_{k}^{\left(  3b\right)  }(\underline{H}\mathbf{)}=\overline
{P}_{k}$, for all $k\in\mathcal{T}$, Theorem \ref{Th_OBRateReg} yields the
capacity region of an ergodic phase fading orthogonal Gaussian\ MARC as
developed in \cite[Theorem 9]{cap_theorems:KGG_IT}.
\end{proposition}

\subsection{Illustration of Results}

We present numerical results for a two-user orthogonal MARC with Rayleigh
fading links. We model the channel fading gains between receiver $m$ and
transmitter $k$, for all $k$ and $m$, as
\begin{equation}
H_{m,k}=\frac{A_{m,k}}{\sqrt{d_{m,k}^{\gamma}}}%
\end{equation}
where $d_{m,k}$ is the distance between the transmitter and receiver, $\gamma$
is the path-loss exponent, and $A_{m,k}$ is a circularly symmetric complex
Gaussian random variable with zero mean and unit variance such that
$\left\vert H_{m,k}\right\vert ^{2}$ is Rayleigh distributed with zero mean
and variance $1/d_{m,k}^{\gamma}$. For the purpose of our illustration, we set
$\gamma=3$.%

%TCIMACRO{\TeXButton{B}{\begin{figure*}[tbp] \centering}}%
%BeginExpansion
\begin{figure*}[tbp] \centering
%EndExpansion%
%TCIMACRO{\FRAME{itbpF}{4.0638in}{2.0591in}{0in}{}{}{one_geometries.eps}%
%{\special{ language "Scientific Word";  type "GRAPHIC";  display "USEDEF";
%valid_file "F";  width 4.0638in;  height 2.0591in;  depth 0in;
%original-width 3.5993in;  original-height 1.5912in;  cropleft "0";
%filename 'One_geometries.eps';file-properties "XNPEU";}}}%
%BeginExpansion
{\includegraphics[
height=2.0591in,
width=4.0638in
]%
{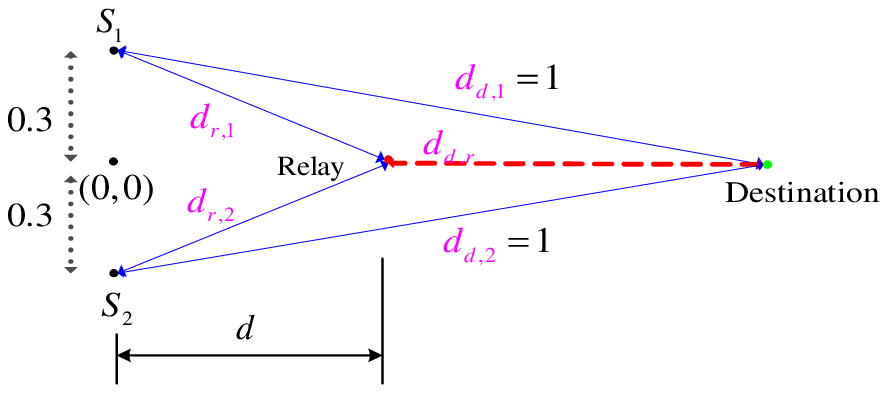}%
}%
%EndExpansion
\caption{A symmetric two-user MARC geometry.}\label{Illus_Geom}%
%TCIMACRO{\TeXButton{E}{\end{figure*}}}%
%BeginExpansion
\end{figure*}%
%EndExpansion

\begin{figure*}[tbp] \centering
%EndExpansion%
%TCIMACRO{\FRAME{itbpF}{4.184in}{3.0692in}{0in}{}{}{sumrate_df_ob_symm.eps}%
%{\special{ language "Scientific Word";  type "GRAPHIC";  display "USEDEF";
%valid_file "F";  width 4.184in;  height 3.0692in;  depth 0in;
%original-width 5.5287in;  original-height 3.8821in;  cropleft "0.0372";
%croptop "0.9701";  cropright "0.9375";  cropbottom "0.0205";
%filename 'Sumrate_DF_OB_Symm.eps';file-properties "XNPEU";}}}%
%BeginExpansion
{\includegraphics[
trim=0.205668in 0.079583in 0.345544in 0.116075in,
height=3.0692in,
width=4.184in
]%
{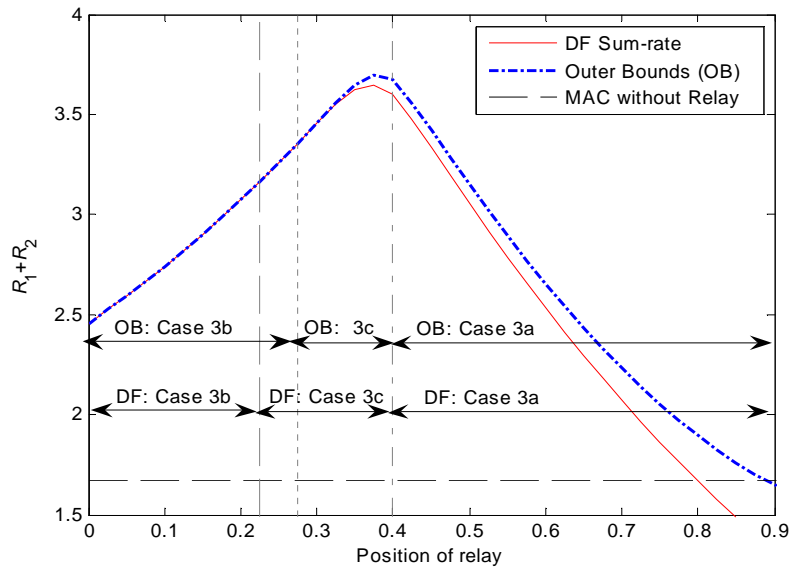}%
}%
%EndExpansion
\caption{Plot of inner and outer bounds on the sum-rate vs. the relay position}\label{Fig_Plot}%
%TCIMACRO{\TeXButton{E}{\end{figure*}}}%
%BeginExpansion
\end{figure*}

Towards illustrating the sum-capacity result, we consider a two-user geometry
shown in Fig. \ref{Illus_Geom}. For this geometry, in Fig. \ref{Fig_Plot} we
plot the inner (DF) and outer cutset bounds on the sum-rate for $\theta=1/2$
as a function of the relay position along the x-axis. As a result of the
symmetric geometry, for every choice of the relay position, both the inner and
outer bounds on the sum-rate are maximized by one of cases $3a$, $3b$, or
$3c$. For each case, we use an iterative algorithm, as described in the
Appendix, to compute the sum-rate maximizing user policies. For cases $3a$ and
$3b,$ the iterative algorithm simplifies to the iterative waterfilling
algorithm developed in \cite{cap_theorems:Yu_Rhee02} in which at each step the
algorithm finds the single-user waterfilling policy for each user while
regarding the signals from the other user as noise. For case $3c$, the optimal
policy at each step is still obtained by regarding the signals from the other
user as noise; however, the user policy at each step is no longer a
waterfilling solution. Finally, the optimality of DF when the sources are
clustered relatively closer to the relay than to the destination is amply
demonstrated in Fig . \ref{Fig_Plot}. The inner and outer bounds are also
compared with the sum-capacity of the fading multiaccess channel without a
relay and $\theta=1$, shown by the dashed line that is a constant independent
of the relay position. Also shown in Fig \ref{Fig_Plot} are the ranges of
relay positions for cases $3a$, $3b,$ and $3c$ for both DF and the cutset bounds.%

%TCIMACRO{\TeXButton{B}{\begin{figure*}[tbp] \centering}}%
%BeginExpansion
%EndExpansion

\section{\label{Sec_End}Concluding Remarks}

We have developed the maximum DF sum-rate and the sum-rate optimal power
policies for an ergodic fading $K$-user half-duplex Gaussian MARC. The MARC is
an example of a multi-terminal network for which the multi-dimensionality of
the policy set, the signal space, and the network topology space contribute to
the complexity of developing capacity results resulting in few, if any, design
rules for real-world communication networks. For a DF relay, the polymatroid
intersection lemma allowed us to simplify the otherwise complicated analysis
of developing the DF sum-rate optimal power policies for the two-user and
$K$-user orthogonal\ MARC and the $K$-user outer bounds. The lemma allowed us
to develop a broad topological classification of fading MARCs into one of
following three types:

i) \textit{partially clustered MARCs} where a subset of all users form a
cluster with the relay while the complementary subset of users form a cluster
with the destination

ii) \textit{clustered MARCs} comprised of either sources-relay or
relay-destination clustered networks, and

iii) \textit{arbitrarily-clustered MARCs} that are a combination of either the
two clustered models or of a clustered and a partially clustered model.

The optimal policies for the inner DF and the outer cutset bounds for the
orthogonal half-duplex MARC model studied here lead to the following observations:

\begin{itemize}
\item that DF achieves the sum-capacity of a class of \textit{source-relay}
\textit{clustered orthogonal MARCs }for which the combined link from all
sources and the relay to the destination, i.e., the link achieving the
$K$-user sum-rate at the destination, is the bottle-neck link. Furthermore, DF
achieves the capacity region when for every weighted sum of user rates, the
limiting bound is the weighted rate-sum achieved at the destination.

\item that for this sum-capacity achieving case, the optimal user policies for
both the orthogonal and non-orthogonal half-duplex MARCs are multi-user
opportunistic waterfilling solutions over their links to the destination and
the optimal relay policy is a water-filling solution over its direct link to
the destination.

\item and that for the remaining classes of MARCs, the optimal users policies
are waterfilling and non-waterfilling solutions for the partially clustered
and arbitrarily clustered models, respectively.
\end{itemize}

For the partially clustered cases, we have showed that the optimal policy for
each user is multiuser waterfilling over its bottle-neck link to one of the
receivers. Thus, the users that are clustered with the destination are forced
to transmit at a lower rate to allow decoding of their signals at the
relatively distant relay. Our results suggest that a useful practical strategy
for the partially clustered topologies may be to allow those distant users
that present little interference at the relay to communicate directly with the destination.

The optimal relaying strategy for all except the capacity achieving clustered
case described above remains open. Given the complexity of finding the optimal
signaling schemes for a given performance metric in multi-terminal networks, a
natural extension to this work could be to understand the gap in spectral
efficiency between DF and the cutset outer bounds for fading MARCs. Such
bounds have been developed recently for time-invariant interference channels
and relay channels in \cite{cap_theorems:Bresler_Tse} and
\cite{cap_theorems:AvestimehrDS_01}, respectively, and for fading Gaussian
broadcast channels with no channel state information at the transmitter in
\cite{cap_theorems:TseYatesLi_01}.

Our analysis can also be extended to study the more general orthogonal
half-duplex MARC model where the sources transmit on both orthogonal channels
while the half-duplex relay is limited to receiving on one and transmitting on
the other. The half-duplex relay receives signals from the sources on one of
the bands while the destination receives it in both bands. For the special
case where the destination can only receive in the band used by the relay to
transmit, we obtain a multiple-access version of the orthogonal relay channel
studied in \cite{cap_theorems:ElGZahedi01}. Irrespective of the receiving
capabilities of the destination, for this more general orthogonal model, each
source transmits two signals, one for each band, subject to an average power
constraint over both bands.

Thus, for this general model, as in \cite{cap_theorems:ElGZahedi01}, one can
consider a more general decoding scheme of \textit{partial decode-and-forward}
(PDF) where each source transmits two independent messages, one on each
orthogonal channel (see also \cite{cap_theorems:SKM02a}). As a result, the
analysis does not simplify to studying an intersection of two polymatroids as
it does for DF. However, analogous to the time-invariant (non-fading) case, we
expect that PDF will simplify to the sum-capacity optimal DF for the special
case in which all the sources and the relay are clustered. While this general
orthogonal model is useful to study for the sake of completeness, the model we
study here abstracts practical multi-hopping architectures and provides
insights into network architectures and topologies where using a
decode-and-forward relay is beneficial.

Finally, a note on complexity: our theoretic analysis distinguishes between
all possible polymatroid intersection cases in determining the optimal policy
for a $K$-user system and therefore has a complexity that grows exponentially
in the number of users. In practice, however, for two intersecting
polymatroids the maximum of a weighted sum of rates and the optimizing
policies can be computed using \textit{strongly polynomial-time} algorithms
\cite[Theorem 47.4]{cap_theorems:Schrijver01}.%

%TCIMACRO{\TeXButton{appendix}{\appendix[Proof of Theorem \ref{DF_Th_OptP}]
%}}%
%BeginExpansion
\appendix[Proof of Theorem \ref{DF_Th_OptP}]
%EndExpansion
{}

\section{\label{App1}Proof of Theorem \ref{DF_Th_OptP}}

The sum-rate maximizing DF power policy $\underline{P}^{\ast}\left(
\underline{H}\right)  $ in Theorem \ref{DF_Th_OptP} is obtained by
sequentially determining the power policies $\underline{P}^{\left(  i\right)
}\left(  \underline{H}\right)  $ and $\underline{P}^{\left(  l,n\right)
}\left(  \underline{H}\right)  $ that maximize the sum-rate for cases $i$ and
$\left(  l,n\right)  $, respectively, over all $\underline{P}\left(
\underline{H}\right)  \in\mathcal{P}$, until one of them satisfies the
conditions for its case. We consider each case separately starting with case
$1$.

\begin{case}
\label{Case_1}This case occurs when the power policy $\underline{P}\left(
\underline{H}\right)  \in\mathcal{B}_{1}$ is such that the intersection of the
relay and destination rate regions belongs to the set of inactive cases (see
Fig \ref{Fig_Case12}). The Lagrangian for this sum-rate maximization is given
by
\begin{equation}
\mathcal{L}^{\left(  1\right)  }=S^{\left(  1\right)  }-\sum\nolimits_{k\in
\mathcal{T}}\nu_{k}\mathbb{E}\left[  P_{k}\left(  \underline{H}\right)
-\overline{P}_{k}\right]  +\sum\nolimits_{k\in\mathcal{T}}\lambda_{k}%
P_{k}\left(  \underline{H}\right)  \label{Case1_sumrate}%
\end{equation}
where, for all $k\in\mathcal{T}$, $\nu_{k}$ are the dual variables associated
with the power constraints in (\ref{GMARC_Pwr_defn}), $\lambda_{k}\geq0$ are
the dual variables associated with the positivity constraints $P_{k}\left(
\underline{H}\right)  \geq0$, and
\begin{multline}
S^{\left(  1\right)  }=R_{\left\{  1\right\}  ,d}\left(  \underline{P}\left(
\underline{H}\right)  \right)  +R_{\left\{  2\right\}  ,r}\left(
\underline{P}\left(  \underline{H}\right)  \right) \\
=\mathbb{E}\left[  \theta C\left(  \left\vert H_{d,1}\right\vert ^{2}\left.
P_{1}\left(  \underline{H}\right)  \right/  \theta\right)  +\overline{\theta
}C\left(  \left\vert H_{d,r}\right\vert ^{2}\left.  P_{r}\left(  \underline
{H}\right)  \right/  \overline{\theta}\right)  \right]  +\mathbb{E}\left[
\theta C\left(  \left\vert H_{r,2}\right\vert ^{2}\left.  P_{2}\left(
\underline{H}\right)  \right/  \theta\right)  \right]  .
\end{multline}
The optimal policy $\underline{P}^{\left(  1\right)  }\left(  \underline
{H}\right)  $ maximizes (\ref{Case1_sumrate}) if it belongs to the open set
$\mathcal{B}_{1}$ defined by the conditions%
\begin{equation}%
\begin{array}
[c]{ccc}%
R_{\left\{  1\right\}  ,d}\left(  \underline{P}^{\left(  1\right)  }\left(
\underline{H}\right)  \right)  <R_{\left\{  1\right\}  ,r}^{\min}\left(
\underline{P}^{\left(  1\right)  }\left(  \underline{H}\right)  \right)  &
\text{and} & R_{\left\{  2\right\}  ,r}\left(  \underline{P}^{\left(
1\right)  }\left(  \underline{H}\right)  \right)  <R_{\left\{  2\right\}
,d}^{\min}\left(  \underline{P}^{\left(  1\right)  }\left(  \underline
{H}\right)  \right)
\end{array}
\label{Case1_Conds}%
\end{equation}
where%
\begin{align}
R_{1,r}^{\min}\left(  \underline{P}\left(  \underline{H}\right)  \right)   &
=\theta I(X_{1};Y_{r}|\underline{H})=\mathbb{E}\left[  \theta C\left(
\frac{\left\vert H_{r,1}\right\vert ^{2}\left.  P_{1}\left(  \underline
{H}\right)  \right/  \theta}{1+\left\vert H_{r,2}\right\vert ^{2}\left.
P_{2}\left(  \underline{H}\right)  \right/  \theta}\right)  \right]
\label{C1_R1rmin}\\
R_{2,d}^{\min}\left(  \underline{P}\left(  \underline{H}\right)  \right)   &
=\theta I\left(  X_{2};Y_{d}|\underline{H}\right)  .=\text{ }\mathbb{E}\left[
\theta C\left(  \frac{\left\vert H_{d,2}\right\vert ^{2}\left.  P_{2}\left(
\underline{H}\right)  \right/  \theta}{1+\left\vert H_{d,1}\right\vert
^{2}\left.  P_{1}\left(  \underline{H}\right)  \right/  \theta}\right)
\right]  . \label{C1_R2dmin}%
\end{align}
The KKT conditions for (\ref{Case1_sumrate}) simplify to
\begin{equation}%
\begin{array}
[c]{cc}%
\frac{\partial\mathcal{L}}{\partial P_{k}\left(  \underline{h}\right)  }%
=f_{k}^{\left(  1\right)  }-\nu_{k}\ln2\leq0, & \text{with equality for }%
P_{k}\left(  \underline{h}\right)  >0\text{, }k=1,2,r
\end{array}
\label{Case1_KKT}%
\end{equation}
where
\begin{align}
&
\begin{array}
[c]{cc}%
f_{k}^{\left(  1\right)  }=\frac{\left\vert h_{m,k}\right\vert ^{2}}{\left(
1+\left\vert h_{m,k}\right\vert ^{2}\left.  P_{k}\left(  \underline{h}\right)
\right/  \theta\right)  } & (k,m)=(1,d),(2,r),\text{ }%
\end{array}
\label{Case1_fk}\\
&
\begin{array}
[c]{cc}%
f_{r}^{\left(  1\right)  }=\frac{\left\vert h_{d,r}\right\vert ^{2}}{\left(
1+\left\vert h_{d,r}\right\vert ^{2}\left.  P_{k}\left(  \underline{h}\right)
\right/  \overline{\theta}\right)  }. &
\end{array}
\label{Case1_fr}%
\end{align}
It is straightforward to verify that these KKT conditions result in%
\begin{equation}%
\begin{array}
[c]{ll}%
P_{k}^{\left(  1\right)  }\left(  \underline{h}\right)  =\left(  \frac{\theta
}{\nu_{k}\ln2}-\frac{\theta}{\left\vert h_{m,k}\right\vert ^{2}}\right)  ^{+}
& (k,m)=(1,d),(2,r)
\end{array}
\label{Case1_OptPolicy}%
\end{equation}
and
\begin{equation}
P_{r}^{\left(  1\right)  }\left(  \underline{h}\right)  =\left(
\frac{\overline{\theta}}{\nu_{r}\ln2}-\frac{\overline{\theta}}{\left\vert
h_{d,r}\right\vert ^{2}}\right)  ^{+}.\text{ } \label{Case1_OptPr}%
\end{equation}
\newline
\end{case}

\begin{case}
With $\nu_{k}$ and $\lambda_{k}$ as the dual variables associated with the
power and positivity constraints on $P_{k}$, respectively, the Lagrangian for
this case is%
\begin{equation}
\mathcal{L}^{\left(  2\right)  }=S^{\left(  2\right)  }-\sum\nolimits_{k\in
\mathcal{T}}\nu_{k}\mathbb{E}\left[  P_{k}\left(  \underline{H}\right)
-\overline{P}_{k}\right]  +\sum\nolimits_{k\in\mathcal{T}}\lambda_{k}%
P_{k}\left(  \underline{H}\right)  \label{Case2_sumrate}%
\end{equation}
where
\begin{align}
S^{\left(  2\right)  }  &  =R_{\left\{  1\right\}  ,r}\left(  \underline
{P}\left(  \underline{H}\right)  \right)  +R_{\left\{  2\right\}  ,d}\left(
\underline{P}\left(  \underline{H}\right)  \right) \\
&  =\mathbb{E}\left[  \theta C\left(  \left\vert H_{r,1}\right\vert
^{2}\left.  P_{1}\left(  \underline{H}\right)  \right/  \theta\right)  +\theta
C\left(  \left\vert H_{d,2}\right\vert ^{2}\left.  P_{2}\left(  \underline
{H}\right)  \right/  \theta\right)  +\overline{\theta}C\left(  \left\vert
H_{d,r}\right\vert ^{2}\left.  P_{r}\left(  \underline{H}\right)  \right/
\overline{\theta}\right)  \right]  .
\end{align}
The optimal policy $\underline{P}^{\left(  2\right)  }\left(  \underline
{H}\right)  $ maximizes (\ref{Case1_sumrate}) if it belongs to the open set
$\mathcal{B}_{2}$ given by the conditions%
\begin{equation}%
\begin{array}
[c]{ccc}%
R_{\left\{  1\right\}  ,r}\left(  \underline{P}\left(  \underline{H}\right)
\right)  <R_{\left\{  1\right\}  ,d}^{\min}\left(  \underline{P}\left(
\underline{H}\right)  \right)  & \text{and} & R_{\left\{  2\right\}
,d}\left(  \underline{P}\left(  \underline{H}\right)  \right)  <R_{\left\{
2\right\}  ,r}^{\min}\left(  \underline{P}\left(  \underline{H}\right)
\right)
\end{array}
\label{Case2_Conds}%
\end{equation}
where $R_{\left\{  2\right\}  ,r}^{\min}$ and $R_{\left\{  1\right\}
,d}^{\min}$ are given by (\ref{C1_R1rmin}) and (\ref{C1_R2dmin}),
respectively, after replacing the user indices $1$ by $2$ and 2 by 1. Note
that $S^{\left(  2\right)  }$ and $\mathcal{L}^{\left(  2\right)  }$ can be
obtained from $S^{\left(  1\right)  }$ and $\mathcal{L}^{\left(  1\right)  }$,
respectively, by interchanging the user indices. Thus, the optimal
$P_{k}^{(2)}\left(  \underline{H}\right)  $ and $P_{r}^{(2)}\left(
\underline{H}\right)  $ are given by (\ref{Case1_OptPolicy}) and
(\ref{Case1_OptPr}), respectively, with $(k,m)=(1,r),(2,d)$ provided
\underline{$P$}$^{\left(  2\right)  }\left(  \underline{H}\right)  $ satisfies
(\ref{Case2_Conds}).
\end{case}

\begin{case}
\label{Case_2}Consider the three cases $3a,$ $3b,$ and $3c$ shown in Fig.
\ref{Fig_Case3abc}. The sum-rate optimization for all three cases is given by
\begin{equation}
\max_{\underline{P}}\min\left(  R_{\mathcal{K},r},R_{\mathcal{K},d}\right)
\label{Case3_sumrate}%
\end{equation}
subject to average power and positivity constraints on $P_{k}$ for all $k$.
Recall that we write $R_{\mathcal{K},j}$ to denote the sum-rate bound at
receiver $j$ where the two bounds at the relay and destination are given by
(\ref{DF_Rates_rel}) and (\ref{DF_Rates_dest}), respectively. We write
$\mathcal{B}_{3}$ to denote the open set consisting of all $\underline
{P}\left(  \underline{H}\right)  \in\mathcal{P}$ that do not satisfy
(\ref{Case1_Conds}) and (\ref{Case2_Conds}) either as strict inequalities,
i.e., do not satisfy the conditions for cases $1$ and $2$, or as a mixture of
equalities and inequalities, where by a mixture we mean that a subset of the
inequalities in (\ref{Case1_Conds}) and (\ref{Case2_Conds}) are satisfied with
equality. We will later show that such sets of mixed equalities and
inequalities in (\ref{Case1_Conds}) and (\ref{Case2_Conds}) corresponds to
conditions for the various boundary cases (see also Figs. \ref{Fig_BC13} and
\ref{Fig_BC23}). Thus, $\underline{P}\left(  \underline{H}\right)
\in\mathcal{B}_{3}$ only when it does not satisfy the conditions for the
inactive and the active-inactive boundary cases. By definition, $\mathcal{B}%
_{3}$ $=$ $\mathcal{B}_{3a}\cup\mathcal{B}_{3b}\cup\mathcal{B}_{3c}$, where
$\mathcal{B}_{i}$, $i=3a,3b,3c,$ is defined for case $i$ below. \newline The
optimization in (\ref{Case3_sumrate}) is a multiuser generalization of the
single-user \textit{max-min} problem studied in
\cite{cap_theorems:Liang_Veeravalli02} (see also \cite[Sec. 3.1]%
{cap_theorems:LiangVP_ResAllocJrnl}) for the orthogonal single-user relay
channel. In \cite{cap_theorems:Liang_Veeravalli02}, the authors use a
technique similar to the minimax detection rule in the two hypothesis testing
problem (see for e.g., \cite[II.C]{cap_theorems:HVPoor01}) to show that the
max-min problem simplifies to optimizing three disjoint cases in which the
maximum rate is achieved either at the relay or at the destination or at both
(\textit{boundary case}). The classical results on minimax optimization also
applies to the multi-user sum-rate optimization in (\ref{Case3_sumrate}), and
thus, the optimal policy $\underline{P}^{\left(  i\right)  }\left(
\underline{H}\right)  $, $i=3a,3b,3c$, satisfies one of following three
conditions%
\begin{align}
&
\begin{array}
[c]{cc}%
\text{\textit{Case 3a}:} & R_{\mathcal{K},r}|_{\underline{P}^{\left(
3a\right)  }\left(  \underline{H}\right)  }<R_{\mathcal{K},d}|_{\underline
{P}^{\left(  3a\right)  }\left(  \underline{H}\right)  }%
\end{array}
\label{Case3aR_Cond}\\
&
\begin{array}
[c]{cc}%
\text{\textit{Case 3b}:} & R_{\mathcal{K},r}|_{\underline{P}^{\left(
3b\right)  }\left(  \underline{H}\right)  }>R_{\mathcal{K},d}|_{\underline
{P}^{\left(  3b\right)  }\left(  \underline{H}\right)  }%
\end{array}
\label{Case3bR_Cond}\\
&
\begin{array}
[c]{cc}%
\text{\textit{Case 3c}:} & R_{\mathcal{K},r}|_{\underline{P}^{\left(
3c\right)  }\left(  \underline{H}\right)  }=R_{\mathcal{K},d}|_{\underline
{P}^{\left(  3c\right)  }\left(  \underline{H}\right)  }.
\end{array}
\label{Case3cR_Cond}%
\end{align}
Note that the conditions in (\ref{Case3aR_Cond})-(\ref{Case3cR_Cond}),
evaluated at any $P\in\mathcal{B}_{3}$, are also conditions defining the sets
$\mathcal{B}_{3a},$ $\mathcal{B}_{3b},$ and $\mathcal{B}_{3c}$, respectively.
Before detailing the optimal solution for each of the above three cases, we
write the Lagrangian $\mathcal{L}^{\left(  i\right)  }$ for case $i$ as%
\begin{align}
&
\begin{array}
[c]{cc}%
\mathcal{L}^{\left(  i\right)  }=S^{\left(  i\right)  }-\sum\limits_{k\in
\mathcal{T}}\nu_{k}\mathbb{E}\left[  P_{k}\left(  \underline{H}\right)
-\overline{P}_{k}\right]  +\sum\limits_{k\in\mathcal{T}}\lambda_{k}%
P_{k}\left(  \underline{H}\right)  & i=3a,3b,3c
\end{array}
\label{Case3Lag}\\
&
\begin{array}
[c]{cc}
& \text{ \ \ \ \ \ }\lambda_{k}P_{k}\left(  \underline{H}\right)  \geq0
\end{array}
\label{Case3_lambda}%
\end{align}
where $\nu_{k}$ and $\lambda_{k}\geq0$ are dual variables associated with the
average power and positivity constraints on $P_{k}$, respectively,
\begin{equation}
S^{\left(  i\right)  }=\left\{
\begin{array}
[c]{ll}%
R_{\mathcal{K},r}=\mathbb{E}\left[  C\left(  \sum\limits_{k=1}^{2}\left\vert
H_{r,k}\right\vert ^{2}\left.  P_{k}\left(  \underline{H}\right)  \right/
\theta\right)  \right]  & i=3a\\
R_{\mathcal{K},d}=\mathbb{E}\left[  C\left(  \sum\limits_{k=1}^{2}\left\vert
H_{d,k}\right\vert ^{2}\left.  P_{k}\left(  \underline{H}\right)  \right/
\theta\right)  \right]  & i=3b\\
\left(  1-\alpha\right)  R_{\mathcal{K},r}+\alpha R_{\mathcal{K},d} & i=3c,
\end{array}
\right.  \label{Case3_Sdef}%
\end{equation}
and $\alpha$ is the dual variable associated with the equality (boundary)
condition in (\ref{Case3cR_Cond}). The resulting KKT conditions are given by%
\begin{equation}%
\begin{array}
[c]{cc}%
\frac{\partial\mathcal{L}}{\partial P_{k}\left(  \underline{h}\right)  }%
=F_{k}^{\left(  i\right)  }=f_{k}^{\left(  i\right)  }-\nu_{k}\ln2\leq0, &
k=1,2,r,i=3a,3b,3c
\end{array}
\label{Cases3_KKT}%
\end{equation}
where (\ref{Cases3_KKT}) holds with equality for $P_{k}\left(  \underline
{H}\right)  >0$ and for $k=1,2,$
\begin{equation}
f_{k}^{\left(  i\right)  }=\left\{
\begin{array}
[c]{ll}%
\left\vert h_{r,k}\right\vert ^{2}\left/  \left(  1+\sum\limits_{k=1}%
^{2}\left\vert h_{r,k}\right\vert ^{2}\left.  P_{k}\left(  \underline
{h}\right)  \right/  \theta\right)  \right.  & i=3a\\
\left\vert h_{d,k}\right\vert ^{2}\left/  \left(  1+\sum\limits_{k=1}%
^{2}\left\vert h_{d,k}\right\vert ^{2}\left.  P_{k}\left(  \underline
{h}\right)  \right/  \theta\right)  \right.  & i=3b\\
\left(  1-\alpha\right)  f_{k}^{\left(  3a\right)  }+\alpha f_{k}^{\left(
3b\right)  } & i=3c.
\end{array}
\right.  \label{Case3_fdef}%
\end{equation}
and $f_{r}^{\left(  3a\right)  }=f_{r}^{\left(  3b\right)  }=f_{r}^{\left(
1\right)  }$, $f_{r}^{\left(  3c\right)  }=\alpha f_{r}^{\left(  1\right)  }$.
We now present the optimal policies and sum-rates for each case in detail.
\newline\textit{Case 3a}: For this case, the KKT conditions in
(\ref{Cases3_KKT}) and (\ref{Case3_fdef}) depend only the sum-rate and
channels gains of the two users at the relay. Thus, the problem simplifies to
that for a MAC channel at the relay and the classic multiuser waterfilling
solution developed in \cite{cap_theorems:Knopp_Humblet,cap_theorems:TH01}
applies. From (\ref{Cases3_KKT}), the optimal user policies are
\begin{equation}%
\begin{array}
[c]{ll}%
\frac{\left\vert h_{r,1}\right\vert ^{2}}{v_{1}}>\frac{\left\vert
h_{r,2}\right\vert ^{2}}{\nu_{2}} & P_{1}^{\left(  3a\right)  }\left(
\underline{h}\right)  =\left(  \frac{\theta}{\nu_{1}\ln2}-\frac{\theta
}{\left\vert h_{r,1}\right\vert ^{2}}\right)  ^{+},P_{2}^{\left(  3a\right)
}=0\\
\frac{\left\vert h_{r,1}\right\vert ^{2}}{v_{1}}<\frac{\left\vert
h_{r,2}\right\vert ^{2}}{\nu_{2}} & P_{1}^{\left(  3a\right)  }\left(
\underline{h}\right)  =0,P_{2}^{\left(  3a\right)  }=\left(  \frac{\theta}%
{\nu_{2}\ln2}-\frac{\theta}{\left\vert h_{r,2}\right\vert ^{2}}\right)  ^{+}\\
\frac{\left\vert h_{r,1}\right\vert ^{2}}{v_{1}}=\frac{\left\vert
h_{r,2}\right\vert ^{2}}{\nu_{2}} & \left\vert h_{r,1}\right\vert ^{2}%
P_{1}^{\left(  3a\right)  }\left(  \underline{h}\right)  +\left\vert
h_{r,2}\right\vert ^{2}P_{2}^{\left(  3a\right)  }\left(  \underline
{h}\right)  =\theta\left(  \frac{\left\vert h_{r,1}\right\vert ^{2}}{\nu
_{1}\ln2}-1\right)  ^{+}.
\end{array}
\label{Case3a_optP}%
\end{equation}
With the exception of the equality condition in (\ref{Case3a_optP}), the
optimal policies are unique, i.e., the optimal $P_{k}^{\left(  3a\right)
}\left(  \underline{H}\right)  $ at user $k$ in (\ref{Case3a_optP}) is an
opportunistic water-filling solution that exploits the fading diversity in a
multiaccess channel from the sources to the relay. If the channel gains are
jointly distributed with a continuous density, the equality condition occurs
with probability 0. Furthermore, even if the distributions were not
continuous, one can choose to schedule one user or the other when the equality
condition is met, thereby maintaining the opportunistic allocation policy.
Finally, the optimal power policy at the relay is not explicitly obtained from
$\mathcal{L}^{\left(  1\right)  }$ in (\ref{Case3Lag}) as for this case
$S^{\left(  1\right)  }$ is the sum-rate achieved by the sources at the relay.
However, since the sum-rate at the relay for this case is smaller than that at
the destination, choosing the optimal waterfilling policy at the relay that
maximizes the relay-destination link preserves the condition for this case,
and thus, $P_{r}^{(3a)}\left(  \underline{H}\right)  $ is given by
(\ref{Case1_OptPr}). When $P^{\left(  3a\right)  }\left(  \underline
{H}\right)  \in\mathcal{B}_{3}$, the requirement of satisfying
(\ref{Case3aR_Cond}), i.e., \underline{$P$}$^{\left(  3a\right)  }\left(
\underline{H}\right)  $ $\in$ $\mathcal{B}_{3a}$, simplifies to a threshold
condition $\overline{P}_{r}$ $>$ $P_{u}\left(  \overline{P}_{1},\overline
{P}_{2}\right)  $ where $\overline{P}_{k}$, $k\in\mathcal{T}$, is defined in
(\ref{GMARC_Pwr_defn}) and the threshold $P_{u}\left(  \overline{P}%
_{1},\overline{P}_{2}\right)  $ is obtained by setting (\ref{Case3aR_Cond}) to
an equality. When \underline{$P$}$^{\left(  3a\right)  }\left(  \underline
{H}\right)  $ $\in$ $\mathcal{B}_{3}$ but \underline{$P$}$^{\left(  3a\right)
}\left(  \underline{H}\right)  $ $\not \in $ $\mathcal{B}_{3a}$, $R_{1}+R_{2}$
is maximized by either \textit{case 3b} or \textit{case 3c}. For
$\underline{P}^{\left(  3a\right)  }\left(  \underline{H}\right)  $
$\not \in $ $\mathcal{B}_{3}$, as argued in Section \ref{Sec_4}, the sum-rate
is not maximized by any $\underline{P}\left(  \underline{H}\right)  $ $\in$
$\mathcal{B}_{3}$. \newline\textit{Case 3b} : The optimal policy
$P_{k}^{\left(  3b\right)  }\left(  \underline{H}\right)  $ at user $k$ for
this case satisfies the KKT conditions in (\ref{Cases3_KKT}) with
$f_{k}^{\left(  i\right)  }=f_{k}^{\left(  3b\right)  }$ in (\ref{Case3_fdef}%
). As with case $3a$, here too, the optimal policy is an opportunistic
water-filling solution and is given by (\ref{Case3a_optP}) with the subscript
`$r$' changed to `$d$' for all $k$ and with the superscript $i=3b$. Further,
for the relay node, the optimal $P_{r}^{\left(  3b\right)  }\left(  H\right)
$ satisfies the KKT conditions in (\ref{Case1_KKT}), i.e., $f_{r}^{\left(
3b\right)  }=f_{r}^{\left(  1\right)  }$, and is given by the water-filling
solution in (\ref{Case1_OptPr}). Finally, for \underline{$P$}$^{\left(
3b\right)  }\left(  \underline{H}\right)  $ $\in$ $\mathcal{B}_{3}$, the
requirement \underline{$P$}$^{\left(  3b\right)  }\left(  \underline
{H}\right)  $ $\in$ $\mathcal{B}_{3b}$ simplifies to satisfying the threshold
condition $\overline{P}_{r}$ $<$ $P_{l}\left(  \overline{P}_{1},\overline
{P}_{2}\right)  $ where $P_{l}\left(  \overline{P}_{1},\overline{P}%
_{2}\right)  $ is determined by setting (\ref{Case3bR_Cond}) to an equality.
\newline\textit{Case 3c }(\textit{equal-rate policy}): The optimal policy
$P_{k}^{\left(  3c\right)  }\left(  \underline{H}\right)  $ at user $k$ for
this case satisfies the KKT conditions in (\ref{Cases3_KKT}) for
$f_{k}^{\left(  i\right)  }=f_{k}^{\left(  3c\right)  }$ in (\ref{Case3_fdef}%
). The function $f_{k}^{\left(  3c\right)  }$ in (\ref{Case3_fdef}) is a
weighted sum of $f_{k}^{\left(  3a\right)  }$ and $f_{k}^{\left(  3b\right)
}$ where the Lagrange multiplier $\alpha$ accounts for the boundary condition
in (\ref{Case3cR_Cond}). Substituting $f_{k}^{(3c)}$ in (\ref{Case3_fdef}) in
(\ref{Cases3_KKT}), we have the following KKT conditions%
\begin{equation}%
\begin{array}
[c]{cc}%
\frac{\alpha\left\vert h_{r,k}\right\vert ^{2}}{1+\sum\limits_{k=1}%
^{2}\left\vert h_{r,k}\right\vert ^{2}\frac{P_{k}\left(  \underline{h}\right)
}{\theta}}+\frac{\left(  1-\alpha\right)  \left\vert h_{d,k}\right\vert ^{2}%
}{1+\sum\limits_{k=1}^{2}\left\vert h_{d,k}\right\vert ^{2}\frac{P_{k}\left(
\underline{h}\right)  }{\theta}}\leq\nu_{k}\ln2 & \text{with equality for
}P_{k}\left(  \underline{h}\right)  >0\text{, }k=1,2
\end{array}
\label{C3cKKT}%
\end{equation}
which implies
\begin{equation}%
\begin{array}
[c]{ll}%
f_{1}^{\left(  3c\right)  }/\nu_{1}>f_{2}^{\left(  3c\right)  }/\nu_{2} &
P_{1}^{\left(  3c\right)  }\left(  \underline{h}\right)  =\left(  \text{root
of }F_{1}^{\left(  3c\right)  }|_{P_{2}=0}\right)  ^{+},P_{2}^{\left(
3c\right)  }\left(  \underline{h}\right)  =0\\
f_{1}^{\left(  3c\right)  }/\nu_{1}<f_{2}^{\left(  3c\right)  }/\nu_{2} &
P_{1}^{\left(  3c\right)  }\left(  \underline{h}\right)  =0,P_{2}^{\left(
3c\right)  }\left(  \underline{h}\right)  =\left(  \text{root of }%
F_{2}^{\left(  3c\right)  }|_{P_{1}=0}\right)  ^{+}\\
f_{1}^{\left(  3c\right)  }/\nu_{1}=f_{2}^{\left(  3c\right)  }/\nu_{2} &
P_{1}^{\left(  3c\right)  }\left(  \underline{h}\right)  \text{ and }%
P_{2}^{\left(  3c\right)  }\left(  \underline{h}\right)  \text{ satisfy }%
f_{k}^{\left(  3c\right)  }=\nu_{k}\ln2
\end{array}
\label{Case3c_OptP}%
\end{equation}
where $F_{k}^{\left(  3c\right)  }$ is defined in (\ref{Cases3_KKT}).
Determining the optimal $P_{k}^{\left(  3c\right)  }\left(  \underline
{h}\right)  $, $k=1,2,$ requires verifying each one of the three conditions in
(\ref{Case3c_OptP}). Note that in contrast to case $3a$ (and case $3b$ with
`$r$' replaced in (\ref{Case3a_optP}) by `$d$'), the opportunistic scheduling
in (\ref{Case3c_OptP}) also depends on the user policies in addition to the
channel states.\ Furthermore, the optimal solutions $P_{k}^{\left(  3c\right)
}\left(  \underline{H}\right)  $ do not take a water-filling form. Thus, for a
given $P_{1}\left(  \underline{h}\right)  $, $P_{2}\left(  \underline
{h}\right)  $ is given by%
\begin{equation}
P_{2}\left(  \underline{h}\right)  =\text{positive root }x\text{ of
(\ref{C3c_Eqn}) if it exists, otherwise }0
\end{equation}
where the root $x$ is determined by the following equation:
\begin{equation}
\frac{\alpha\left\vert h_{r,2}\right\vert ^{2}}{1+\left\vert h_{r,1}%
\right\vert ^{2}\frac{P_{1}\left(  \underline{h}\right)  }{\theta}+\left\vert
h_{r,2}\right\vert ^{2}\frac{x}{\theta}}+\frac{\left(  1-\alpha\right)
\left\vert h_{d,2}\right\vert ^{2}}{1+\left\vert h_{d,k}\right\vert ^{2}%
\frac{P_{1}\left(  \underline{h}\right)  }{\theta}+\left\vert h_{d,k}%
\right\vert ^{2}\frac{x}{\theta}}=\nu_{2}\ln2. \label{C3c_Eqn}%
\end{equation}
Using $P_{2}\left(  \underline{h}\right)  $ given by (\ref{C3c_Eqn}),
$P_{1}\left(  \underline{h}\right)  $ is obtained as the root of
\begin{equation}
\frac{\alpha\left\vert h_{r,1}\right\vert ^{2}}{1+\left\vert h_{r,1}%
\right\vert ^{2}\frac{P_{1}\left(  \underline{h}\right)  }{\theta}+\left\vert
h_{r,2}\right\vert ^{2}\frac{P_{2}\left(  \underline{h}\right)  }{\theta}%
}+\frac{\left(  1-\alpha\right)  \left\vert h_{d,1}\right\vert ^{2}%
}{1+\left\vert h_{d,k}\right\vert ^{2}\frac{P_{1}\left(  \underline{h}\right)
}{\theta}+\left\vert h_{d,k}\right\vert ^{2}\frac{P_{2}\left(  \underline
{h}\right)  }{\theta}}=\nu_{1}\ln2.
\end{equation}
Thus, for all $\underline{h}$, starting with an initial $P_{1}\left(
\underline{h}\right)  $, we iteratively obtain $P_{1}\left(  \underline
{h}\right)  $ and $P_{2}\left(  \underline{h}\right)  $ until they converge to
$P_{1}^{\left(  3c\right)  }\left(  \underline{H}\right)  $ and $P_{2}%
^{\left(  3c\right)  }\left(  \underline{H}\right)  $. The proof of
convergence is detailed below. Finally, the optimal policies are determined
over all $\alpha\in\lbrack0,1]$ to find an $\alpha^{\ast}$ that satisfies the
equal rate condition in (\ref{Case3cR_Cond}).

\textit{Proof of Convergence}:\ The proof follows along the same lines as that
detailed in \cite[p. 3440]{cap_theorems:LiangVP_ResAllocJrnl} and relies on
the fact that the maximizing function $S^{\left(  3c\right)  }$ in
(\ref{Case3_Sdef}) is a strictly concave function of $P_{1}\left(
\underline{H}\right)  $ and $P_{2}\left(  \underline{H}\right)  $ and is
bounded from above because of the power constraints at the source and relay
nodes. In each iteration, the optimal $P_{1}\left(  \underline{H}\right)  $
and $P_{2}\left(  \underline{H}\right)  $ are the KKT solutions that maximize
the objective function. Thus, after each iteration, the objective function
either increases or remains the same. It is easy to check that for a given
$P_{1}\left(  \underline{H}\right)  $ the objective function is a strictly
concave function of $P_{2}\left(  \underline{H}\right)  $, and thus,
(\ref{C3c_Eqn}) yields a unique value of $P_{2}\left(  \underline{H}\right)
$. Furthermore, the objective function is also a strictly concave function of
$P_{1}\left(  \underline{H}\right)  $ for a fixed $P_{2}\left(  \underline
{H}\right)  $. Thus, as the objective function converges, $\left(
P_{1}\left(  \underline{H}\right)  ,P_{2}\left(  \underline{H}\right)
\right)  $ also converge. Finally, $P_{1}\left(  \underline{H}\right)  $ and
$P_{2}\left(  \underline{H}\right)  $ converge to the solutions of the KKT
conditions, which is sufficient for $\left(  P_{1}\left(  \underline
{H}\right)  ,P_{2}\left(  \underline{H}\right)  \right)  $ to be optimal since
the objective function is concave over all \underline{$P$}$\left(
\underline{H}\right)  \in\mathcal{P}$.

Finally, since $f_{r}^{\left(  3c\right)  }=\alpha f_{r}^{\left(  1\right)  }%
$, the relay's optimal policy simplifies to the water-filling solution given
by
\begin{equation}
\text{ }P_{r}^{\left(  3c\right)  }\left(  \underline{H}\right)  =\left(
\frac{\alpha\overline{\theta}}{\nu_{r}\ln2}-\frac{\overline{\theta}%
}{\left\vert h_{d,r}\right\vert ^{2}}\right)  ^{+}.\text{ }
\label{Case3c_OptPr}%
\end{equation}

\end{case}

\begin{case}
(\textit{Boundary Cases)}: Recall that we define the sets $B_{i}$,
$i=1,2,3a,3b,3c$, as open sets to ensure that an optimal \underline{$P$%
}$^{\ast}$ maximizes the sum-rate for a case only if it satisfies the
conditions for that case. Since an optimal policy can lie on the boundary of
any two such cases, we also consider six additional cases that lie at the
boundary of an inactive and an active case. These boundary cases result when
the conditions for an inactive case $l$, $l=1,2$, and an active case $n$,
$n=3a,3b,3c$, are such that the sum-rate is the same for both cases. We
consider each of the six boundary cases separately and develop the optimal
\underline{$P$}$^{\left(  l,n\right)  }\left(  \underline{H}\right)  $ for
each case. The requirement that the optimal \underline{$P$}$^{\left(
l,n\right)  }\left(  \underline{H}\right)  $ satisfies the condition
$S^{\left(  l\right)  }=S^{\left(  n\right)  }$ for the boundary case $\left(
l,n\right)  $ simplifies to%
\begin{align}
&
\begin{array}
[c]{ll}%
\text{case }\left(  1,3a\right)  & R_{\left\{  1\right\}  ,d}+R_{\left\{
2\right\}  ,r}=R_{\mathcal{K},r}<R_{\mathcal{K},d}%
\end{array}
\label{C13a_Cond}\\
&
\begin{array}
[c]{ll}%
\text{case }\left(  1,3b\right)  & R_{\left\{  1\right\}  ,d}+R_{\left\{
2\right\}  ,r}=R_{\mathcal{K},d}<R_{\mathcal{K},r}%
\end{array}
\label{C13b_Cond}\\
&
\begin{array}
[c]{ll}%
\text{case }\left(  1,3c\right)  & R_{\left\{  1\right\}  ,d}+R_{\left\{
2\right\}  ,r}=R_{\mathcal{K},r}=R_{\mathcal{K},d}%
\end{array}
\label{C13c_Cond}\\
&
\begin{array}
[c]{ll}%
\text{case }\left(  2,3a\right)  & R_{\left\{  1\right\}  ,r}+R_{\left\{
2\right\}  ,d}=R_{\mathcal{K},r}<R_{\mathcal{K},d}%
\end{array}
\label{C23a_Cond}\\
&
\begin{array}
[c]{ll}%
\text{case }\left(  2,3b\right)  & R_{\left\{  1\right\}  ,r}+R_{\left\{
2\right\}  ,d}=R_{\mathcal{K},d}<R_{\mathcal{K},r}%
\end{array}
\label{C23b_Cond}\\
&
\begin{array}
[c]{ll}%
\text{case }\left(  2,3c\right)  & R_{\left\{  1\right\}  ,r}+R_{\left\{
2\right\}  ,d}=R_{\mathcal{K},d}=R_{\mathcal{K},r}%
\end{array}
\label{C23c_Cond}%
\end{align}
where the conditions in (\ref{C13a_Cond})-(\ref{C23c_Cond}) are evaluated at
the appropriate \underline{$P$}$^{\left(  l,n\right)  }\left(  \underline
{H}\right)  $. Note that the conditions in (\ref{C13a_Cond})-(\ref{C23c_Cond})
also define the conditions for the sets $\mathcal{B}_{\left(  1,3a\right)  }$
through $\mathcal{B}_{\left(  2,3c\right)  }$, respectively. Using
(\ref{C13a_Cond})-(\ref{C23c_Cond}), we write the Lagrangian for all boundary
cases except cases $\left(  1,3c\right)  $ and $\left(  2,3c\right)  $ as
\begin{align}
&
\begin{array}
[c]{cc}%
\mathcal{L}^{\left(  l,n\right)  }=\alpha S^{\left(  l\right)  }+\left(
1-\alpha\right)  S^{\left(  n\right)  }-\sum\limits_{k\in\mathcal{T}}\nu
_{k}\mathbb{E}\left[  P_{k}\left(  \underline{H}\right)  -\overline{P}%
_{k}\right]  +\sum\limits_{k\in\mathcal{T}}\lambda_{k}P_{k}\left(
\underline{H}\right)  & l=1,2,n=3a,3b
\end{array}
\\
&  \lambda_{k}P_{k}\left(  \underline{H}\right)  \geq0
\end{align}
and the Lagrangian for cases $\left(  1,3c\right)  $ and $\left(  2,3c\right)
$ as
\begin{align}
&  \mathcal{L}^{\left(  l,3c\right)  }=\alpha_{1}S^{\left(  l\right)  }%
+\alpha_{2}S^{\left(  3a\right)  }+\left(  1-\alpha_{1}-\alpha_{2}\right)
S^{\left(  3b\right)  }-\sum\limits_{k\in\mathcal{T}}\nu_{k}\mathbb{E}\left[
P_{k}\left(  \underline{H}\right)  -\overline{P}_{k}\right] \nonumber\\
&
\begin{array}
[c]{ccc}
& \text{ \ \ \ \ \ }+\sum\limits_{k\in\mathcal{T}}\lambda_{k}P_{k}\left(
\underline{H}\right)  , & l=1,2
\end{array}
\\
&  \lambda_{k}P_{k}\left(  \underline{H}\right)  \geq0
\end{align}
where $\nu_{k}$ and $\lambda_{k}\geq0$ are dual variables associated with the
average power and positivity constraints on $P_{k}$, respectively. The
variable $\alpha$ is the dual variable associated with all boundary cases with
a single boundary condition while $\alpha_{1}$ and $\alpha_{2}$ are the dual
variables associated with cases $\left(  1,3c\right)  $ and $\left(
2,3c\right)  $. The resulting KKT conditions, one for each $P_{k}\left(
\underline{h}\right)  $, $k=1,2,r,$ are
\begin{align}
&
\begin{array}
[c]{cc}%
\text{Case }\left(  l,n\not =3c\right)  : & \frac{\partial\mathcal{L}^{\left(
l,n\right)  }}{\partial P_{k}\left(  \underline{h}\right)  }=f_{k}^{\left(
l,n\right)  }=\alpha f_{k}^{\left(  l\right)  }+\left(  1-\alpha\right)
f_{k}^{\left(  n\right)  }\leq\nu_{k}\ln2
\end{array}
\label{CBC_KKT}\\
&
\begin{array}
[c]{cc}%
\text{Case }\left(  l,n=3c\right)  : & \frac{\partial\mathcal{L}^{\left(
l,n\right)  }}{\partial P_{k}\left(  \underline{h}\right)  }=f_{k}^{\left(
l,n\right)  }=\alpha_{1}f_{k}^{\left(  l\right)  }+\alpha_{2}f_{k}^{\left(
3a\right)  }+\left(  1-\alpha_{1}-\alpha_{2}\right)  f_{k}^{\left(  3b\right)
}\leq\nu_{k}\ln2
\end{array}
\label{C3cK}%
\end{align}
where $f_{k}^{\left(  l\right)  }$ and $f_{k}^{\left(  n\right)  }$ are as
defined earlier for cases $l$ and $n$ and equality holds in (\ref{CBC_KKT})
and (\ref{C3cK}) for $P_{k}\left(  \underline{h}\right)  >0$, for all
$\underline{h}$. We now present the optimal policies for each case separately.
\newline\textit{Case }$\left(  1,3a\right)  $: From (\ref{CBC_KKT}), the KKT
conditions for this case are
\begin{align}
&
\begin{array}
[c]{cc}%
f_{1}^{\left(  1,3a\right)  }=\frac{\alpha\left\vert h_{d,1}\right\vert ^{2}%
}{1+\left\vert h_{d,1}\right\vert ^{2}P_{1}\left(  \underline{h}\right)
/\theta}+\frac{\left(  1-\alpha\right)  \left\vert h_{r,1}\right\vert ^{2}%
}{1+\sum\nolimits_{j=1}^{2}\left\vert h_{r,j}\right\vert ^{2}P_{j}\left(
\underline{h}\right)  /\theta}\leq\nu_{1}\ln2 & \text{with equality if }%
P_{1}\left(  \underline{h}\right)  >0
\end{array}
\label{C13a_g1}\\
&
\begin{array}
[c]{cc}%
f_{2}^{\left(  1,3a\right)  }=\frac{\alpha\left\vert h_{r,2}\right\vert ^{2}%
}{1+\left\vert h_{r,2}\right\vert ^{2}P_{2}\left(  \underline{h}\right)
/\theta}+\frac{\left(  1-\alpha\right)  \left\vert h_{r,2}\right\vert ^{2}%
}{1+\sum\nolimits_{j=1}^{2}\left\vert h_{r,j}\right\vert ^{2}P_{j}\left(
\underline{h}\right)  /\theta}\leq\nu_{2}\ln2 & \text{with equality if }%
P_{2}\left(  \underline{h}\right)  >0
\end{array}
\label{C13a_g2}\\
&
\begin{array}
[c]{cc}%
f_{r}^{\left(  1,3a\right)  }=\frac{\alpha\left\vert h_{d,r}\right\vert ^{2}%
}{1+\left\vert h_{d,r}\right\vert ^{2}P_{r}\left(  \underline{h}\right)
/\overline{\theta}}\leq\nu_{r}\ln2 & \text{with equality if }P_{r}\left(
\underline{h}\right)  >0
\end{array}
\label{C13a_gr}%
\end{align}
which implies
\begin{equation}%
\begin{array}
[c]{ll}%
\frac{f_{1}^{\left(  1,3a\right)  }}{\nu_{1}}>\frac{f_{2}^{\left(
1,3a\right)  }}{\nu_{2}} & P_{1}\left(  \underline{h}\right)  =\left(
\text{root of }F_{1}^{\left(  1,3a\right)  }|_{P_{2}=0}\right)  ^{+}%
,P_{2}\left(  \underline{h}\right)  =0\\
\frac{f_{1}^{\left(  1,3a\right)  }}{\nu_{1}}<\frac{f_{2}^{\left(
1,3a\right)  }}{\nu_{2}} & P_{1}\left(  \underline{h}\right)  =0,P_{2}\left(
\underline{h}\right)  =\left(  \text{root of }F_{2}^{\left(  1,3a\right)
}|_{P_{1}=0}\right)  ^{+}\\
\frac{f_{1}^{\left(  1,3a\right)  }}{\nu_{1}}=\frac{f_{2}^{\left(
1,3a\right)  }}{\nu_{2}} & P_{1}\left(  \underline{h}\right)  \text{ and
}P_{2}\left(  \underline{h}\right)  \text{ satisfy }\frac{f_{1}^{\left(
1,3a\right)  }}{\nu_{1}}=\frac{f_{2}^{\left(  1,3a\right)  }}{\nu_{2}}%
\end{array}
\label{CB13aKKT}%
\end{equation}
where
\begin{equation}%
\begin{array}
[c]{cc}%
F_{k}^{\left(  l,n\right)  }=f_{k}^{\left(  l,n\right)  }-\nu_{k}\ln2\leq0, &
\text{for all }\left(  l,n\right)  .
\end{array}
\end{equation}
As in case $3c$, the optimal policies take an opportunistic non-waterfilling
form and in fact can be obtained by the iterative algorithm described for that
case. Finally, from (\ref{C13a_gr}), the optimal $P_{r}^{\left(  1,3a\right)
}\left(  \underline{H}\right)  $ is given by (\ref{Case3c_OptPr}).
\newline\textit{Case }$\left(  1,3b\right)  $: The analysis for this case
mirrors that for case $\left(  1,3a\right)  $ and the optimal user policies
are opportunistic non-water-filling solution given by (\ref{CB13aKKT}) with
$f_{k}^{\left(  3a\right)  }$ replaced by $f_{k}^{\left(  3b\right)  }$,
$k=1,2$. On the other hand in contrast to case $\left(  1,3a\right)  $ where
$f_{r}^{\left(  3a\right)  }=0$, since both $f_{r}^{\left(  1\right)  }$ and
$f_{r}^{\left(  3b\right)  }$ are non-zero, the optimal relay policy
$P_{r}^{\left(  2,3a\right)  }=$ $P_{r}^{\left(  1\right)  }$. \newline%
\textit{Case }$\left(  1,3c\right)  $: For this case, the KKT conditions in
(\ref{C3cK}) involves a weighted sum of $f_{k}^{\left(  l\right)  }$,
$f_{k}^{\left(  3a\right)  }$, and $f_{k}^{\left(  3b\right)  }$. Thus, for
$k=1,2$, $(k,m)=(1,d),(2,r)$, we have the KKT conditions
\begin{align}
&
\begin{array}
[c]{cc}%
f_{1}^{\left(  1,3c\right)  }=\alpha_{1}f_{1}^{\left(  1\right)  }+\alpha
_{2}f_{1}^{\left(  3a\right)  }+\left(  1-\alpha_{1}-\alpha_{2}\right)
f_{1}^{\left(  3b\right)  }\leq\nu_{1}\ln2 & \text{with equality if }%
P_{1}\left(  \underline{h}\right)  >0
\end{array}
\label{C13cP1}\\
&
\begin{array}
[c]{cc}%
f_{2}^{\left(  1,3c\right)  }=\alpha_{1}f_{2}^{\left(  1\right)  }+\alpha
_{2}f_{2}^{\left(  3a\right)  }+\left(  1-\beta\right)  \left(  1-\alpha
\right)  f_{2}^{\left(  3b\right)  }\leq\nu_{2}\ln2 & \text{with equality if
}P_{2}\left(  \underline{h}\right)  >0
\end{array}
\label{C13cP2}\\
&
\begin{array}
[c]{cc}%
f_{r}^{\left(  1,3c\right)  }=\left(  1-\alpha_{2}\right)  \left\vert
h_{d,r}\right\vert ^{2}\left/  C\left(  \left\vert h_{d,r}\right\vert
^{2}P_{r}/\overline{\theta}\right)  \right.  \leq\nu_{r}\ln2 & \text{with
equality if }P_{r}\left(  \underline{h}\right)  >0
\end{array}
\end{align}
where $\alpha_{1}$ and $\alpha_{2}$ are the dual variables associated with the
equalities $R_{\mathcal{K},d}=R_{\left\{  1\right\}  ,d}+R_{\left\{
2\right\}  ,r}$ and $R_{\mathcal{K},d}=R_{\mathcal{K},r}$, respectively, in
(\ref{C13c_Cond}). From (\ref{C13cP1}) and (\ref{C13cP2}), one can verify that
the optimal user policies are opportunistic non-waterfilling solutions given
by (\ref{CB13aKKT}) with the superscript $\left(  1,3a\right)  $ replaced by
$\left(  1,3c\right)  $. Finally, $P_{r}^{\left(  1,3c\right)  }\left(
\underline{H}\right)  $ is given by the water-filling solution in
(\ref{Case1_OptPr}) with $\alpha$ replaced by $\left(  1-\alpha_{2}\right)  $.
\newline\textit{Case }$\left(  2,3a\right)  $: The optimal user policies for
this case and the KKT conditions they satisfy are given by (\ref{C13a_g1}),
(\ref{C13a_g2}), and (\ref{CB13aKKT}) when $f_{k}^{\left(  1\right)  }$ is
replaced by $f^{\left(  2\right)  }$, for all $k$, and $g_{k}^{\left(
\cdot\right)  }$ is superscripted by $\left(  2,3a\right)  $. Thus, here too,
the optimal user policies are opportunistic non-water-filling solutions. The
optimal relay policy $P_{r}^{\left(  2,3a\right)  }\left(  \underline
{H}\right)  $ is the same as that obtained in case $\left(  1,3a\right)  $.
\newline\textit{Case }$\left(  2,3b\right)  $: The optimal user policies
$P_{k}^{\left(  2,3b\right)  }\left(  \underline{H}\right)  $, $k=1,2,$ are
again opportunistic non-water-filling solutions and are given by
(\ref{C13a_g1}), (\ref{C13a_g2}), and (\ref{CB13aKKT}) when $f_{k}^{\left(
1\right)  }$ and $f_{k}^{\left(  3a\right)  }$ are replaced by $f^{\left(
2\right)  }$ and $f^{\left(  3b\right)  }$, respectively, for all $k$, and
$g_{k}^{\left(  \cdot\right)  }$ is superscripted by $\left(  2,3b\right)  $.
The optimal relay policy $P_{r}^{\left(  2,3b\right)  }\left(  \underline
{H}\right)  $ is the same as that for case $\left(  1,3b\right)  $.
\newline\textit{Case }$\left(  2,3c\right)  $: The optimal policy vector
\underline{$P$}$^{(2,3c)}\left(  \underline{H}\right)  $ is the same as that
for \textit{case} $\left(  1,3c\right)  $ with $f_{k}^{\left(  1\right)  }$ is
replaced by $f^{\left(  2\right)  }$, for all $k$, and with the superscript
$\left(  2,3c\right)  $.
\end{case}

\bibliographystyle{IEEEtran}
\bibliography{MARC_refs}

\end{document}